\documentclass[manuscript,longauthor]{aastex63}

\usepackage{upgreek}
\usepackage{amsmath}
\usepackage{graphicx}
\usepackage{placeins}
\usepackage{multirow}
\usepackage{parskip}
\usepackage{lineno}

\shorttitle{Gemini-TEXES observations of Jupiter's mid-infrared aurora}
\shortauthors{Sinclair et al.}

\begin{document}

\title{A high spatial and spectral resolution study of Jupiter's mid-infrared auroral emissions and their response to a solar wind compression}

\correspondingauthor{James A. Sinclair}
\email{james.sinclair@jpl.nasa.gov}

\author{James A. Sinclair}
\affiliation{Jet Propulsion Laboratory/California Institute of Technology\\
MS 183-601, 4800 Oak Grove Dr \\
Pasadena, CA 91109, United States}

\author{Thomas K. Greathouse}
\affiliation{Southwest Research Institute \\
6220 Culebra Road, San Antonio, TX 78238, United States}

\author{Rohini S. Giles}
\affiliation{Southwest Research Institute \\
6220 Culebra Road, San Antonio, TX 78238, United States}

\author{John Lacy}
\affiliation{The University of Texas at Austin, Department of Astronomy
Austin, TX 78712, United States}

\author{Julianne Moses}
\affiliation{Space Science Institute\\
4765 Walnut St, Suite B, Boulder, CO 80301, United States}

\author{Vincent Hue}
\affiliation{Southwest Research Institute \\
6220 Culebra Road, San Antonio, TX 78238, United States}
\affiliation{Aix-Marseille Université, CNRS, CNES, Institut Origines, LAM, Marseille, France}

\author{Denis Grodent}
\affiliation{Université de Liège, STAR Institute, \\
Laboratoire de Physique Atmosphérique et Planétaire, \\
Quartier Agora allée du six Août 19 c 4000 Liège 1, Belgium}

\author{Bertrand Bonfond}
\affiliation{Université de Liège, STAR Institute, \\
Laboratoire de Physique Atmosphérique et Planétaire, \\
Quartier Agora allée du six Août 19 c 4000 Liège 1, Belgium}

\author{Chihiro Tao}
\affiliation{National Institute of Information and Communications Technology, \\
4-2-1, Nukui-Kitamachi, Koganei, Tokyo 184-8795, Japan}

\author{Thibault Cavali\'e}
\affiliation{Laboratoire d’Astrophysique de Bordeaux, \\
All. Geoffroy Saint-Hilaire, 33600 Pessac, France}

\author{Emma K. Dahl}
\affiliation{Jet Propulsion Laboratory/California Institute of Technology\\
MS 183-601, 4800 Oak Grove Dr \\
Pasadena, CA 91109, United States}

\author{Glenn S. Orton}
\affiliation{Jet Propulsion Laboratory/California Institute of Technology\\
MS 183-601, 4800 Oak Grove Dr \\
Pasadena, CA 91109, United States}

\author{Leigh N. Fletcher}
\affiliation{School of Physics \& Astronomy, University of Leicester \\
University Road, Leicester, LE1 7RH, United Kingdom}

\author{Patrick G. J. Irwin}
\affiliation{Atmospheric, Oceanic \& Planetary Physics, University of Oxford \\
Parks Road, Oxford, OX1 3PU, United Kingdom}




\begin{abstract}

We present mid-infrared spectroscopy of Jupiter’s mid-to-high latitudes using Gemini-North/TEXES (Texas Echelon Cross Echelle Spectrograph) on March 17-19, 2017.  These observations capture Jupiter's hydrocarbon auroral emissions before, during and after the arrival of a solar wind compression on March 18th, which highlights the coupling between the polar stratosphere and external space environment.  In comparing observations on March 17th and 19th, we observe a brightening of the CH$_4$, C$_2$H$_2$ and C$_2$H$_4$ emissions in regions spatially coincident with the northern, duskside main auroral emission (henceforth, MAE).   In inverting the spectra to derive atmospheric information, we determine that the duskside brightening results from an upper stratospheric (p $<$ 0.1 mbar/z $>$ 200 km) heating (e.g. $\Delta T$ = 9.1 $\pm$ 2.1 K at 9 $\upmu$bar at 67.5$^\circ$N, 162.5$^\circ$W) with negligible heating at deeper pressures.  Our interpretation is that the arrival of the solar wind enhancement drove magnetospheric dynamics through compression and/or viscous interactions on the flank.  These dynamics accelerated currents and/or generated higher Poynting fluxes, which ultimately warmed the atmosphere through Joule heating and ion-neutral collisions.  Poleward of the southern MAE, temperature retrievals demonstrate that auroral-related heating penetrates as deep as the 10-mbar level, in contrast to poleward of the northern MAE, where heating is only observed as deep as $\sim$3 mbar.  We suggest this results from the south having higher Pedersen conductivities, and therefore stronger currents and acceleration of the neutrals, as well as the poleward heating overlapping with the apex of Jupiter’s circulation thereby inhibiting efficient horizontal mixing/advection.

\end{abstract}


\section{Introduction}

Jupiter has the largest planetary magnetosphere in our solar system.  This is further augmented by volcanic emissions from Io, which load the magnetosphere with sulfur and oxygen ions and modulate magnetospheric dynamics (e.g. \citealt{bonfond_2012}, \citealt{yoshikawa_2017}).  These dynamics as well as interactions of the magnetosphere with the solar wind ultimately drive ions and electrons into Jupiter's polar atmosphere.  Precipitating particles deposit their energies at high-southern and high-northern latitudes and ultimately produce auroral emissions over a large range of wavelengths including the X-ray (e.g. \citealt{gladstone_2002}, \citealt{dunn_2017}, \citealt{mori_2022}), ultraviolet (e.g. \citealt{nichols_2017}, \citealt{grodent_2018}, \citealt{greathouse_2021}, \citealt{hue_2021}), near-infrared (e.g. \citealt{moore_2017}, \citealt{johnson_2018}) and mid-infrared (e.g. \citealt{caldwell_1980}, \citealt{livengood_1993}).  

Jupiter's auroral regions are divided into three main sub-regions based on the morphology of the ultraviolet emissions.  First, the main auroral oval or main auroral emission (henceforth ``MAE'') is a persistent strip of emission around the magnetic poles and is thought to be generated by the breakdown of corotation in the middle magnetosphere (e.g. \citealt{grodent_2015}).  Second, discrete spots or ``footprints'' of emission, which are slightly equatorward of the MAE, and result from the electromagnetic interactions of Io, Europa and Ganymede with the rotating magnetospheric plasma (e.g. \citealt{bonfond_2012}).  Third, the diffuse and highly-variable ``polar'' emissions, which are enclosed or poleward of the MAE.  These are considered to be driven by outer magnetospheric dynamics related to the Vasilyunas and Dungey cycles (e.g. \citealt{cowley_2003}).   Jupiter's stratospheric hydrocarbon species, including CH$_4$, C$_2$H$_2$, C$_2$H$_4$ and C$_6$H$_6$ exhibit enhanced, mid-infrared emissions, which are generally coincident with the ultraviolet polar emissions described above (e.g. \citealt{kim_1985}, \citealt{kostiuk_1993}, \citealt{drossart_1993}, \citealt{flasar_2004_jet}).  One component of the enhanced mid-infrared emissions results from heating of the atmosphere at microbar pressures (e.g. \citealt{sinclair_2017b,sinclair_2020b}), at similar altitudes as those inferred of the ultraviolet emissions,  and therefore presumably driven by processes such as chemical heating, H$_2$ dissociation from excited states and Joule heating from Pedersen currents (e.g. \citealt{grodent_2001,yates_2014,badman_2015}).   A second component is the enrichment of the aforementioned hydrocarbon species by higher rates of ion-neutral and electron-recombination reactions due to the influx of ions and electrons (e.g. \citealt{sinclair_2017a,sinclair_2018a,sinclair_2019b}).  A third component is a discrete layer of lower stratospheric auroral-related heating ($\sim$1 mbar), which we consider driven by different mechanisms compared to that in the upper stratosphere.  A recent analysis of ALMA observations demonstrated the presence of eastward and westward jets at $\sim$0.1 mbar, horizontally coincident with the southern MAE and possibly generated by ion-neutral collisions \citep{cavalie_2021}. They suggest the jets highlight the presence of a counterrotating vortex with the enclosed atmospheric subsidence heating the lower stratosphere through adiabatic compression.   Such a vortex would confine aurorally-heated and hydrocarbon-enriched gas to the region until advected to deeper pressures where the vortex dissipates.  This describes an extreme example of space weather where the magnetosphere and solar wind can modulate the thermal structure, chemistry and dynamics as deep as the middle atmosphere. 

In \citet{sinclair_2019a}, a series of Subaru-COMICS (Cooled Mid-Infrared Camera and Spectrograph, \citealt{kataza_2000}) broadband,  7.8-$\upmu$m images of Jupiter's CH$_4$ emission were presented.   In comparing images recorded on January 11, 12 and 14, 2017, a brightening and subsequent dimming, of the south polar emissions was observed.  Images recorded on January 12 demonstrated a brightening of the CH$_4$ emission in a region that was spatially coincident with the duskside MAE. This duskside feature was absent $\sim$19 hours later, which suggested it was also transient and presumably driven by the same mechanism driving the variable southern poleward CH$_4$ emissions.  The arrival of a solar wind compression early on January 12th (though with a potential $\sim$48 hour uncertainty) was predicted using OMNI measurements of solar wind conditions at Earth (e.g. \citealt{thatcher_2011}) and a 1-dimensional solar wind propagation model \citep{tao_2005} to calculate the flow out to Jupiter's orbit.  The interpretation was that the solar wind compression perturbed the Jovian magnetosphere through either magnetic reconnection (e.g. \citealt{masters_2021}) and/or velocity shears on the nightside magnetospheric flank (e.g. \citealt{zhang_2018}).  This ultimately accelerated energetic particles into the atmosphere, which deposited their energy in such a way to produce either: 1) heating of the atmosphere and/or 2) vertical winds that advect CH$_4$ and its photochemical by-products to higher altitudes and/or 3) excitement of CH$_4$ to higher ro-vibrational states, and thereby enhancing the 7.8-$\upmu$m CH$_4$ emissions.  The daily timescales over which the variability was observed implied these changes occurred at the highest altitudes (10 - 1 $\upmu$bar) sensed by Jupiter's 7.8-$\upmu$m CH$_4$ emission, where the thermal inertia timescales are shortest.  

\begin{figure}[!t]
\begin{center}
\includegraphics[width=0.65\textwidth]{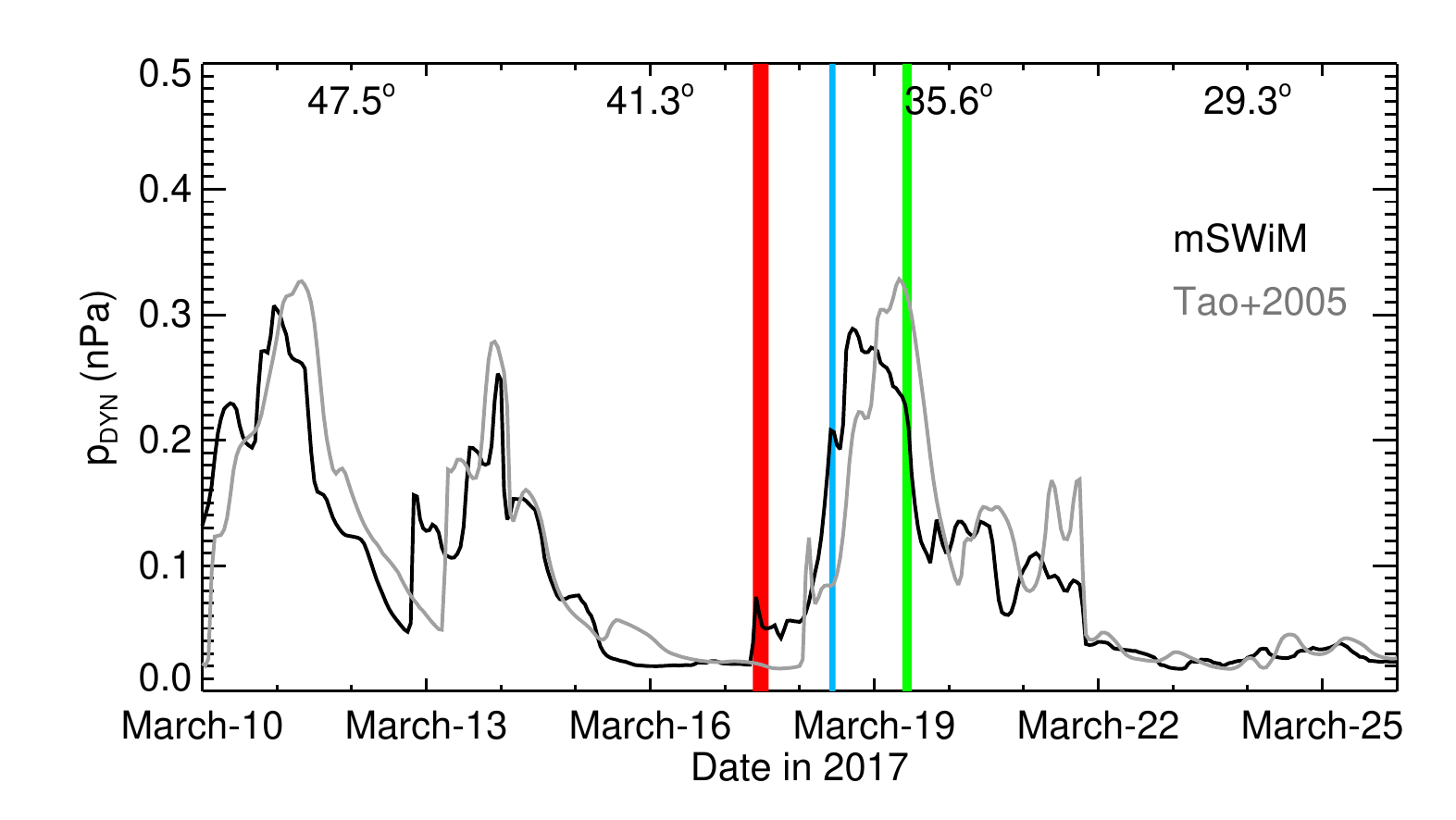}
\caption{Predicted solar wind dynamical pressures at Jupiter in March 2017 using OMNI-measured conditions at Earth \citep{thatcher_2011} and two solar wind propagation models - mSWiM (Michigan Solar Wind Model, \citealt{zieger_2008}) in black, the \citealt{tao_2005} model in grey - to calculate the solar wind flow at Jupiter's orbit.  Dates/times are UTC.  Red, blue and green ranges denote the time periods on March 17th, 18th and 19th, respectively, where the Gemini-TEXES spectra presented in this study were recorded.  The Earth-Jupiter-Sun angles are shown near the top of the panel. }
\label{fig:sw}
\end{center}
\end{figure}

In this work, we present evidence of a similar event where the duskside mid-infrared MAEs brighten seemingly in response to the arrival of a solar wind compression.  However, the observations and analysis presented in this work provide several advantages over those presented in \citet{sinclair_2019a} and allow this phenomenon to be studied with less ambiguity and greater detail.  First, the observations presented in this work were recorded when Jupiter was within 50$^\circ$ of opposition, when the uncertainties on magnitude and timing of solar wind propagation results are smaller.  Solar-wind propagation models, such as the mSWiM \citep[Michigan Solar Wind Model,][]{zieger_2008}, and a similar model first presented in \citet{tao_2005}, adopt empirical solar-wind conditions at Earth and use a 1-dimensional magnetohydrodynamic calculation to model the solar wind flow out to the orbit of the target planet.  Given variations in solar wind flow as a function of heliocentric longitude, the model results are most accurate when the target planet is close to opposition (i.e. the Sun - Target - Earth angle is small). \citet{zieger_2008} quantified the uncertainty on predicted solar wind conditions as a function of angle from opposition and found uncertainties on solar wind conditions to be $<$20 hours in timing and $<$38\% in dynamical pressure when Jupiter was within 50$^\circ$ of opposition.  Figure \ref{fig:sw} shows predicted solar wind pressures at Jupiter during the period of observations presented in this work.  A solar wind compression arrived at Jupiter's magnetosphere on March 18th, 2017.  Even with a timing uncertainty as high as 20 hours, we can still confidently interpret observations recorded on March 17th, 18th and 19th, 2017 to have captured Jupiter's high latitudes before, during and after the arrival of this solar wind compression.  In contrast, the data presented in \citet{sinclair_2019a} were recorded when Jupiter was $\sim$80$^\circ$ from opposition, where uncertainty could be as high as 48 hours \citep{zieger_2008}.  This introduced ambiguity into the timing of the solar wind compression arrival with respect to the observed mid-infrared main auroral emission brightening.  Second, in this work, we present high-resolution spectroscopy (65000 $<$ R $<$ 80000) in discrete 5 - 7 cm$^{-1}$ wide bands, which capture Jupiter's H$_2$ S(1), CH$_4$, C$_2$H$_2$, C$_2$H$_4$, C$_2$H$_6$ emission features.  Further details of the observations are discussed in Section \ref{sec:obs}.  This allows the variability of CH$_4$ emission as well as its photochemical by-products to be measured, which can better disentangle whether any observed brightening results from a heating and/or chemistry effect.  The high spectral resolution resolves individual strong and weak emission lines of the aforementioned hydrocarbons, which in turn sounds a large range of altitudes in Jupiter's stratosphere and mesosphere (Section \ref{sec:contr_fns}) . By inverting these spectra using radiative transfer software, changes in atmospheric temperatures and/or abundances can be quantified and the altitudes at which they occur better determined.   In contrast, the data presented in \citet{sinclair_2019a} were broadband ($\Delta \lambda \sim $ 0.7 $\upmu$m) 7.8 $\upmu$m imaging of Jupiter's CH$_4$ emission, which do not allow retrieval of atmospheric information without significant degeneracy.

\section{Observations}\label{sec:obs}

High-resolution spectra were recorded using TEXES (Texas Echelon Cross Echelle Spectrograph, \citealt{lacy_2002}) on the 8-m Gemini (North) telescope on March 17th, 18th and 19th, 2017.  Spectra were recorded in discrete settings centered at 587, 730, 819, 950 and 1248 cm$^{-1}$, which respectively capture stratospheric H$_2$ S(1) quadrupole, C$_2$H$_2$, C$_2$H$_6$, C$_2$H$_4$ and CH$_4$ emissions.  These observations were performed when the relative Earth-Jupiter velocity was 10 - 12 km/s, which produces a Doppler shift sufficient to disentangle Jovian CH$_4$ emission lines from telluric CH$_4$ absorption.

The 4'' x 0.5'' slit of the spectrograph was aligned parallel to the direction of Jupiter's central meridian.  Starting from dark sky west of the planet, the slit was stepped across Jupiter's high-northern latitudes in steps of a half slit width for Nyquist sampling until the slit reached 4 planet-free steps east of the limb.  Successive scans were repeated with the slit offset in a direction parallel to Jupiter's central meridian in order to build up latitudinal coverage.   The same process was repeated at Jupiter's high-southern latitudes.  Looping through the five aforementioned spectral settings, scans of both high-southern and high-northern latitudes were repeated over the course of each night with Jupiter's rotation used to build up longitudinal coverage.   For the 587 cm$^{-1}$ spectral setting, a wider, 0.8'' slit was used due to the larger angular diffraction at longer wavelengths. Further details of individual scans are provided in Table \ref{tab:mar17_obs}.   
%
%

\begin{figure}[!th]
\begin{center}
\includegraphics[width=0.37\textwidth]{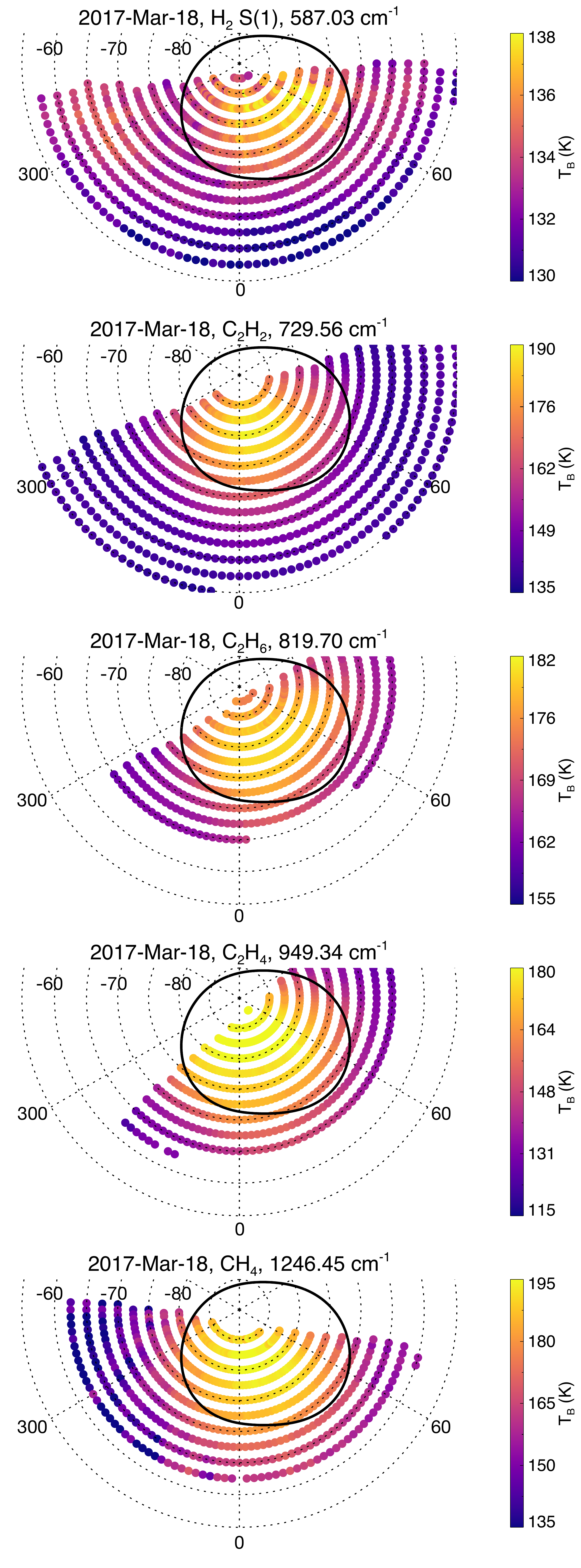}
\caption{The distribution of coadded TEXES observations of H$_2$ S(1) quadrupole (1st row), C$_2$H$_2$ (2nd), C$_2$H$_6$ (3rd), C$_2$H$_4$ (4th) and CH$_4$ emission on March 18th, 2017.  Points are colored by brightness temperature of an emission line at the indicated wavenumber according to the right-hand colorbar.  The statistical-mean auroral oval position is shown as solid, black \citep{bonfond_2017}.}
\label{fig:spx_south_coadd}
\end{center}
\end{figure}

\begin{figure*}[!t]
\begin{center}
\includegraphics[width=0.65\textwidth]{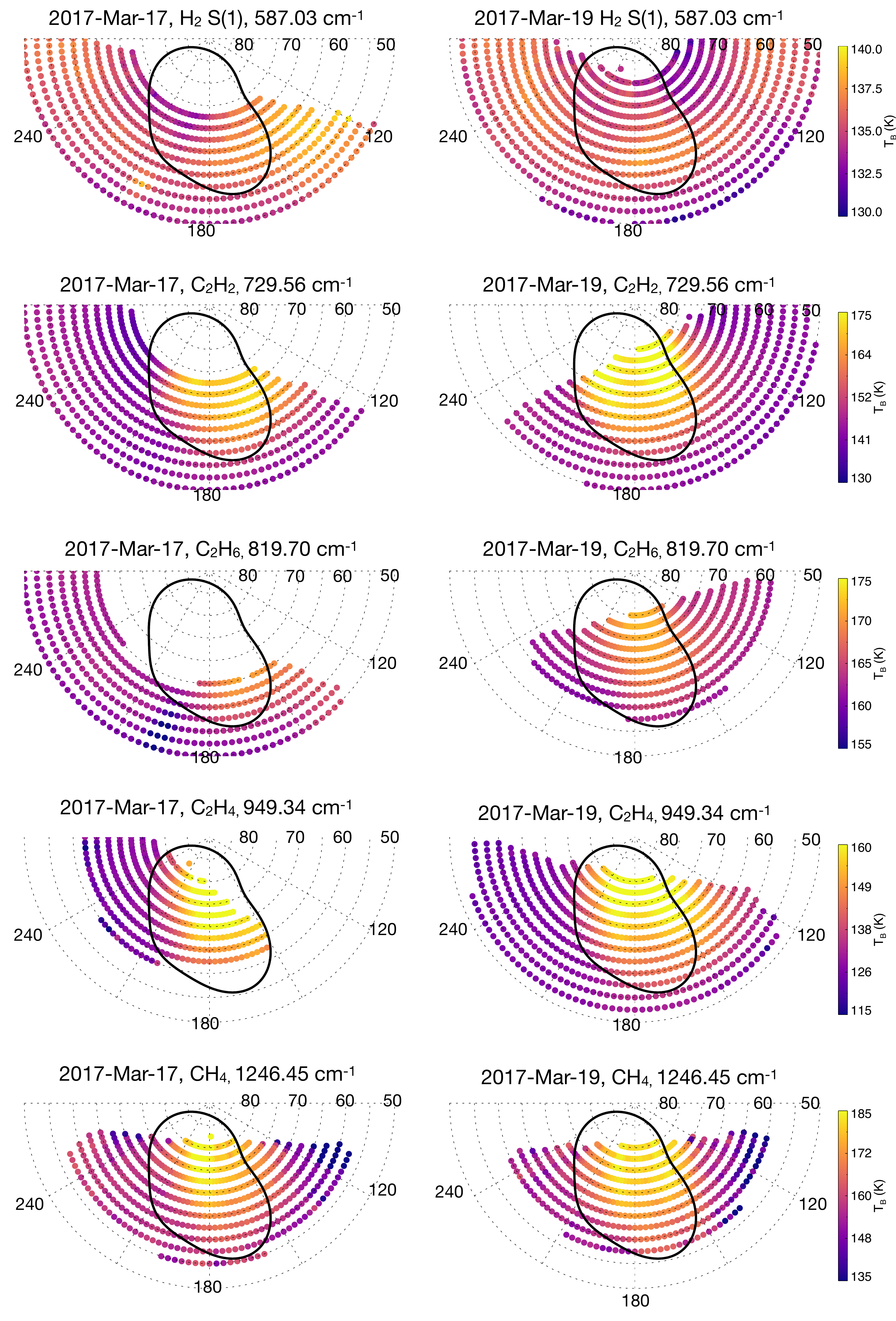}
\caption{The distribution of coadded TEXES observations of H$_2$ S(1) quadrupole (1st row), C$_2$H$_2$ (2nd), C$_2$H$_6$ (3rd), C$_2$H$_4$ (4th) and CH$_4$ emission on March 17th (left column) and March 19th (right column), 2017.  Points are colored by brightness temperature of an emission line at the indicated wavenumber according to the right-hand colorbar.  The statistical-mean auroral oval position is shown as solid, black \citep{bonfond_2017}.}
\label{fig:spx_north_coadd}
\end{center}
\end{figure*}

Sky emission sampled at the start and end of each scan allowed for subtraction of sky emission from the measured spectra of the target. The wavelength-dependent absolute calibration for each spectrum was performed using corresponding exposures of a blackbody card of known temperature mounted above the instrument's window.  Custom IDL software and the NAIF (the Navigation and Ancillary Information Facility) SPICY toolkit \citep{acton_1996} were used perform a limb fitting procedure in order to calculate the latitude, longitude, emission angle and velocity (combining Jupiter's radial and rotational velocity) of each pixel.  In this work, we use planetocentric latitude and the System III longitude system.    Each individual spectrum was corrected for Doppler shift.  Figures \ref{fig:spx_587_north} to \ref{fig:spx_1248_south} show the distributions of individual observations sorted by setting, hemisphere and in chronological order, with each spectrum colored according to the radiance of an emission line of the relevant species. 

From these individual scans alone, we noticed significant variability in the strength and morphology of the hydrocarbon emissions within the main ovals over the three days of observations.    For example, on March 17th, observations of the northern oval in scans 7017.01, 7018.01 and 7018.02 show a single hotspot of the strongest CH$_4$ emissions at a longitude of approximately 180$^\circ$W, whereas two days later, observations in scans 9011.01 and 9011.02 show the strongest CH$_4$ emissions are in two distinct regions, the first on the eastern or dusk side of the main oval, the second at longitudes of approximately 190-200 $^\circ$W (Figure \ref{fig:spx_1248_north}).   Similar changes in morphology inside the northern, main oval between March 17th and 19th are also evident in the 730 cm$^{-1}$ spectra of the C$_2$H$_2$ emission (Figure \ref{fig:spx_730_north}).  Two regions of stronger C$_2$H$_4$ emissions within northern main oval are also evident in the 950 cm$^{-1}$ observations recorded on March 19th 2017 (see Figure \ref{fig:spx_950_north}).  Although the same regions were not sufficiently sampled at 950 cm$^{-1}$ on March 17th to indicate whether this morphology is transient, the fact that a similar morphology is observed in the CH$_4$ and C$_2$H$_2$ emissions strongly suggests the C$_2$H$_4$ features are also transient features that emerged sometime between March 17th and March 19th, 2017.  

In the south, we also observe variability of the poleward emissions between dates.  For example, C$_2$H$_2$ emission poleward of the main auroral oval decreases from March 17th (see scans 7044.01, 7044.02, Figure \ref{fig:spx_730_south}) to 18th.  A similar marginal decrease is also observed of the C$_2$H$_6$ emissions between March 17th and 18th (Figure \ref{fig:spx_819_south}) whereas an increase in C$_2$H$_4$ emissions is observed between March 17th and 18th (Figure \ref{fig:spx_950_south}).  For the CH$_4$ emissions, our observations did not sufficiently sample the same regions on both nights to determine whether any similar variability occurred.  This possibly suggests the southern poleward hydrocarbon emissions exhibit variability on timescales of hours. 

In previous studies using similar TEXES spectra (e.g. \citealt{fletcher_2016,sinclair_2018a,melin_2018,fletcher_2020,sinclair_2020b}), all individual spectra recorded on one night or across several nights were combined to extend longitudinal coverage and then coadded into latitude-longitude bins to increase the signal-to-noise ratio (SNR) for the radiative transfer analysis.  Given the longitudinal overlap between scans and the temporal variability of the hydrocarbon emissions evident in the northern main oval, we chose to coadd only a subset of the scans such that temporal variability was not diluted or averaged out entirely.   The scans chosen for coaddition and further analysis are shown in bold in Table \ref{tab:mar17_obs}.  For observations on March 17th and 19th, we coadded spectra only at high-northern latitudes to focus on variability of the northern auroral oval between these dates.  At high-southern latitudes, spectra on March 17th and 19th did not sufficiently sample the same locations in all five spectral settings.  Similarly, for observations on March 18th, we coadded observations only at high-southern latitudes since they did not sufficiently sample the northern auroral oval.   Our analysis therefore provides a only a snapshot of the southern auroral oval but we note to readers that we do observe evidence of variability in the southern poleward emissions between March 17th and 18th.  Individual spectra were coadded into 5$^\circ$ x 5$^\circ$ latitude-longitude bins, stepped in increments of 2.5$^\circ$ for Nyquist sampling.  $\sim$5$^\circ$ corresponds to the diffraction-limited spatial resolution in latitude-longitude at 60$^\circ$ and the expected accuracy of the spatial registration of the spectra. Figure \ref{fig:spx_south_coadd} shows the distributions of coadded observations at high-southern latitudes on March 18th, 2017.  Figure \ref{fig:spx_north_coadd} shows the distributions of coadded observations at high-northern latitudes on March 17th and March 19th, 2017, which further highlight the variability of the hydrocarbon emissions over $\sim$2 days.  

\section{Radiative Transfer Model}\label{sec:rad_model}

The NEMESIS radiative transfer model \citep{irwin_2008} was adopted for all forward modelling and retrievals of atmospheric information.    NEMESIS assumes local thermodynamic equilibrium (LTE) and adopts the optimal estimation technique, where an initial guess or \textit{a priori}, $\mathbf{x_a}$, of an atmospheric parameter (e.g. the vertical temperature profile) is initially adopted.  A synthetic spectrum, $\mathbf F(\hat{x})$, is computed, the cost function ($\phi$, Equation \ref{eq:cost}) is evaluated and the variable parameter, $\mathbf{x}$, is iteratively adjusted until it converges on a solution that minimises the cost function (Equation \ref{eq:cost}).

\begin{equation}\label{eq:cost}
\begin{split}
\phi &= (\mathbf y - \mathbf F(\hat{x}))^T {\mathbf{S_{\epsilon}}}^{-1} (\mathbf y - \mathbf F(\hat{x})) \\
& + (\mathbf{\hat{x}} - \mathbf{x_a})^T {\mathbf{S_a}}^{-1} (\mathbf{\hat{x}} - \mathbf{x_a})
\end{split}
\end{equation}
Here, $\mathbf y$ and $\mathbf{S_{\epsilon}}$ are the observed spectrum and uncertainty on the observed spectrum, respectively and $\mathbf{\hat{x}}$ and $\mathbf{S_a}$ are the retrieved value and \textit{a priori} error, respectively. 

\subsection{Spectroscopic line data}

 We chose to perform forward modelling and retrievals using the correlated-k treatment (e.g. \citealt{correlated_k}) of spectroscopic line data for computational efficiency.  The spectroscopic line data for the H$_2$ S(1) quadrupole line feature, NH$_3$, PH$_3$, CH$_4$, CH$_3$D, $^{13}$CH$_4$, C$_2$H$_2$, C$_2$H$_4$ and C$_2$H$_6$ in \citet{fletcher_poles_2018} were adopted in this work.  K-distributions for the aforementioned species were calculated from 728 - 733 cm$^{-1}$, 815 - 824 cm$^{-1}$, 947 - 953 cm$^{-1}$ and 1244 - 1252 cm$^{-1}$ using a 4 km/sec sinc-squared convolution.  Due to the effects of diffraction a wider slit was used for the 587 cm$^{-1}$ observations compared to the other wavelength settings.  Modeling of this observation required a 6 km/s sinc-squared convolution.

\begin{figure}[!t]
\begin{center}
\includegraphics[width=0.7\textwidth]{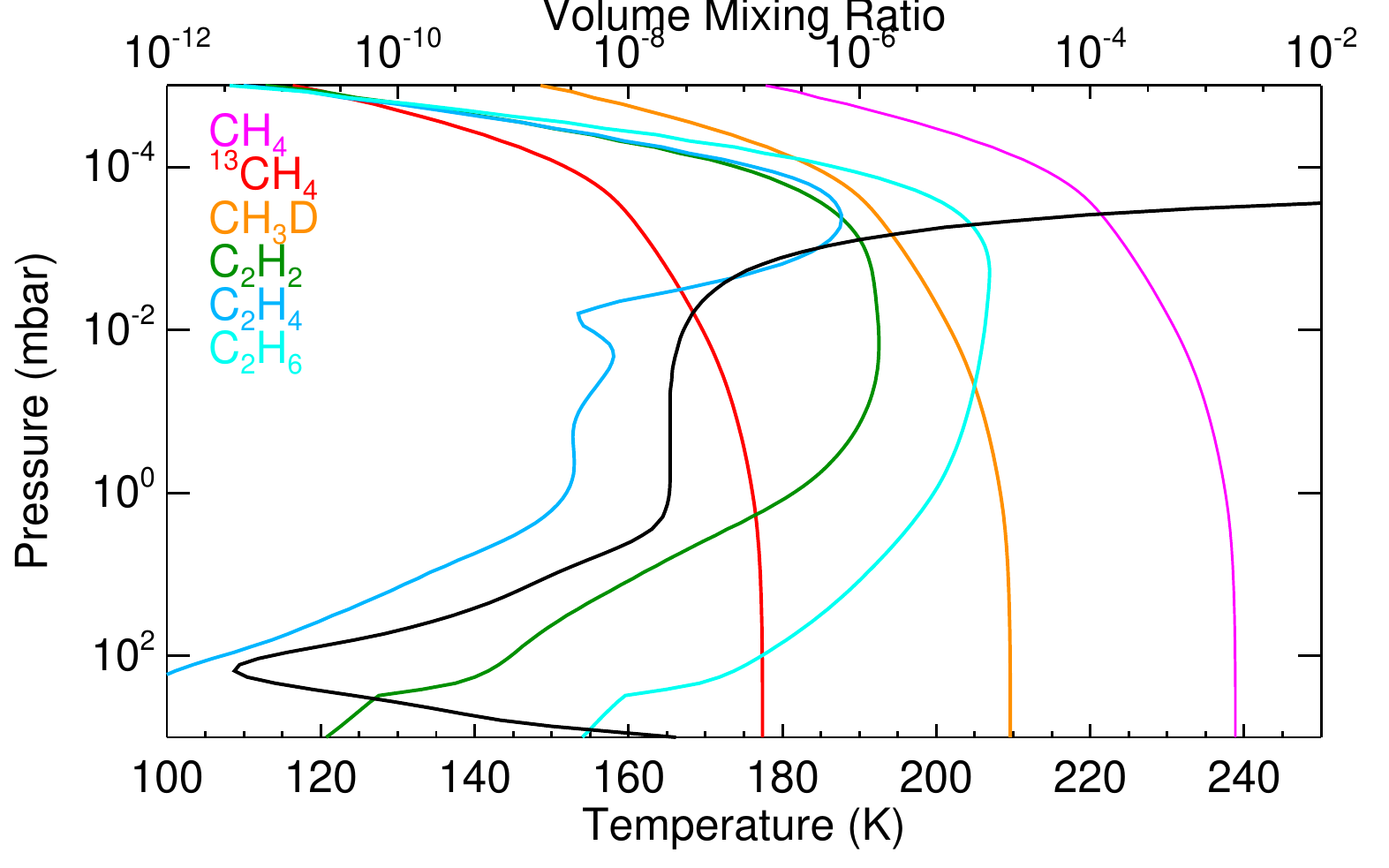}
\caption{The vertical profile of temperature (black) according to the lower axis and the vertical profiles of CH$_4$ (pink), $^{13}$CH$_4$ (red), CH$_3$D (orange), C$_2$H$_2$ (green), C$_2$H$_4$ (blue) and C$_2$H$_6$ (cyan) according to the upper axis.  This is Model 7 from \citet{sinclair_2020b}, which is based on the photochemical model presented in \citet{moses_2017}. }
\label{fig:photo_model}
\end{center}
\end{figure}
\subsection{Atmospheric model}\label{sec:atmos_model}

\begin{figure}[!h]
\begin{center}
\includegraphics[width=0.6\textwidth]{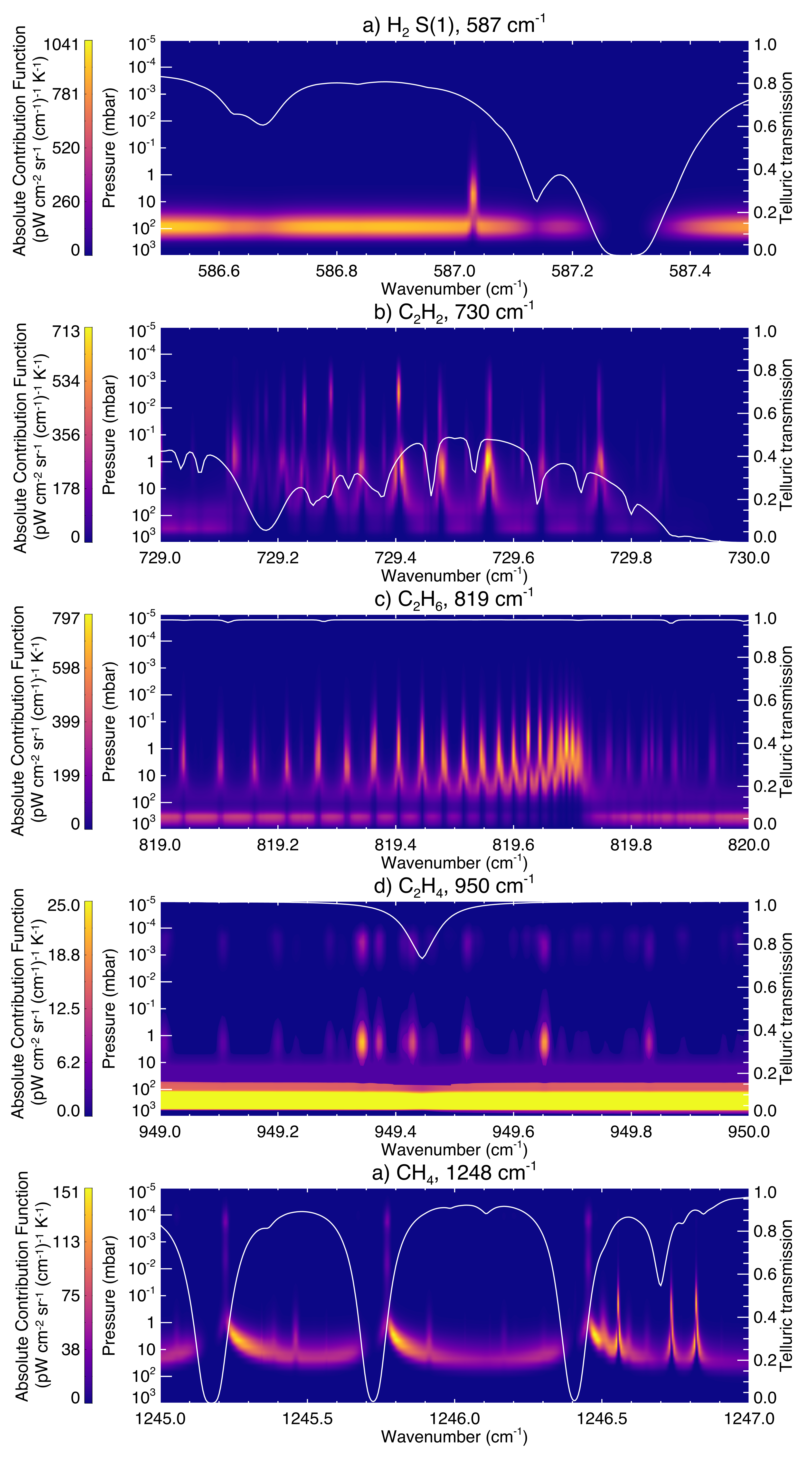}
\caption{The vertical, absolute contribution functions with respect to temperature for a subset of the a) 587 cm$^{-1}$, b) 730cm$^{-1}$, c) 819cm$^{-1}$ d) 950cm$^{-1}$ and e) 1248 cm$^{-1}$ settings.  The contribution functions have been convolved with the telluric transmission spectrum shown as white solid line and according to the right-hand axis. }
\label{fig:contr_functions}
\end{center}
\end{figure}

For the vertical temperature profile, we assumed the profile adopted in \citet{moses_2017}, which uses ISO-SWS (Short Wave Spectrometer on the Infrared Space Observatory) results (\citealt{lellouch_2001} and references therein) at pressures greater than 1 mbar and Galileo-ASI (Atmospheric Structure Instrument) measurements \citep{seiff_1998} at pressures lower than 1 mbar.   For the vertical profiles of all hydrocarbons, we adopted `Model 7' presented in \citet{sinclair_2020b} but re-calculated at a latitude of 60$^\circ$N.  These profiles were derived from a photochemical model similar to that described in Moses and Poppe (2017), except for the adoption of a larger vertical gradient in the eddy diffusion coefficient, such that the CH$_4$ homopause resides at ~7 nbar.  In \citet{sinclair_2020b}, a 7-nbar homopause pressure was found to be the best-fitting model to observations of H$_2$ S(1), CH$_4$ and CH$_3$ emission inside Jupiter's northern main oval.  Figure \ref{fig:photo_model} shows the vertical profiles of temperature, CH$_4$ and its isotopologues, C$_2$H$_2$, C$_2$H$_4$ and C$_2$H$_6$ from Model 7, which we adopted in this work.  The vertical profiles of temperature, C$_2$H$_2$, C$_2$H$_4$ and C$_2$H$_6$ serve as initial guess or \textit{a priori} profiles in performing inversions or retrievals in Section \ref{sec:temp} and \ref{sec:cxhy}.

The vertical profiles of H$_2$, He, NH$_3$, PH$_3$ were adopted from \citet{sinclair_2018a}.  For tropospheric aerosols, we assumed a 0.7-bar NH$_3$ ice cloud with a 10-$\upmu$m optical depth of 0.5 and fractional scale height of 0.5 and a 4.0-bar NH$_4$SH cloud with a 10-$\upmu$m optical depth of 1.0 and a fractional scale height of 0.05.  Although the physical properties of clouds and concentrations of disequilibrium species vary with latitude and longitude on Jupiter (e.g. \citealt{fletcher_2016,giles_2017,blain_2018}), such spatial variations modulate only the tropospheric continuum at the wavelengths presented in this work, and have negligible effect on retrieved stratospheric properties.

\subsection{Vertical sensitivity}\label{sec:contr_fns}

Adopting the model atmosphere described in Section \ref{sec:atmos_model}, the vertical functional derivatives with respect to temperature were calculated over the ranges of the five spectral settings.  The vertical functional derivatives with respect to temperature, or contribution functions, quantify the contribution of each atmospheric level to the total observed radiance at the top of the atmosphere at a given wavelength and therefore also quantify the range of pressures/altitudes over which the observations sound.

In each spectral setting, the functional derivatives were computed and then convolved with a telluric transmission spectrum doppler shifted by +11 km/s to simulate Jupiter's radial velocity from -12 to -10 km/s at the time of the observations (see Table \ref{tab:mar17_obs}).  The telluric transmission spectra were calculated using ATRAN \citep{lord_1992} assuming an altitude of 13800 ft (the altitude of Mauna Kea), a precipitable water vapor of 1 mm, an airmass of 1.4 (or zenith angle of 45$^\circ$).   Figure \ref{fig:contr_functions} shows the resulting vertical functional derivatives.   In the 587 cm$^{-1}$ spectral setting, the continuum sounds the upper troposphere at $\sim$100 mbar and the H$_2$ S(1) quadrupole line sounds the lower stratosphere from $\sim$50 mbar to $\sim$1 mbar.   CH$_4$ emission lines sampled in the 1248 cm$^{-1}$ setting sound the atmosphere from $\sim$20 mbar to 0.1 $\upmu$bar.  As in previous studies, we invert the spectra in the 587 cm$^{-1}$ and 1248 cm$^{-1}$ simultaneously in order to constrain the vertical temperature profile at $\sim$100 mbar and from $\sim$50 mbar to 0.1 $\upmu$bar.  

In the 730 cm$^{-1}$ setting, a mixture of weak and strong C$_2$H$_2$ emission lines sound the atmosphere from $\sim$50 mbar to $\sim$1 $\upmu$bar.  C$_2$H$_4$ emission lines measured in the 950 cm$^{-1}$ setting exhibit a double-peaked contribution function with sensitivity peaking at approximately 1 mbar and 0.5 $\upmu$bar.  In the 819 cm$^{-1}$ setting, C$_2$H$_6$ emission sounds the atmosphere from $\sim$20 to $\sim$0.08 mbar.  The vertical sensitivities to the C$_2$ hydrocarbons does differ and we discuss the impact this has on the analysis in Section \ref{sec:cxhy}.

\section{Temperature results}\label{sec:temp}
\begin{figure}[!t]
\begin{center}
\includegraphics[width=0.45\textwidth]{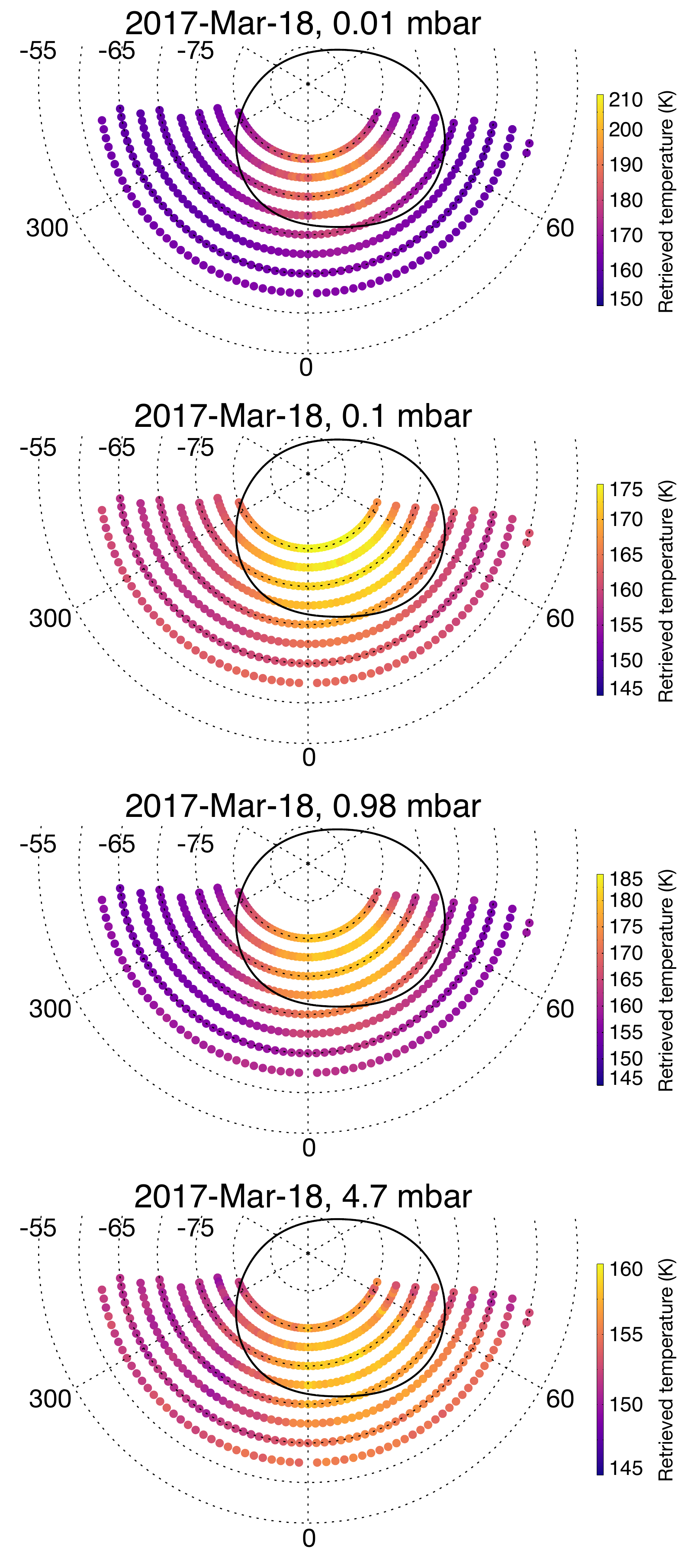}
\caption{Retrieved temperature distributions on March 18th 2017 at 0.01 mbar (1st row), 0.1 mbar (2nd row), 0.98 mbar (3rd row) and 4.7 mbar (4th row).  Solid, black lines denote the statistical-mean position of the ultraviolet main auroral emission \citep{bonfond_2017}.}
\label{fig:t_retr_south}
\end{center}
\end{figure}
The vertical temperature profile was retrieved by simultaneously inverting the H$_2$ S(1) quadrupole emission line and surrounding continuum in the 587 cm$^{-1}$ setting and CH$_4$ emission in the 1248 cm$^{-1}$ setting.  The vertical profile of temperature shown in Figure \ref{fig:photo_model} was adopted as the \textit{a priori} profile.  

\subsection{High-southern latitudes}\label{sec:T_south}

\begin{figure}[!t]
\begin{center}
\includegraphics[width=0.45\textwidth]{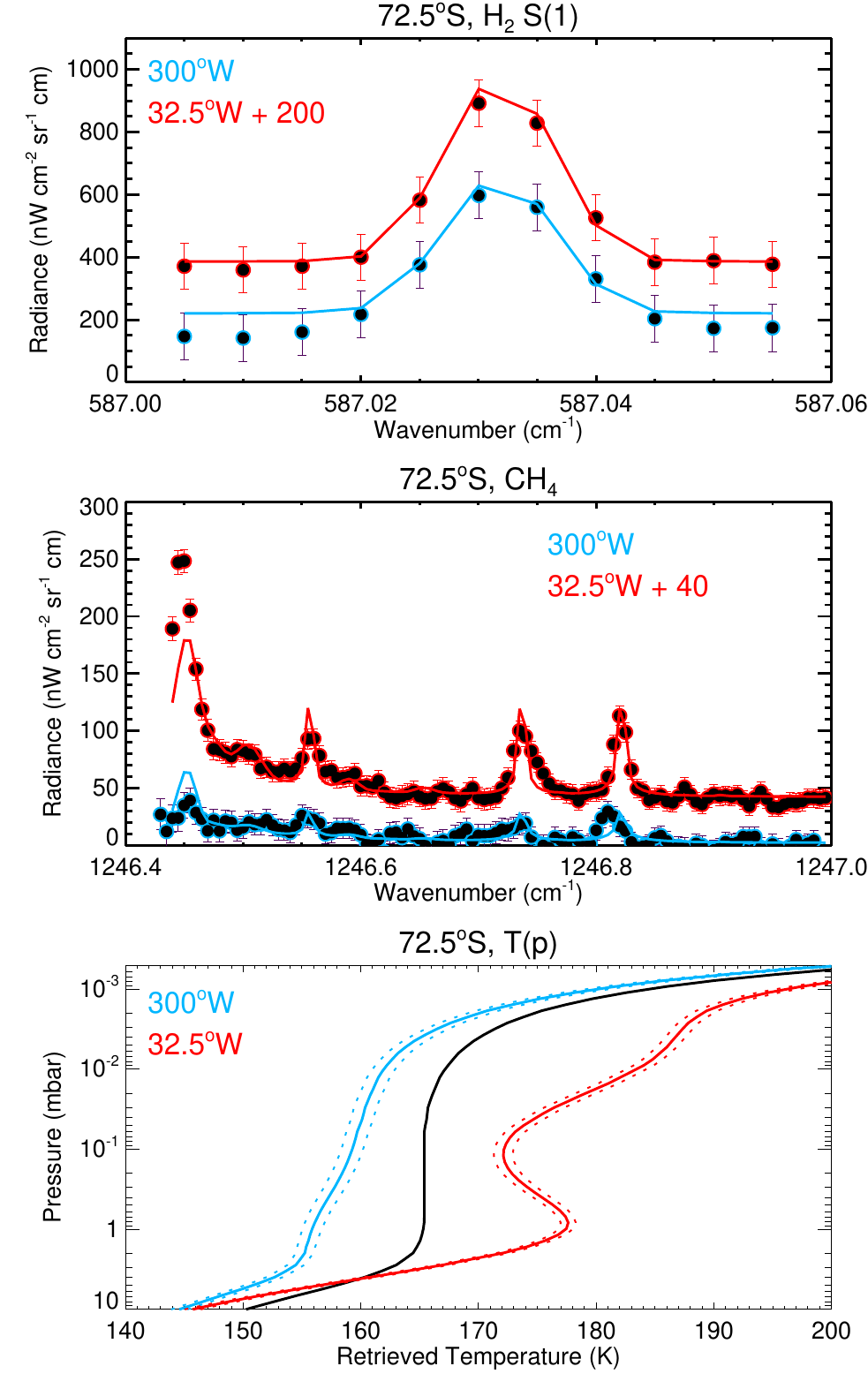}
\caption{Comparison of observed (points with error bars) and modelled spectra (solid lines) for H$_2$ S(1) (top panel) and CH$_4$ emission (middle panel) at 72.5$^\circ$S.  Red and blue respectively denote results at 32.5$^\circ$W (a longitude poleward of Jupiter's southern main auroral oval) and 300$^\circ$W (equatorward of the main auroral emission).  An offset has been added to red spectra for clarity.  The corresponding retrieved profiles of temperature (solid, colored) and uncertainty (dotted, colored) are shown in the bottom panel.  The black, solid profile denotes the initial guess or \textit{a priori} profile. }
\label{fig:spx_retr_south}
\end{center}
\end{figure}


\begin{figure*}[!t]
\begin{center}
\includegraphics[width=0.7\textwidth]{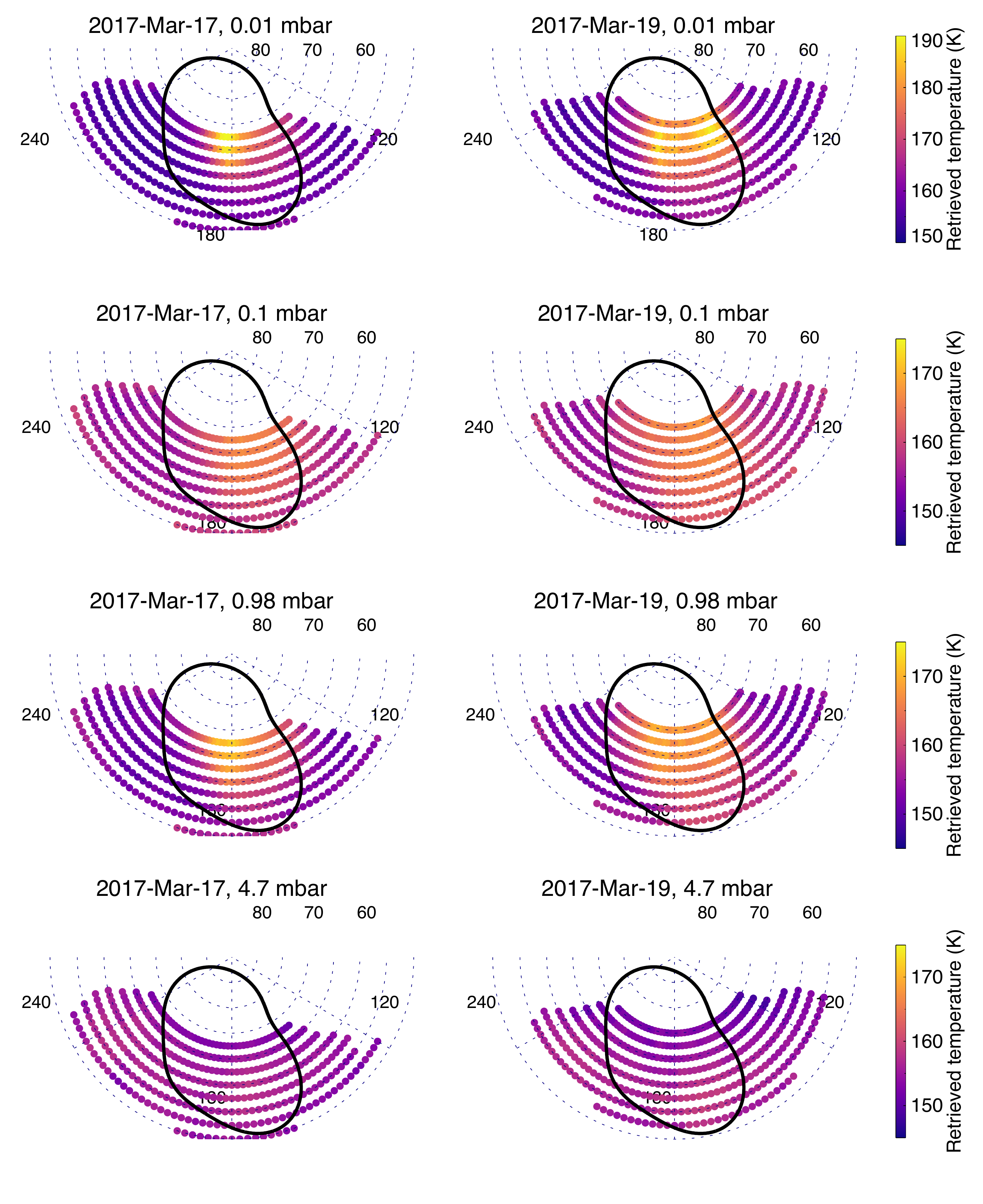}
\caption{Retrieved temperature distributions on March 17th 2017 (left column) and March 19th 2017 (right column) at 0.01 mbar (1st row), 0.1 mbar (2nd row), 0.98 mbar (3rd row) and 4.7 mbar (4th row).  Solid, black lines denote the statistical-mean position of the ultraviolet oval \citep{bonfond_2017}.}
\label{fig:t_retr_17_19}
\end{center}
\end{figure*}

Figure \ref{fig:t_retr_south} shows retrieved temperature distributions at high-southern latitudes using observations recorded on March 18th, 2017 and demonstrates that heating associated with auroral processes is observed predominantly at longitudes inside the southern auroral oval (e.g. \citealt{sinclair_2017b,sinclair_2018a}).  In comparison to longitudes outside the southern auroral oval, temperatures are elevated at pressures as deep as $\sim$10 mbar.  This is also demonstrated in Figure \ref{fig:spx_retr_south}, which compares the observed and modelled spectra and retrieved vertical profiles of temperature at 72.5$^\circ$S, 300$^\circ$W (equatorward of the main auroral emission) and 32.5 $^\circ$W (poleward of the main auroral emission).   Auroral-related heating as deep as 10 mbar is in  contrast to the results of Jupiter's northern auroral region (Section \ref{sec:T_north}, \citealt{sinclair_2017a,sinclair_2018a,sinclair_2020b}), where no significant auroral-related heating was observed at pressures higher than 2-3 mbar.   Either the heating as deep as 10 mbar is a persistent feature of the southern auroral region that was not observed in previous IRTF-TEXES observations due to the poorer spatial resolution insufficiently sampling such high latitudes, or the deeper extent of heating is a transient response to the arrival of the solar wind compression.  We favor the former explanation as detailed further in Section \ref{sec:discuss}.  

\begin{figure}[!t]
\begin{center}
\includegraphics[width=0.47\textwidth]{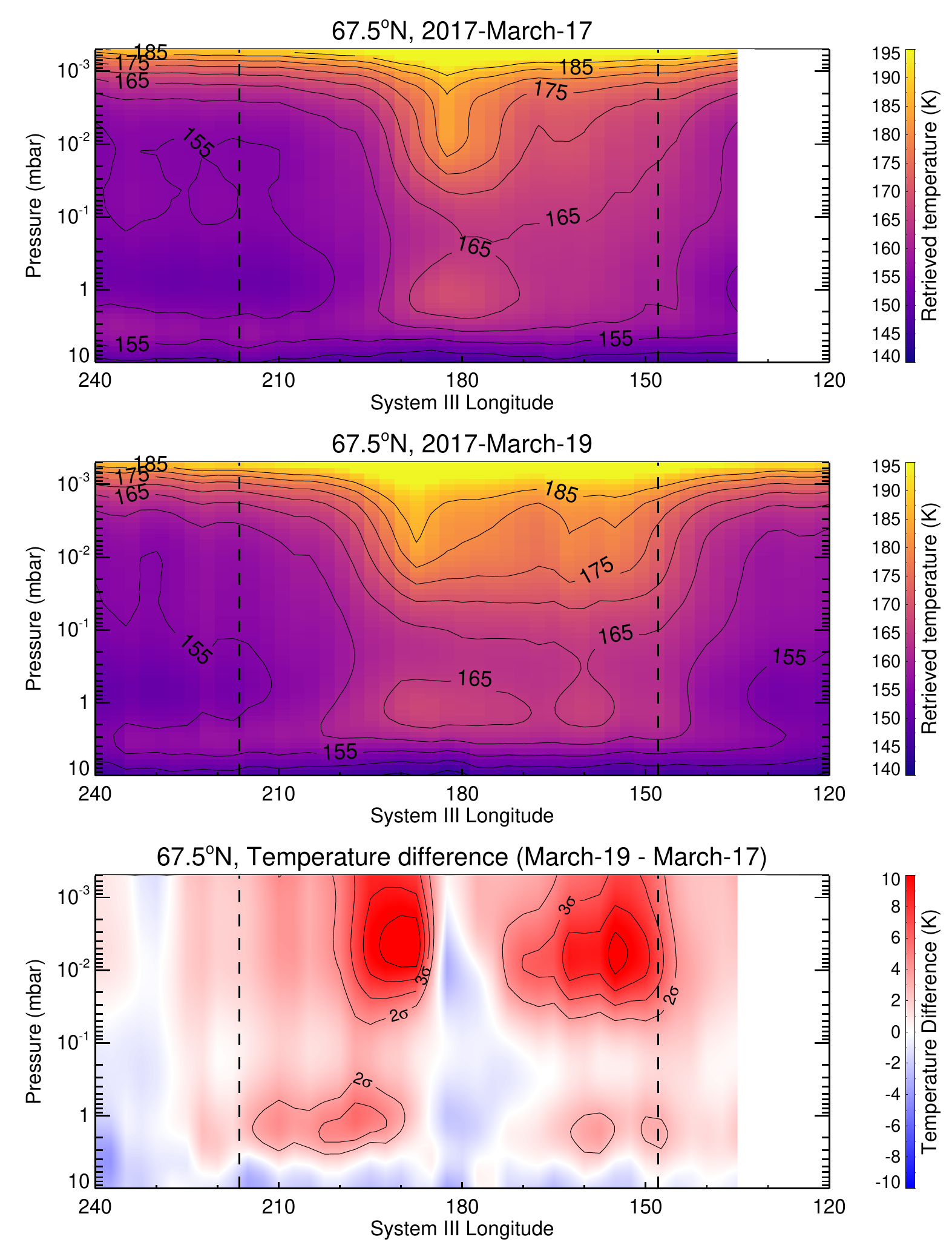}
\caption{Longitude-pressure cross-section of retrieved temperatures at 67.5$^\circ$N.  Results for March 17th 2017 are shown in the top panel and for March 19th in the middle panel.  The temperature difference (March 19th - March 17th) is shown according to the colorbar and line contours mark the 2-, 3-, 4-, 5- and 6-$\sigma$ levels to demonstrate where temperature changes are significant with respect to uncertainty. Vertical, dashed lines correspond to the longitudinal boundaries of the auroral oval at 67.5$^\circ$N.  }
\label{fig:contour_vs_long}
\end{center}
\end{figure}

The retrieved temperature profile inside the auroral region exhibits a local maximum at $\sim$1 mbar, which is approximately 25 K warmer than temperatures derived outside the auroral oval, where no such maximum in temperature is observed.  Temperatures inside the auroral oval decrease from $\sim$1 mbar until 0.1 mbar, then continue to rise towards lower pressures.  At pressures lower than 0.1 mbar, atmospheric heating is inferred to result from processes directly related to the deposition of energy from the magnetosphere and external space environment, including chemical heating, H$_2$ dissociation from excited states \citep{grodent_2001}, joule heating from Pedersen currents (e.g. \citealt{badman_2015}) and ion drag.  These processes essentially move the base of the thermosphere to lower altitudes, as has been shown using general circulation modelling of Jupiter's thermosphere \citep{bougher_2005}.   A localised maximum in temperature at $\sim$1 mbar has been observed in previous work, using different datasets and appears to be a persistent feature of both the northern and southern auroral regions \citep{kostiuk_agu_2016,sinclair_2017a,sinclair_2017b,sinclair_2018a,sinclair_2020b}.   In earlier studies, we suggested this heating resulted from short-wave solar heating of haze particles produced by the complex chemistry prevalent in Jupiter's auroral regions (e.g. \citealt{friedson_2002},\citealt{wong_2003}).   However, temperatures inside the northern auroral region at 1 mbar have been observed to vary erratically on timescales of weeks to months \citep{sinclair_epsc_2018}, which cannot be explained by longer-term variations in solar insolation.  We also considered pumping of the CH$_4$ 3-$\upmu$m band by overlapping H$_3^+$ emission lines from higher in the atmosphere.  This would in turn affect the transitions responsible for the 8-$\upmu$m band, a component of which originate from the $\sim$1 mbar level.  However, we have also ruled this out as a dominant mechanism since the strongest H$_3^+$ emissions are spatially coincident with the main auroral emission, whereas the strongest 1-mbar heating is generally observed coincident with the poleward emissions.  We also consider it very unlikely that the 1-mbar heating is driven directly by precipitation of magnetospheric electrons and ions.  For example, \citet{gustin_2016} demonstrate that electrons do not precipitate deeper than $\sim$200 km/0.1 mbar.  Similarly, precipitation of ions, and the resulting secondary electrons, is not expected at pressures higher than $\sim$200 km/0.1 mbar (e.g. \citealt{ozak_2013,houston_2020}).  Instead, we favour the explanation presented in \citet{cavalie_2021}, who observed anticyclonic rotation at $\sim$0.1 mbar inside the auroral ovals, which implies the presence of atmospheric subsidence and adiabatic heating.  We discuss this further in Section \ref{sec:discuss}.

\begin{figure}[!th]
\begin{center}
\includegraphics[width=0.47\textwidth]{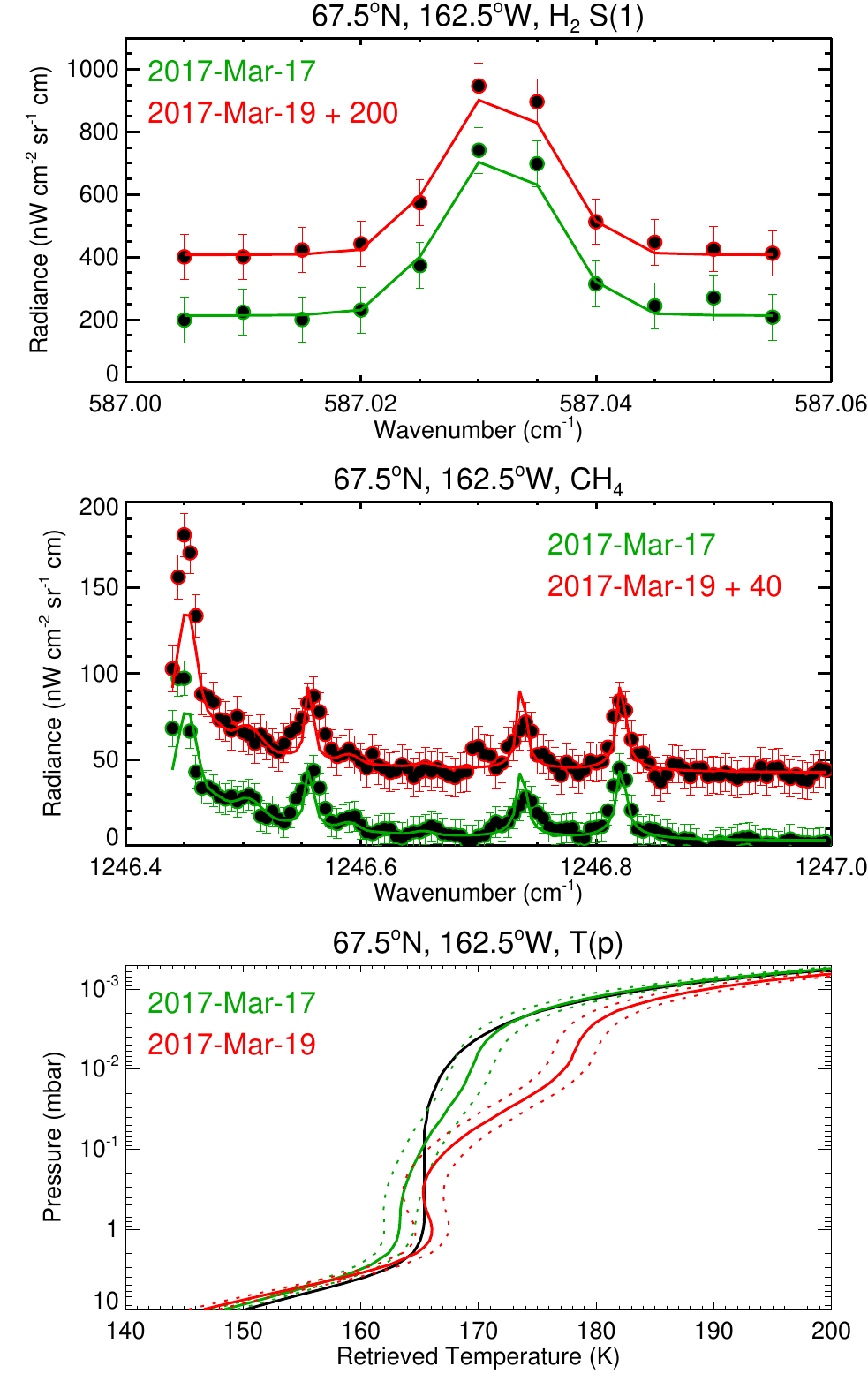}
\caption{Comparisons of observations (points with error bars) and synthetic spectra at 67.5$^\circ$N, 155$^\circ$W in the 587 cm$^{-1}$ (1st row) and 1248 cm$^{-1}$ (2nd row) spectral setting, which respectively capture H$_2$ S(1) and CH$_4$ emissions.  The \textit{a priori} (solid, black) and retrieved temperature profiles (solid, colored lines) and uncertainty (dotted, colored lines) are shown in the bottom panel.  Green and red results denote March-17th and March-19th results, respectively.  }
\label{fig:compare_Tp_17_19}
\end{center}
\end{figure}

\subsection{High-northern latitudes}\label{sec:T_north}

Figure \ref{fig:t_retr_17_19} shows retrieved temperature distributions at high-northern latitudes using observations on March 17th and 19th, 2017.  As for the southern auroral region, temperatures are predominantly elevated at $\sim$1 mbar and at pressures lower than 0.1 mbar.  As noted previously, we observe no significant heating at pressures higher than 2 - 3 mbar, in contrast to what is observed of the southern auroral region.   In comparing the results on March 17th and March 19th, morphological changes in the temperature field are apparent.  On March 17th, the warmest temperatures are in a single region centered at $\sim$183$^\circ$W whereas on March 19th, this region and an additional region of warm temperatures at $\sim$162.5$^\circ$W (the duskside of the northern auroral oval) are apparent.  This is demonstrated further in Figure \ref{fig:contour_vs_long}, which shows longitudinal cross sections of temperature at 67.5$^\circ$N on March 17th and 19th, and their difference.  

The western region of heating extends from approximately 185$^\circ$W to 200$^\circ$W with a $\sim$5 K heating at the 1-mbar level and a $\sim$9 K heating at the 0.01-mbar level.  This results from an (apparent) eastward shift of a region of warmer temperatures.  On March 17th, 0.01-mbar temperatures at 67.5$^\circ$N peak at a longitude of 182.5$^\circ$W with a temperature of 183.0 $\pm$ 1.3 K, whereas on March 19th, the peak temperature (183.4 $\pm$ 1.3 K) instead occurs at 187.5$^\circ$W. This apparent 5$^\circ$ longitude shift is comparable with the diffraction-limited spatial resolution at this latitude ($\sim$5$^\circ$ in longitude) and so we cannot determine whether this apparent westward shift is physical or a result of uncertainty introduced by the spatial registration.  

However, we believe the eastern region of heating centered at $\sim$162.5$^\circ$W is physical and colocated with the MAE within uncertainty. Figure \ref{fig:compare_Tp_17_19} shows observed and synthetic spectra and the corresponding vertical profiles of temperature at 67.5$^\circ$N, 162.5$^\circ$W, where the largest change in temperature between March 17th and 19th occured.   Over the range of vertical sensitivity of the observations (100 mbar $<$ p $<$ 1 $\upmu$bar), the strongest duskside heating occurs over the 40 to 1 $\upmu$bar pressure range.  For example, at 162.5$^\circ$W, the temperature at 9.0 $\upmu$bar increases from 170.0 $\pm$ 1.5 K to 179.1 $\pm$ 1.4 K, or a difference of 9.1 $\pm$ 2.1 K.   We discuss the possible mechanisms by which the arrival of the solar wind compression between March 17th and 19th (Figure \ref{fig:sw}) drove the observed heating in Section \ref{sec:discuss}.

Marginal temperature increases are also observed at $\sim$1 mbar  at $\sim$160$^\circ$W when comparing results on March 17th and 19th.  For example, at 157.5$^\circ$W, we find a 1-mbar temperature difference of 3.3 $\pm$ 1.6 K or $\sim$2.1$\sigma$.  Given only two locations in the 145 - 170$^\circ$W range are just barely over the 2-$\sigma$ level, we consider the apparent lower-stratospheric heating to be a tentative result.


\section{Hydrocarbon Retrievals}\label{sec:cxhy}

The retrieved temperature distributions presented in Section \ref{sec:temp} were adopted as the temperature profile and the 730, 950 and 819 cm$^{-1}$ spectra were inverted in order to retrieve the vertical profiles of C$_2$H$_2$, C$_2$H$_4$ and C$_2$H$_6$, respectively.  The vertical profiles of C$_2$H$_2$, C$_2$H$_4$ and C$_2$H$_6$ shown in Figure \ref{fig:photo_model} were adopted as the \textit{a priori} and nominally, a fractional uncertainty of 18\% was adopted at all altitudes. 

\subsection{C$_2$H$_4$ retrieval artefacts}\label{sec:bad_c2h4}

In our first attempts at retrieving the vertical profile of C$_2$H$_4$, we noticed unphysical discontinuities in abundance at a given altitude between neighboring spatial bins.  This issue seemed to arise at locations where a significant enhancement of C$_2$H$_4$ was required with respect to \textit{a priori} values in order to fit the observations.  The discontinuity results from the different altitudes at which the retrieval chooses to enhance the abundance.  This is exemplified in Figure \ref{fig:bad_c2h4} which shows retrievals of temperature and C$_2$H$_4$ at 72.5$^\circ$N, 165$^\circ$W and 162.5$^\circ$W using observations recorded on March 19th, 2017.  At 165$^\circ$W, the retrieved C$_2$H$_4$ profile exhibits the largest departure from the \textit{a priori} at $\sim$2 mbar whereas at 162.5$^\circ$W, this instead occurs at $\sim$0.3 $\upmu$bar.  In plotting latitude-longitude distributions of C$_2$H$_4$ abundance at a given altitude, this results in several order-of-magnitude changes in C$_2$H$_4$ between neighboring spatial bins, which cannot be physical since the two locations are separated by less than the diffraction-limited spatial resolution.  
\begin{figure*}[!th]
\begin{center}
\includegraphics[width=0.85\textwidth]{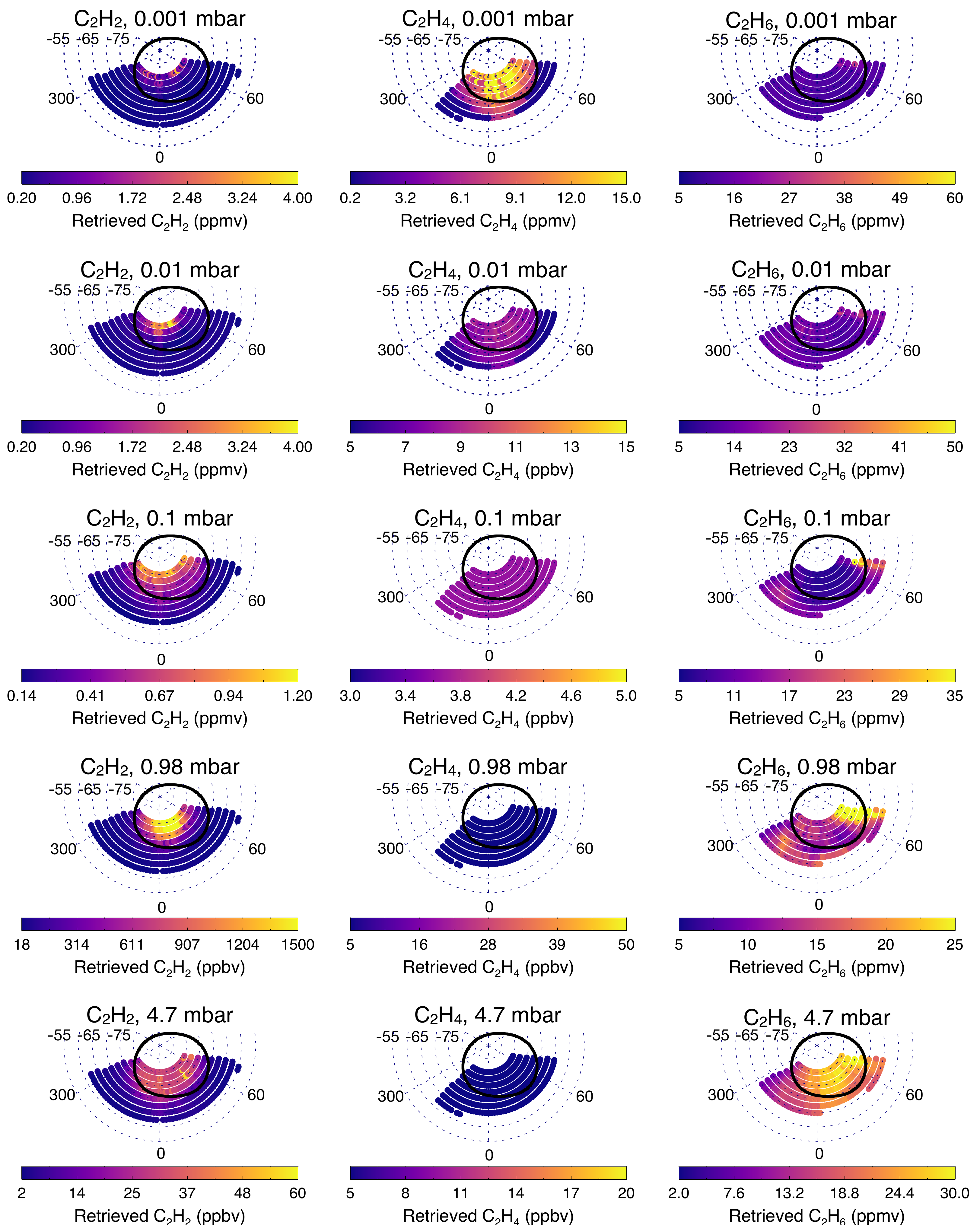}
\caption{Retrieved abundances of C$_2$H$_2$ (left column), C$_2$H$_4$ (middle column) and C$_2$H$_6$ (right column) at high-southern latitudes on March 18th, 2017.  Results are shown at 0.001 mbar (top row), 0.01 mbar (2nd row), 0.1 mbar (3rd row), 0.98 mbar (4th row) and 4.7 mbar (5th row).   The solid, black line denotes the statistical-mean position of the ultraviolet auroral oval \citep{bonfond_2017}. }
\label{fig:cxhy_mar18}
\end{center}
\end{figure*}

As shown in Figure \ref{fig:bad_c2h4}, the retrieved temperature profiles at both locations are very similar and so this cannot be driving the different locations of where the retrieval chooses to enhance C$_2$H$_4$.   We also note that the spectral fit to the 950 cm$^{-1}$ spectrum, particularly to the cores of the C$_2$H$_4$ lines, is significantly better at 162.5$^\circ$W ($\chi^2$/n $\sim$ 1.3), where the C$_2$H$_4$ enhancement occurs in the upper stratosphere, compared to 165$^\circ$W ($\chi^2$/n $\sim$ 2.2), which exhibits a lower-stratospheric enhancement.  Our interpretation is that a subset of retrievals converge on a secondary $\chi^2$ minimum, corresponding to a peak at 2 mbar, and a subset converge on a better-fitting, primary $\chi^2$ minimum corresponding to the peak at 0.3 $\upmu$bar.  C$_2$H$_4$ is predicted to be more sensitive to temperature in the lower stratosphere than C$_2$H$_2$ or C$_2$H$_6$ \citep{moses_2015}, so a C$_2$H$_4$ peak in the lower stratosphere is not unexpected.  However, larger temporal and spatial changes in C$_2$H$_4$ in the upper stratosphere might also be expected given the shorter photochemical lifetimes at $\sim$0.3 $\upmu$bar compared to $\sim$2 mbar (e.g. \citealt{moses_2005,hue_2018}). 

 We tested whether the observations at 72.5$^\circ$N, 165$^\circ$W could still be adequately fit ($\chi^2$/n $\sim$ 1) by allowing the retrieved C$_2$H$_4$ profile to depart from the \textit{a priori} only in the upper stratosphere.  We performed a set of retrievals where the uncertainty on the \textit{a priori} profile was reduced to 1\% at pressures higher than a given ``cutoff'' level, $p_0$, and set to 18\% at pressures lower than and including the cutoff level.  This thereby forces the retrieval to fit the spectra by varying abundances at lower pressures.  We tested values of $p_0$ of 3, 1, 0.3, 0.1, 0.03, 0.01 mbar, which are approximately spaced evenly in logarithmic pressure.  Figure \ref{fig:c2h4_vs_apriori} shows the results of these tests.  In allowing C$_2$H$_4$ to vary at all altitudes or fixing the abundance at pressures higher than 3 mbar, the retrieval chooses to modify the lower-stratospheric (1 - 2 mbar) C$_2$H$_4$ abundance with relatively little change at lower pressure levels.  If the C$_2$H$_4$ abundance is fixed at pressures higher than 1, 0.3, 0.1, 0.03 or 0.01 mbar, the retrieval instead modifies the upper stratospheric C$_2$H$_4$ abundance and the quality of fits are significantly improved ($\chi^2/n$ $\sim$ 1.1) compared to the former ($\chi^2/n$ $\sim$ 2.2 - 2.5).  There is negligible/zero difference in the quality of fit to the spectra in using $p_0$ = 0.3 mbar compared to p$_0$ = 0.01 mbar.

 \begin{figure}[h!]
\begin{center}
\includegraphics[width=0.41\textwidth]{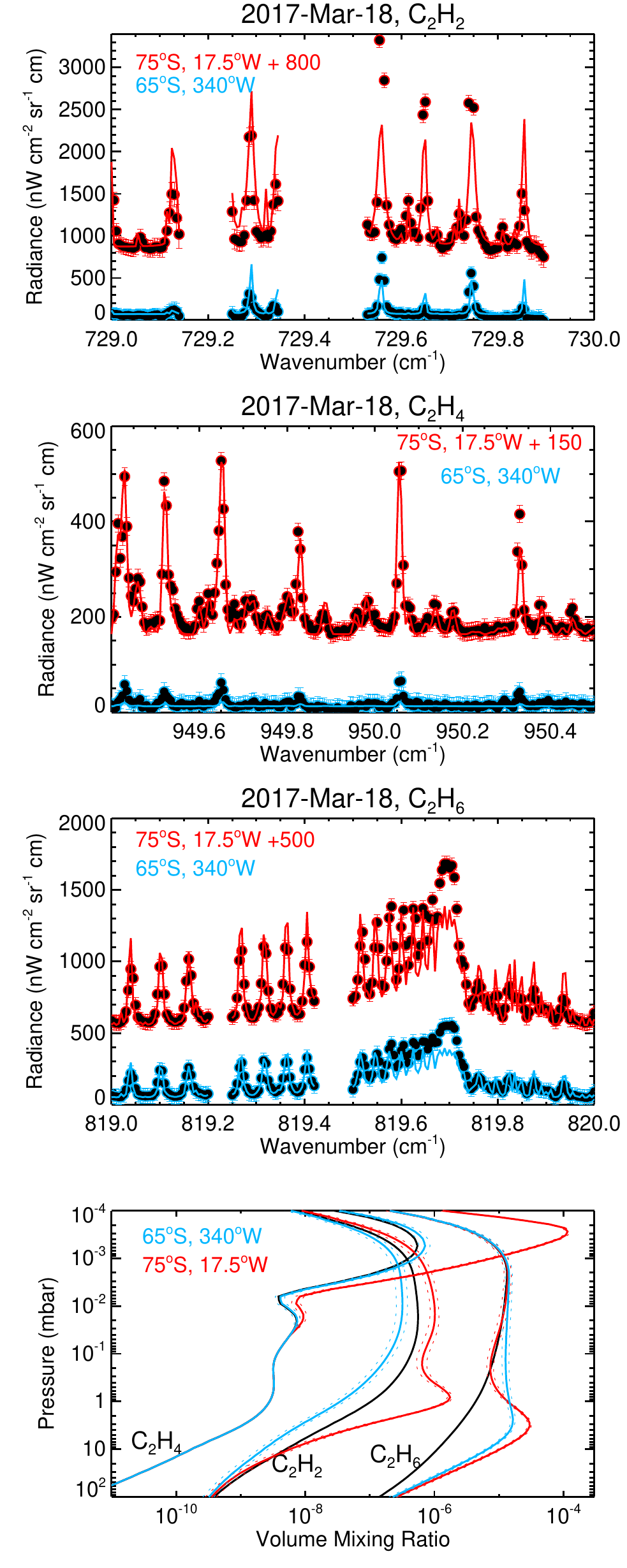}
\caption{Comparisons of observations (points with error bars) and synthetic spectra on March 18th, 2017 in the 730 cm$^{-1}$ (1st row), 950 cm$^{-1}$ (2nd row) and 819 cm$^{-1}$ (3rd row) spectral settings, which respectively capture C$_2$H$_2$, C$_2$H$_4$ and C$_2$H$_6$ emissions.   The \textit{a priori} (solid, black) and retrieved profiles (solid, colored lines) and uncertainty (dotted, colored lines) are shown in the bottom panel.  Blue results denote spectra/results for 65$^\circ$S, 340$^\circ$W and red results denote 72.5$^\circ$S, 17.5$^\circ$W.  }
\label{fig:cxhy_south}
\end{center}
\end{figure}

In summary, we find that allowing the retrieval of C$_2$H$_4$ to vary only at pressures lower than 0.3 mbar significantly improves the quality of fit compared to adopting the nominal approach of a constant, fractional uncertainty at all altitudes.  This also was found to remove the discontinuities in abundances between neighboring spatial bins as noted above.  We therefore adopt the \textit{a priori}, where the uncertainty is set to 1\% at pressures higher than 0.3 mbar, in all C$_2$H$_4$ retrievals presented in the remainder of this work.

\subsection{High-southern latitudes}\label{sec:cxhy_south}

Figure \ref{fig:cxhy_mar18} shows the retrieved distributions of C$_2$H$_2$, C$_2$H$_4$ and C$_2$H$_6$ at high-southern latitudes using observations recorded on March 18th, 2017.  We find that all three hydrocarbons exhibit enhancements in abundance inside the southern auroral oval, though at apparently different altitudes.  This is also demonstrated in Figure \ref{fig:cxhy_south}, which compares the observed and synthetic spectra and corresponding retrieved vertical profiles at 75$^\circ$S, 17.5$^\circ$W and 340$^\circ$W (a location inside and outside the southern auroral oval, respectively).  

In general, we were able to fit the C$_2$H$_4$ emission spectra adequately well ($\chi^2/n$ $\sim$1) using the \textit{a priori} profile detailed in Section \ref{sec:bad_c2h4}.  However, we found it challenging to adequately fit the strong emission lines of C$_2$H$_2$ and C$_2$H$_6$ in the 730 cm$^{-1}$ and 819 cm$^{-1}$ settings, respectively.  For example, as shown in Figure \ref{fig:cxhy_south}, the synthetic spectra struggle to fit the cores of the strong C$_2$H$_2$ emission lines at 729.56 cm$^{-1}$ and 729.75 cm$^{-1}$ and the strong C$_2$H$_6$ emission line at 819.7 cm$^{-1}$ respectively.  This was an issue also observed in a similar analysis of IRTF-TEXES observations recorded in December 2014 \citep{sinclair_2018a}, which was attributed to non-LTE (non local thermodynamic equilibrium) effects.  We discuss both possibilities in further detail in Section \ref{sec:discuss}.   

For C$_2$H$_2$, the largest change in abundance between non-auroral and auroral locations occurs at $\sim$1 mbar.  At 75$^\circ$S, 17.5$^\circ$W (inside the southern auroral oval), we retrieve an abundance of 1.61 $\pm$ 0.13 ppmv compared to 0.04 $\pm$ 0.01 ppmv retrieved at 65$^\circ$S, 340$^\circ$W.  For C$_2$H$_4$, the largest change in abundance between non-auroral and auroral regions occurs significantly higher in the atmosphere at $\sim$1 $\upmu$bar, which is expected given we modified the C$_2$H$_4$ \textit{a priori} uncertainty to favor varying the upper stratospheric abundance - see Section \ref{sec:bad_c2h4}).  At 1 $\upmu$bar, we retrieve an abundance of 9.68 $\pm$ 1.53 ppmv at 75$^\circ$S, 17.5$^\circ$W compared to 0.54 $\pm$ 0.11 ppmv at 65$^\circ$S, 340$^\circ$W.  In contrast, C$_2$H$_6$ exhibits the largest change between non-auroral and auroral locations several decades of pressure deeper - at $\sim$4 mbar.  For example, at 4.7 mbar, the retrieved abundance of C$_2$H$_6$ exhibits an increase from 13.7 $\pm$ 1.1 ppmv at 65$^\circ$S, 340$^\circ$W to 25.1 $\pm$ 1.7 ppmv at 75$^\circ$S, 17.5$^\circ$W.  

The apparent variability of C$_2$H$_2$, C$_2$H$_4$ and C$_2$H$_6$ at very different altitudes, between Jupiter's southern auroral region and a nearby non-auroral location, is puzzling.  At first, we considered whether this was a retrieval artefact.  However, as detailed further in Section \ref{sec:cxhy_vs_apriori}, varying C$_2$H$_6$ abundances only in the upper stratosphere, at similar altitudes as those where C$_2$H$_2$ and C$_2$H$_4$ vary, leads to poorer fits to the spectra.   We discuss this issue further in the Discussion.  

\begin{figure*}[!th]
\begin{center}
\includegraphics[width=0.8\textwidth]{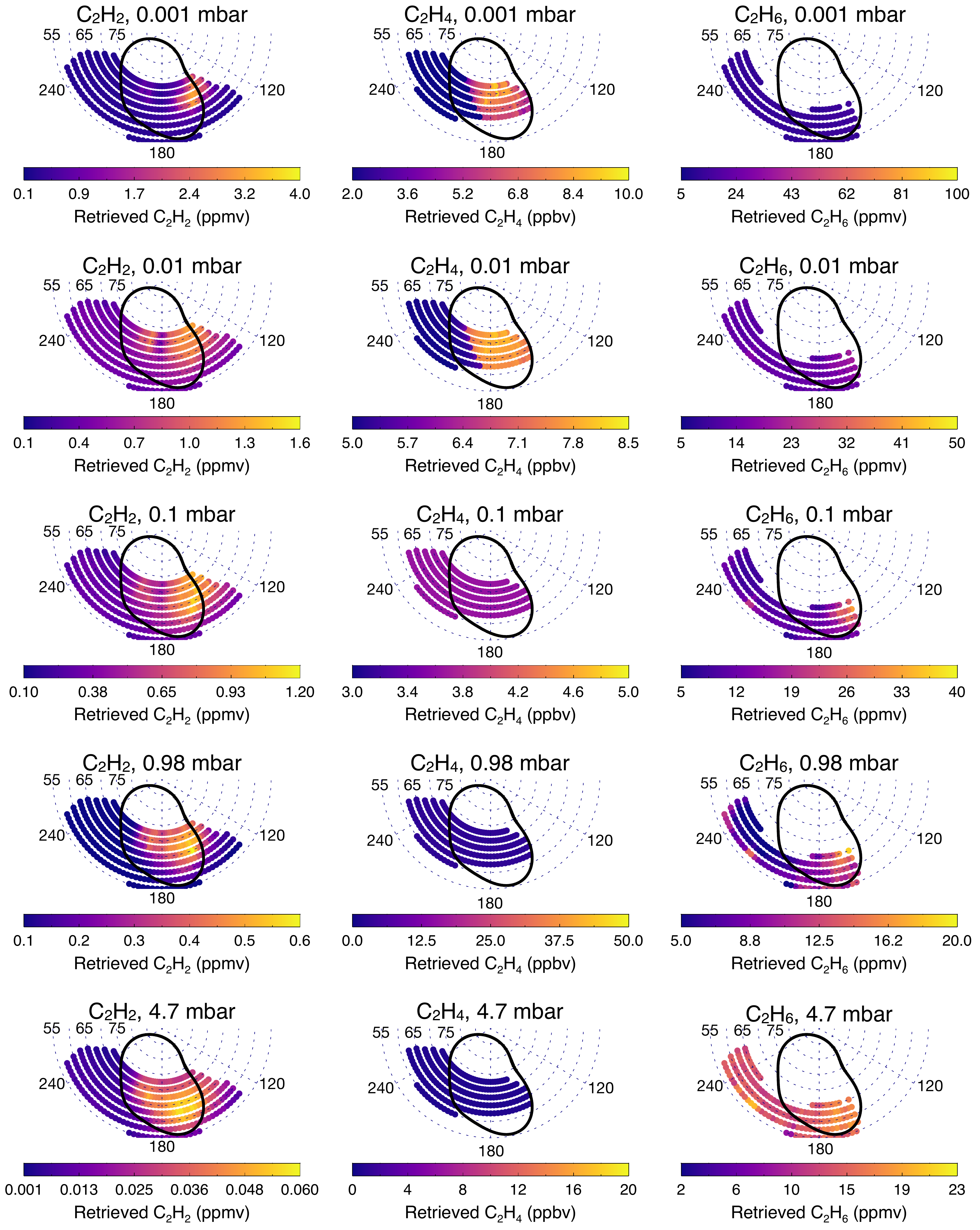}
\caption{Retrieved abundances of C$_2$H$_2$ (left column), C$_2$H$_4$ (middle column) and C$_2$H$_6$ (right column) at high-southern latitudes on March 17th, 2017.  Results are shown at 0.001 mbar (top row), 0.01 mbar (2nd row), 0.1 mbar (3rd row), 0.98 mbar (4th row) and 4.7 mbar (5th row).   The solid, black line denotes the statistical-mean position of the ultraviolet auroral oval \citep{bonfond_2017}.                                                      }
\label{fig:cxhy_mar17}
\end{center}
\end{figure*}

\begin{figure*}[!th]
\begin{center}
\includegraphics[width=0.8\textwidth]{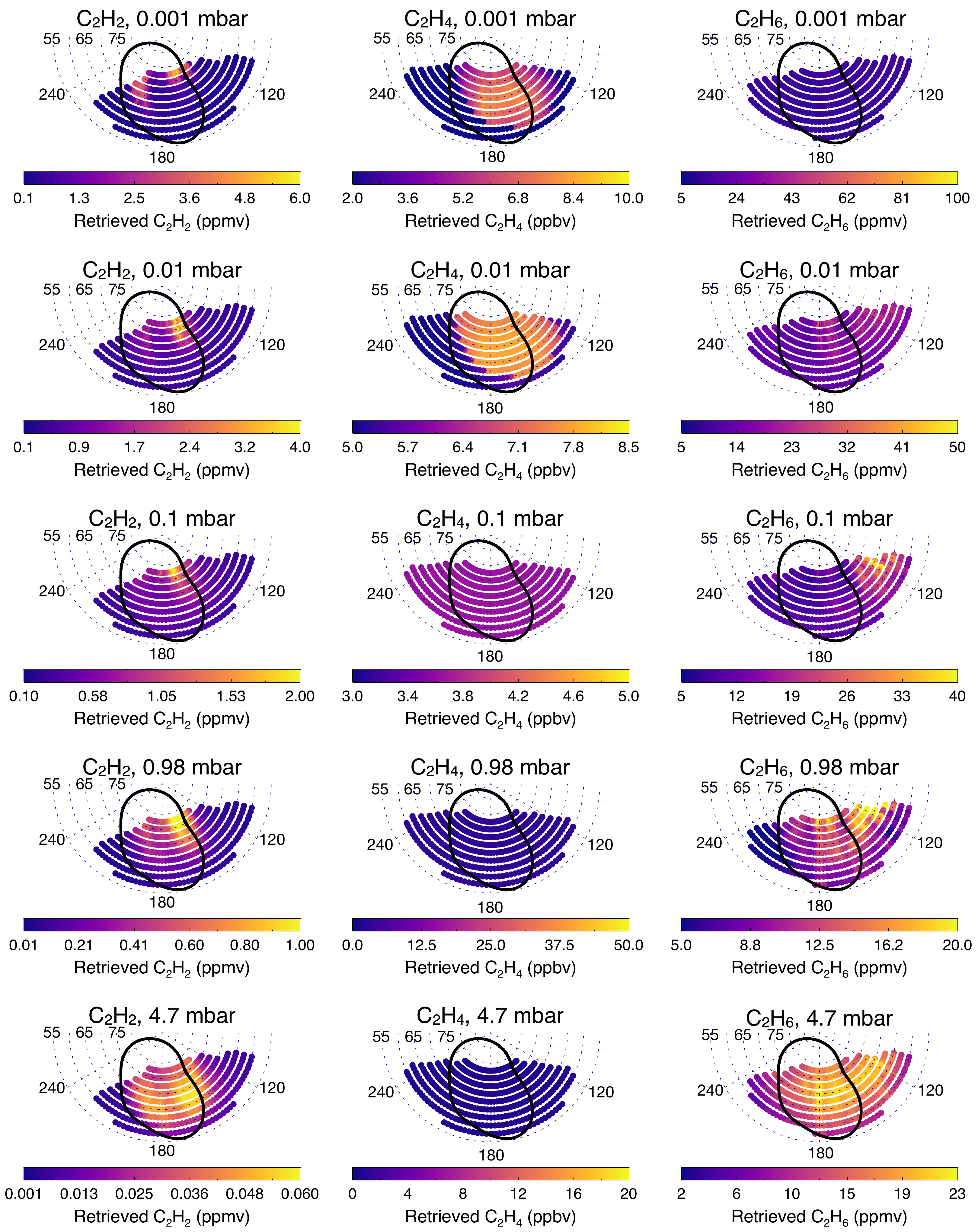}
\caption{As in Figure \ref{fig:cxhy_mar17} but instead using observations on March 19th, 2017 \vspace{5pt}                                }
\label{fig:cxhy_mar19}
\end{center}
\end{figure*}

\subsection{High-northern latitudes}\label{sec:cxhy_north}

Figures \ref{fig:cxhy_mar17} and \ref{fig:cxhy_mar19} show the retrieved hydrocarbon distributions at high-northern latitudes on March 17th and March 19th, respectively.  On March 17th, C$_2$H$_2$ exhibits two discrete regions of enriched abundances within the main auroral oval.  The first region appears on the duskside of the main oval with the largest 0.01-mbar abundance of 1.31 $\pm$ 0.25 ppmv recorded at 72.5$^\circ$N and 147.5$^\circ$W. The second region appears slightly west from the center of the auroral region with a peak 0.01-mbar abundance of 0.93 $\pm$ 0.17 ppmv retrieved at 67.5$^\circ$N, 192.5$^\circ$W. In contrast to C$_2$H$_2$, C$_2$H$_4$ on March 17th exhibits a single region of higher abundances inside the northern auroral oval and is significantly enriched compared to longitudes outside the auroral oval.  For example, at 70$^\circ$N, the retrieved 0.01-mbar abundances increase from 5.08 $\pm$ 0.75 ppbv at 240$^\circ$W to 8.06 $\pm$ 1.18 ppbv at 175$^\circ$W.  Observations of C$_2$H$_6$ emission at 819 cm$^{-1}$ on March 17th did not sample latitudes north of $\sim$65$^\circ$N.  Nevertheless, for the southern part of the northern aurora that were recorded on March 17th, we retrieve higher abundances towards the duskside of the main oval.  For example, at 60$^\circ$N, the 0.98-mbar abundance of C$_2$H$_6$ increases from 6.23 $\pm$ 0.97 ppmv at 240$^\circ$W to 13.40 $\pm$ 1.91 ppmv at 170$^\circ$W.   

\begin{figure}[!th]
\begin{center}
\includegraphics[width=0.45\textwidth]{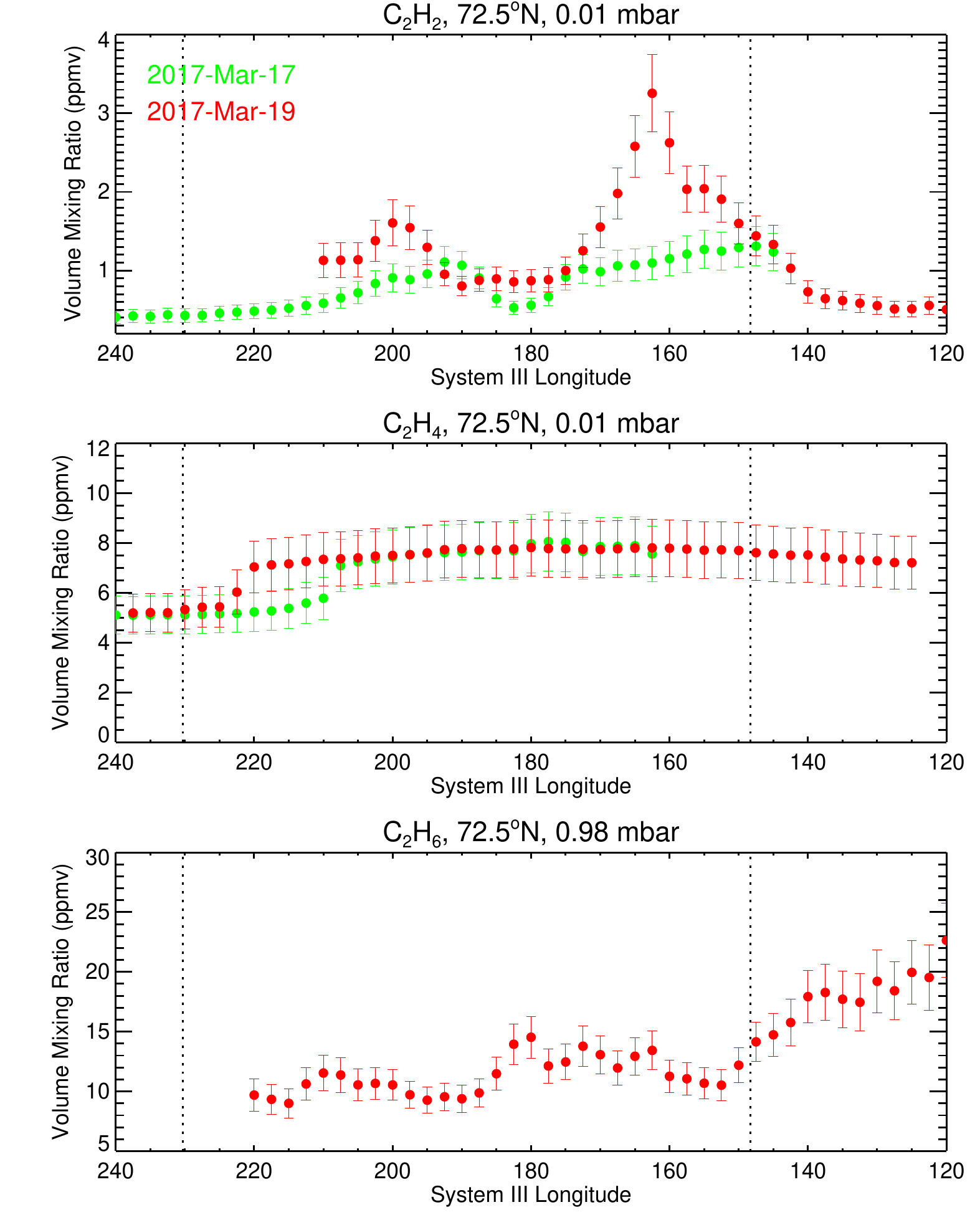}
\caption{Longitudinal variations of retrieved C$_2$H$_2$ at 0.01 mbar (1st row), C$_2$H$_4$ at 0.01 mbar (2nd row) and C$_2$H$_6$ at 0.98 mbar (3rd row) at 72.5$^\circ$N. Results are shown at altitudes where the observations are more sensitive and thus where longitudinal/temporal variations are strongest.  Results for March 17th and 19th are shown in green and red, respectively.  Observations at 819 cm$^{-1}$ on March 17th did not capture the 72.5$^\circ$N and so no C$_2$H$_6$ results are shown for this date.  The vertical, dotted lines denote the statistical-mean position of the ultraviolet main auroral oval \citep{bonfond_2017}.  }
\label{fig:cxhy_vs_long}
\end{center}
\end{figure}
\begin{figure}[ht!]
\begin{center}
\includegraphics[width=0.4\textwidth]{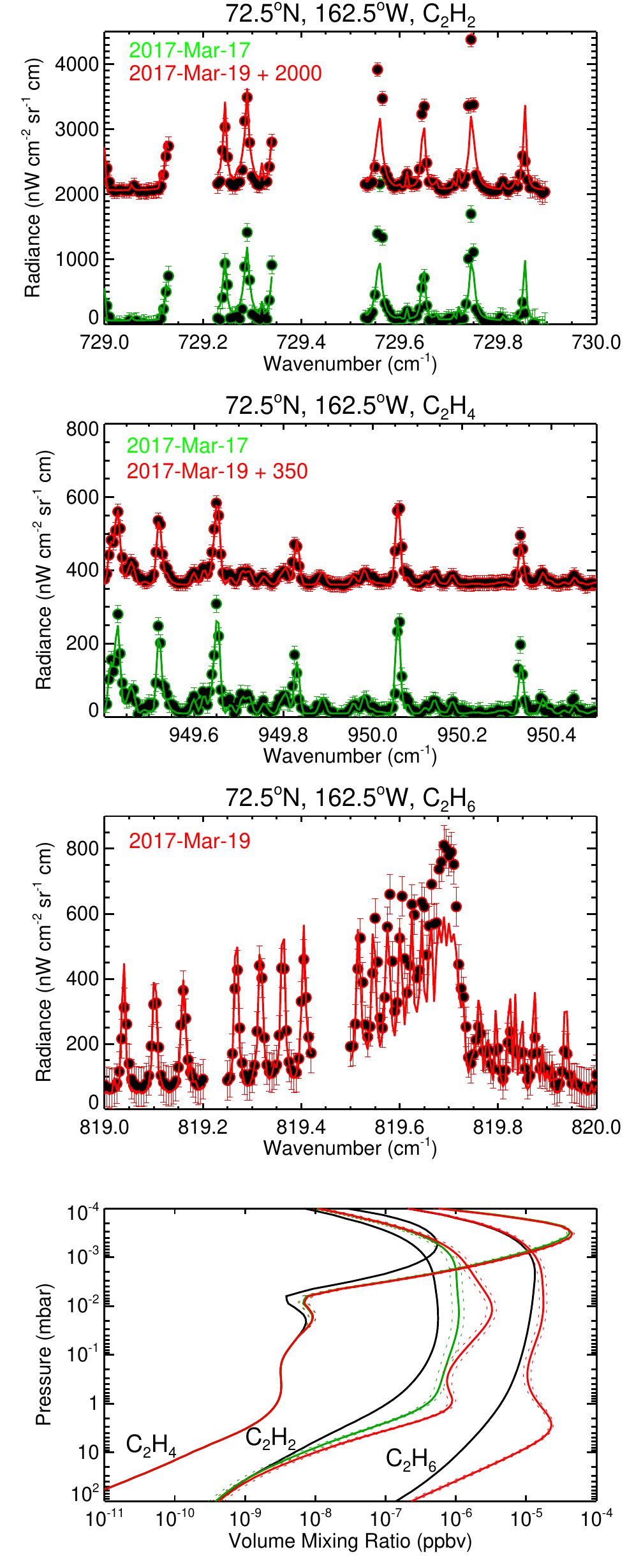}
\caption{Comparisons of observations (points with error bars) and synthetic spectra at 72.5$^\circ$N, 162.5$^\circ$W in the 730 cm$^{-1}$ (1st row), 950 cm$^{-1}$ (2nd row) and 819 cm$^{-1}$ (3rd row) spectral settings, which respectively capture C$_2$H$_2$, C$_2$H$_4$ and C$_2$H$_6$ emissions.  Only a subset of the spectral ranges inverted are shown for clarity. The \textit{a priori} (solid, black) and retrieved profiles (solid, colored lines) and uncertainty (dotted, colored lines) are shown in the bottom panel.  Green and red results denote spectra/results for March 17th and March 19th, respectively. }
\label{fig:cxhy_17mar_19mar}
\end{center}
\end{figure}

In comparing the retrieved distributions of C$_2$H$_4$ and C$_2$H$_6$ between March 17th and March 19th, we do not observe any statistically-significant variability in abundance.  This is also demonstrated in Figure \ref{fig:cxhy_vs_long}, which shows longitudinal variations in C$_2$H$_2$, C$_2$H$_4$ and C$_2$H$_6$ at 72.5$^\circ$N on March 17th and 19th.  However, we do observe significant variability in the retrieved distributions of C$_2$H$_2$ between March 17th and 19th, particularly on the dusk side of the main oval.  For example, at 72.5$^\circ$N and 162.5$^\circ$W, we retrieve an 0.01-mbar abundance of 1.10 $\pm$ 0.21 ppmv on March 17th and an abundance of 3.25 $\pm$ 0.49 ppmv on March 19th, which represents an increase by approximately a factor of 3.  This is also coincident spatially and vertically with the $\sim$9 K heating observed between March 17th and 19th, as discussed previously in Section \ref{sec:T_north}, and presumably driven by a common mechanism.  As shown in Figure \ref{fig:cxhy_17mar_19mar}, lower-stratospheric abundances of C$_2$H$_2$ at 72.5$^\circ$N, 162.5$^\circ$W also exhibit a smaller yet still statistically-significant increase between the two dates. For example, at 0.98 mbar, the abundance increases from 397.1 $\pm$ 48.7 ppbv on March 17th to 873.6 $\pm$ 101.8 ppmv on March 19th. In Section \ref{sec:cxhy_vs_apriori}, we tested whether the observations can be adequately fit by varying only upper-stratospheric C$_2$H$_2$ abundances.  However, we find that the best fits to the observations are achieved when 1-mbar and 0.01-mbar abundances are both allowed to vary. 

At 200$^\circ$W in the same latitude band, we also observe a smaller but still significant increase in the 0.01-mbar C$_2$H$_2$ abundance from 0.91 $\pm$ 0.18 ppmv on March 17th to 1.61 $\pm$ 0.29 ppmv on March 19th.  This is also coincident spatially with an apparent region of heating observed between March 17th and 19th although, as discussed in Section \ref{sec:temp}, we are uncertain whether this specific region of heating is real or an artefact of uncertainties on spatial registration.  We therefore consider this region of an apparent C$_2$H$_2$ abundance increase at 200$^\circ$W a tentative result.

\subsection{A priori testing}\label{sec:cxhy_vs_apriori}

We performed additional retrievals of C$_2$H$_6$ at 72.5$^\circ$S, 17.5$^\circ$W (a location inside the southern auroral oval) and C$_2$H$_2$ at 72.5$^\circ$N, 162.5$^\circ$W (the duskside of the northern oval) to explore the robustness of results presented in Sections \ref{sec:cxhy_south} and \ref{sec:cxhy_north}.  While retrievals of C$_2$H$_2$, C$_2$H$_4$ and C$_2$H$_6$ show the southern auroral oval is enriched in all three hydrocarbons compared to a location outside the oval, the enrichment of C$_2$H$_6$ is strongest in the lower stratosphere ($\sim$4 mbar) compared to that of C$_2$H$_2$ and C$_2$H$_4$, which is stronger in the upper stratosphere (1 - 10 $\upmu$bar).  Our goal was to test whether similar or improved fits to the observations of C$_2$H$_6$ could be achieved by allowing the enrichment of C$_2$H$_6$ to occur at the same altitudes as that of C$_2$H$_2$ and C$_2$H$_4$.  Secondly, a comparison of retrievals of C$_2$H$_2$ at 72.5$^\circ$N, 162.5$^\circ$W between March 17th and 19th show an increase in abundance at both $\sim$1 mbar and $\sim$0.01 mbar.  Our goal was to test whether increasing the abundance only in the upper stratosphere could fit the observations adequately.  

\begin{figure}[!th]
\begin{center}
\includegraphics[width=0.45\textwidth]{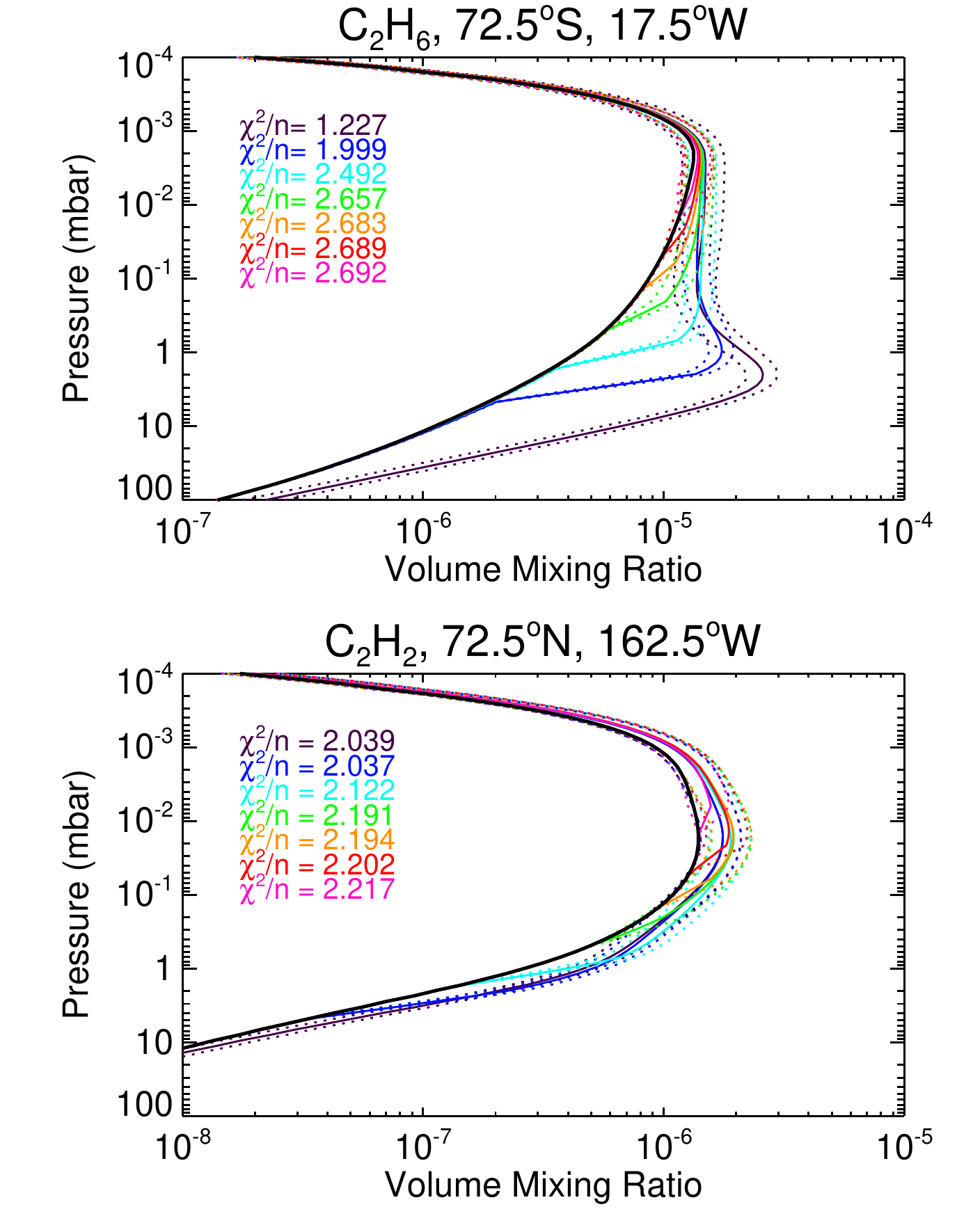}
\caption{Test retrievals of C$_2$H$_6$ at 72.5$^\circ$S, 17.5$^\circ$W (top panel) and C$_2$H$_2$ at 72.5$^\circ$N, 162.5$^\circ$W (bottom panel).  Black denotes the \textit{a priori}, retrieved profiles are solid, colored lines and uncertainties  on retrieved profiles are dotted, colored lines.  Dark, purple denotes the profile retrieved when an 18\% fractional uncertainty is adopted at all altitudes.  The remaining profiles are those retrieved when the \textit{a priori} uncertainty was decreased to 1\% at pressures higher than 3 mbar (blue), 1 mbar (cyan), 0.3 mbar (green), 0.1 mbar (orange), 0.03 mbar (red), 0.01 mbar(pink).  The corresponding reduced $\chi^2$ values are also shown in each plot. }
\label{fig:cxhy_vs_apriori}
\end{center}
\end{figure}

In order to perform these tests, we use the same approach adopted in Section \ref{sec:bad_c2h4}, where the uncertainty on the \textit{a priori} abundance is decreased significantly at pressures higher than a cutoff pressure, $p_0$.  Again, we tested values of $p_0$ = 3, 1, 0.3, 0.1, 0.03, 0.01 mbar.  The results of these tests are shown in Figure \ref{fig:cxhy_vs_apriori}.  

For C$_2$H$_6$ in the southern auroral region, the profile retrieved using the nominal approach, where a constant fractional uncertainty is adopted for the \textit{a priori} at all altitudes, is indeed the best-fitting solution ($\chi^2/n$ = 1.23).  In reducing the fractional uncertainty to 1\% at pressures higher than 3 mbar, such that the retrieval is forced to vary C$_2$H$_6$ at lower pressures, the quality of fit to the spectra degrades significantly ($\chi^2/n \sim$ 2) and even poorer fits are obtained using even lower $p_0$ pressures.  Thus, the contrast in C$_2$H$_6$ abundance between a non-auroral and auroral location at predominantly $\sim$4 mbar does appear to be most consistent with the observations.   This is in contrast to results for C$_2$H$_2$ and C$_2$H$_4$, which exhibit the largest spatial variation at significantly lower pressures ($\sim$0.01 mbar).

For C$_2$H$_2$ on the duskside of the northern oval, profiles retrieved using a constant fractional uncertainty at all altitudes or where the fractional uncertainty was set to 1\% are pressures higher than 3 mbar result in very similar quality of fits to the observations ($\chi^2/n$ = 2.04).  The fact that $\chi^2/n$ $\sim$2 represents our best-fitting solution again denotes the challenge in fitting both the weak and strong emission lines of C$_2$H$_2$.  We attribute this to either non-LTE effects and/or temporal variability of the atmosphere between measurements in different spectral settings, as discussed further in Section \ref{sec:discuss}.  In fixing the retrieved profile at pressures higher than 1 mbar, 0.3 mbar and so on, the quality of fit to the observations degrades ($\Delta \chi^2/n \sim$0.2).  The best-fitting retrieval corresponds to an absolute $\chi^2$ value of 265.08.  Retrieved profiles resulting in absolute $\chi^2$ values of 265.08 + 1 = 266.08, 265.08 + 4 = 269.08 and 265.08 + 9 = 275.08 would signify the 1-$\sigma$, 2-$\sigma$ and 3-$\sigma$ confidence levels, respectively \citep{press_1992}.  In fixing the \textit{a priori} at pressures higher than 1, 0.3, 0.1, 0.03, 0.01 mbar, the absolute $\chi^2$ are 276.98, 285.97, 286.43, 287.34, 289.34, which are all outside the 3-$\sigma$ confidence level relative to the best-fitting retrieved profile.  Thus, the increase in C$_2$H$_2$ abundance at 72.5$^\circ$N, 162.5$^\circ$W at both the 1-mbar and 0.01-mbar levels does appear to best reproduce the observations.

\section{Discussion}\label{sec:discuss}

Spectra of Jupiter's H$_2$ S(1), CH$_4$, C$_2$H$_2$, C$_2$H$_4$ and C$_2$H$_6$ emission were recorded at Jupiter's mid-to-high latitudes using TEXES \citep{lacy_2002} on Gemini-North on March 17-19th, 2017.  These measurements provided a rare combination of high spectral and spatial resolution observations that allowed the 3-dimensional (latitude, longitude, altitude) distributions of temperature and hydrocarbon abundances, and their variability, to be constrained.   Using OMNI measurements at Earth and the results of two different solar wind propagation models \citep{tao_2005,zieger_2008} that calculated the flow out to Jupiter's orbit, a solar wind compression arrived at Jupiter on March 18th resulting in an increase in solar wind dynamic pressure of $\Delta p_{dyn} $=0.25 - 0.45 (depending on the model adopted).  While the 1-D nature of such solar wind propagation models can introduce uncertainties on timing and magnitude of downstream solar wind properties, these are expected to be $<$20 hours and $<$38\%, respectively, given Jupiter was within 50$^\circ$ of opposition \citep{zieger_2008}.  Our analysis therefore captures how the stratospheric thermal structure and composition were modulated by the arrival of a solar-wind compression and its perturbing effects on Jupiter's magnetosphere.  

\subsection{Upper stratospheric heating}
\label{sec:discuss_upper_heat}

At high-northern latitudes, spectra of H$_2$ S(1) and CH$_4$ emission recorded on March 17th and 19th 2017 were inverted to derive temperature distributions and their net variability over $\sim$2 days (Figure \ref{fig:t_retr_17_19}). The difference between the temperature distributions on March 17th and 19th (Figure \ref{fig:compare_Tp_17_19}) reveal two regions of heating, predominantly in the upper stratosphere (1 - 10 $\upmu$bar).  The first appears at $\sim$68$^\circ$N , $\sim$190$^\circ$W with $\sim$5-K heating at the 1-mbar level and $\sim$9-K heating between 1 - 10 $\upmu$bar and appears to result from a 5$^\circ$N westward shift of the longitude-temperature distribution between March 17th and 19th, 2017.  Given 5$^\circ$ in longitude is similar to the diffraction-limited spatial resolution achieved at this latitude, we consider this first region of heating to likely be an artefact of uncertainties on spatial registration of the spectral cubes. However, we observe a second region of transient heating on the duskside of the northern oval, which we believe to be physical.  The heating occurs predominantly in the upper stratosphere with a temperature increase at 9 $\upmu$bar of 9.1 $\sim$ 2.1 K at 67.5$^\circ$N, 162.5$^\circ$W, as an example location/altitude.  This location is spatially coincident with the MAE (within uncertainty).  

In \citet{sinclair_2019a}, broadband 7.8-$\upmu$m images of Jupiter's CH$_4$ emission on January 12th, 2017, also captured a bright, elongated feature colocated with the northern duskside MAE.  This feature was absent from images recorded less than 24 hours later, which demonstrated its transience.  However, without images recorded before January 12th, it could not be determined how long the feature had been present.  From the same dataset, Jupiter's southern auroral oval exhibited a brightening, then dimming of CH$_4$ emission between January 11th - 14th 2017 and contemporaneous with the northern, duskside feature, which suggested both phenomena were driven by a common mechanism.   Both southern and northern phenomena were tentatively linked to the predicted arrival of a solar wind compression on approximately January 12th, with the 48-hour timing uncertainty on the solar wind propagation model results hindering a more conclusive link.  In this work, we believe our measurements have captured a similar brightening of the CH$_4$ emission on the duskside of the MAE between March 17th and 19th, 2017.  However, in contrast to \citet{sinclair_2019b}, we can more confidently link this phenomenon to the arrival of a solar-wind compression on March 18th since the propagation model timing uncertainty was less than 20 hours.  A further advance of this work is that the high-resolution spectra recorded by TEXES allowed retrievals of vertical temperature profiles, which allow us to disentangle the altitudes at which temperature changes occured

Given the duskside, upper stratospheric heating observed in this work is horizontally and vertical coincident with the Lyman-$\alpha$ ultraviolet main auroral emissions, we believe their variability in response to solar wind compressions are driven by similar mechanisms.  A leading theory for the generation of Jupiter's ultraviolet main auroral emissions is by the production of field-aligned currents, driven by the breakdown of corotation in the magnetospheric plasma at 20 - 30 R$_J$ (e.g. \citealt{cowley_2003,hill_2001,southwood_2001}).  The currents of charged particles bombard and excite molecular hydrogen in Jupiter's atmosphere, which subsequently de-excite through Lyman-$\alpha$ emission at $\sim$0.12 $\upmu$m.  The same currents will also ultimately warm the upper atmosphere through processes including Joule heating, ion drag and chemical heating (e.g. \citealt{grodent_2001,yates_2014}).  However, the corotation breakdown theory would predict a dimming of the main auroral emissions in response to solar wind forcing (e.g. \citealt{bonfond_2020b}), whereas there is overwhelming evidence to the contrary (e.g. \citealt{clarke_2009,kita_2016,nichols_2017,kita_2018,odonoghue_2021,yao_2022}).  Alternative mechanisms for coupling the solar wind to the main auroral emissions have been suggested.  This includes Kelvin-Helmholtz instabilities at the flanks of the magnetosphere due to the increased velocity shears during a solar wind enhancement \citep{delamere_2010}. This allows solar wind plasma to enter the magnetosphere, drive magnetospheric flows which ultimately accelerates the currents driving the main auroral emissions and heating.  In addition, \citet{pan_2021} found a positive correlation between ultralow-frequency wave activity and ultraviolet auroral power, which suggests Alfv\'enic waves could also be a significant mechanism in coupling the magnetosphere and auroral emissions \citep{saur_2018}.   \citet{yao_2019} presented near-simultaneous Juno, HST and Hisaki measurements of Jupiter over the same period as the observations in this work.  They also observed variability in the ultraviolet and kilometric wave emissions over the March 17 - 22, 2017 time period, which they attribute to cycles of magnetic loading and unloading of the magnetosphere at 60 - 80 R$_J$.  They suggest that either auroral intensification or a current loop couples the loading and unloading events in the outer magnetosphere to the middle magnetosphere (20 - 30 R$_J$), which is the expected magnetospheric origin of the main auroral emissions.

\subsection{Lower stratospheric auroral-related heating}\label{sec:lower_strato_heating}

Jupiter's northern auroral region is host to localized heating at $\sim$1 mbar (e.g. Figure \ref{fig:t_retr_17_19}), as demonstrated in previous studies (e.g. \citealt{kostiuk_agu_2016,sinclair_2017a,sinclair_2017b,sinclair_2020b}).  A temperature minimum occurs at $\sim$0.1 mbar, which suggests the 1-mbar heating is driven by a different mechanism compared to the upper stratospheric heating (Section \ref{sec:discuss_upper_heat}).  This interpretation is supported by the analyses of \citet{ozak_2013}, \citet{gustin_2016} and \citet{houston_2020} which demonstrates that energy from magnetospheric particles is not deposited deeper than 0.1 mbar or $\sim$200 km.  Indeed, at locations where significant upper stratospheric heating was observed in response to the arrival of a solar-wind compression (e.g. 67.5$^\circ$N, 162.5$^\circ$W, Figure \ref{fig:compare_Tp_17_19}), there was only marginal/tentative evidence of heating deeper than 0.1 mbar in comparing March 17th and 19th results.  We discuss the possible mechanisms for the 1-mbar auroral-related heating below.

The spatial resolving power provided by Gemini-North's 8-meter primary together with the relatively high southern sub-observer latitude ($\sim$3.5$^\circ$S) at the time of observations has allowed for rare, high spatial- and spectral-resolution observations of the southern auroral oval.  The southern auroral oval is otherwise challenging to sample from smaller Earth-based telescopes (with lower diffraction-limited resolutions) since it is at a relatively higher latitude compared to the northern auroral oval.  In inverting spectra of H$_2$ S(1) quadrupole line and CH$_4$ emission to retrieve the vertical temperature profile, we also find a deeper, discrete level of heating associated with the aurora.  However, unlike the lower stratospheric heating in the northern auroral region which extends to 2 - 3 mbar (Figure \ref{fig:compare_Tp_17_19}), lower stratospheric heating of the southern auroral region is evident almost a decade of pressure higher: as deep as the $\sim$10-mbar level (Figure \ref{fig:spx_retr_south}).  This result can also be inferred simply by comparing high-southern and high-northern latitude maps of the H$_2$ S(1) quadrupole line emission (Figure \ref{fig:spx_587_north} and \ref{fig:spx_587_south}), which sounds the 50- to 5-mbar level (Figure \ref{fig:contr_functions}a).  Heating as deep as 10 mbar was not found in previous analysis of IRTF-TEXES and Cassini-CIRS observations with limited views of the southern auroral oval \citep{sinclair_2017a,sinclair_2018a,sinclair_2020b}.  This suggests the presence of 10-mbar heating is either a transient feature, perhaps related to the arrival of the solar wind compression at Jupiter $\sim$6 hours earlier (Figure \ref{fig:sw}) and/or the limited spatial resolution of previous observations simply did not resolve smaller-scale regions at high latitudes where such 10-mbar heating occurs.  We favor the latter explanation since the temperature profile at $\sim$10 mbar is constrained predominantly by H$_2$ S(1) quadrupole emission at 587 cm$^{-1}$, which is the longest wavelength setting in this work and the most limited in spatial resolution by diffraction. Our working theory for the mechanism of the lower-stratospheric heating would also rule out the former hypothesis, as detailed below.

The mechanisms responsible for the lower stratospheric auroral-related heating have proven elusive.  Previous studies have suggested shortwave solar heating of haze particles (e.g. \citealt{sinclair_2017a,sinclair_2018a}) generated by the unique chemistry inside Jupiter's auroral ovals \citep{wong_2000,friedson_2002,wong_2003}.  However, we have ruled this out as the dominant mechanism responsible for the lower stratospheric auroral-related heating since 1-mbar temperatures have been observed to vary over a larger temperature range ($\sim$20 K) and on timescales too short ($<$3 months) to be explained by temporal variations in solar insolation \citep{sinclair_epsc_2018}.  We also considered pumping of the CH$_4$ 3-$\upmu$m band by overlapping H$_3^+$ emission lines from higher in the atmosphere.  This would in turn affect the transitions responsible for the 8-$\upmu$m band, a component of which originate from the $\sim$1 mbar level.  However, we have also ruled this out as a dominant mechanism since the strongest H$_3^+$ emissions are spatially coincident with the main auroral emission, whereas the strongest 1-mbar heating is generally observed coincident with the poleward emissions. 

Instead, we favour the explanation presented in a recent analysis of ALMA observations by \citet{cavalie_2021}, where the doppler shift of atmospheric lines on the limb of the planet was used to derive zonal wind velocities.  Using the HCN lines, which sound $\sim$0.1 mbar, they observed both an eastern and western jet horizontally-coincident with the southern main auroral emission. These winds were inferred to have been generated from acceleration of the neutral atmosphere by energetic ions and possibly an extension of a counterotating electrojet observed in the ionosphere (e.g. \citep{achilleos_2001,johnson_2017}).  \citet{cavalie_2021} interpreted that counterrotation would result in atmospheric subsidence enclosed within the jet boundary.  The compression of atmospheric gas as it descends would yield adiabatic heating deeper in the atmosphere and could be the mechanism responsible for the auroral-related lower stratospheric heating.  The adiabatically-heated gas would be confined inside the vortex until deeper altitudes where the vortex dissipates and horizontal mixing can occur.  Magnetospheric flows driven by internal processes or solar wind perturbations would accelerate currents into the neutral atmosphere.  This would in turn accelerate the vortex through ion-neutral collisions, a higher rate of subsidence and adiabatic heating.  Lower-stratospheric temperatures poleward of the main auroral emissions would therefore be expected to be modulated by magnetospheric events/solar wind conditions but with a phase lag compared to the upper stratospheric heating and ultraviolet main auroral emissions.  We believe this working hypothesis accounts for how lower stratospheric temperatures in Jupiter's auroral regions have been observed to vary on monthly timescales \citep{sinclair_epsc_2018} but not on the daily timescales demonstrated for upper stratospheric temperatures (see Section \ref{sec:discuss_upper_heat}).  \citet{cavalie_2021} also observed a strong westward jet at mid-northern latitudes, which may be one component of a similar, counterotating jet coincident with the northern main auroral emission, however, future ALMA observations would be required to conclusively determine its presence/absence.

In order to explain the stronger and deeper heating in the southern auroral oval compared to the north, we suggest the following.  First, the southern auroral oval spans a smaller range in latitude/longitude and thus energy deposited there is concentrated to a much smaller region.  Additionally, the southern auroral oval overlaps with the rotational axis and so adiabatically-heated gas at $\sim$1 mbar would be less efficiently diffused/advected horizontally and so vertical mixing would allow the warm gas to be transported deeper in the atmosphere.  Second, \citet{kotsiaros_2019} demonstrated that Pedersen conductivities are higher in the southern auroral oval compared to the north.  Higher conductivities would allow stronger currents, greater acceleration of the neutrals and spin-up of the vortex and ultimately stronger lower stratospheric heating.   ALMA observations of the stratospheric winds simultaneously, in both auroral regions, would help to check whether or not southern auroral winds are stronger than northern ones.

\subsection{Hydrocarbon results}

The magnitude of the spectral emission features of C$_2$H$_2$, C$_2$H$_4$ and C$_2$H$_6$ depend both on the vertical temperature profile and the vertical profile of the emitting molecule.  Retrievals of their abundance were required in order to disentangle whether spatial/temporal variations in emission were the result of temperature changes alone or both changes in temperature and abundance.  Adopting the temperature distributions inverted from the H$_2$ S(1) and CH$_4$ emission spectra, 3-D (longitude, latitude, altitude) distributions of C$_2$H$_2$, C$_2$H$_4$ and C$_2$H$_6$ abundances on March 17th, 18th and 19th were retrieved.  As detailed further in Section \ref{sec:bad_c2h4}, we modified the \textit{a priori} profile of C$_2$H$_4$ such that solutions varying only the upper stratospheric abundance would be favoured in order to avoid unphysical, spatial discontinuities. 

In comparing retrieved hydrocarbon abundances between March 17th and 19th, we found a statistically-significant increase in the abundance of C$_2$H$_2$ in both the lower and upper stratosphere spatially coincident with the duskside MAE.  For example, at 72.5$^\circ$N, 162.5$^\circ$W, we retrieve a 0.01-mbar C$_2$H$_2$ abundance of 1.10 $\pm$ 0.21 ppmv on March 17th and 3.25 $\pm$ 0.49 ppmv on March 19th: approximately a three-fold increase (Figure \ref{fig:cxhy_mar17}, \ref{fig:cxhy_mar19}, \ref{fig:cxhy_17mar_19mar}).  At the same locations and dates, the 1-mbar C$_2$H$_2$ abundance increased from 397.1 $\pm$ 48.7 ppbv to 873.6 $\pm$ 101.8 ppbv, which is a smaller fractional increase compared to that at 0.01 mbar, yet still statistically-significant.  At the intermediate level of 0.1 mbar, there was negligible change in C$_2$H$_2$ abundance with respect to uncertainty. As demonstrated in Section \ref{sec:cxhy_vs_apriori}, we could not achieve the same quality of fits to the C$_2$H$_2$ emission spectra in varying only the upper-stratospheric abundances: a bifurcated change in both the lower and upper stratosphere does optimize the fit to the observations (Figure \ref{fig:cxhy_vs_apriori}).  At this same location (72.5$^\circ$N, 162.5$^\circ$W), we found no statistically-significant change in the C$_2$H$_4$ abundance between March 17th and 19th (Figure \ref{fig:cxhy_17mar_19mar}) and so the duskside brightening of C$_2$H$_4$ emission appears to result from heating of the atmosphere alone. Observations of C$_2$H$_6$ emission at 819 cm$^{-1}$ at this location were only recorded on March 19th and so we cannot determine the variability of the C$_2$H$_6$ abundance between the two dates.

Observations on March 18th of high-southern latitudes provided a snapshot of the hydrocarbon abundances inside and outside the southern MAE.  We found that C$_2$H$_2$, C$_2$H$_4$ and C$_2$H$_6$ all exhibit enrichments in abundance inside the southern auroral oval in comparison to a location equatorward of the southern auroral region (Figure \ref{fig:cxhy_south}).  However, the altitudes at which the enrichment occurs differs for each hydrocarbon. 

It is very challenging to reconcile these results for several reasons.  First, photochemical models of Jupiter demonstrate that the production timescale for C$_2$H$_2$ is on the order of $\sim$100 days at 1 - 10 $\upmu$bar, increasing to $\sim$300 days at 1 mbar \citep{nixon_2007,moses_2005,hue_2018}.  It is possible that the production timescales in the upper stratosphere may be shorter in Jupiter's auroral regions since these models do not account for the higher rates of ion-neutral and electron-recombination reactions expected in this region \citep{sinclair_2017a,sinclair_2019a}.   Nevertheless, the apparent three-fold increase in C$_2$H$_2$ abundance at 0.01 mbar, and two-fold increase at $\sim$1 mbar over a 2-day timescale seems unphysically large.  While the vortex we suggest as the mechanism for the lower-stratospheric heating (Section \ref{sec:lower_strato_heating}) would also advect hydrocarbon-richer to lower altitudes, we would expect such a process to be phase-lagged and occur over longer timescales than $\sim$2 days.  Second, C$_2$H$_2$, C$_2$H$_4$ and C$_2$H$_6$ are strongly coupled photochemically and so it is challenging to explain the apparent increase in C$_2$H$_2$ without a corresponding increase in C$_2$H$_4$ and C$_2$H$_6$. 

We suggest the following reasons to explain the physically-counterintuitive hydrocarbon results described above.  First, the altitudes ranges of sensitivity in the 587 and 1248 cm$^{-1}$ spectral settings used to constrain temperature, and those in the 730, 819 and 950 cm$^{-1}$ used to constrain the C$_2$ hydrocarbon abundances do differ (Figure \ref{fig:contr_functions}).  The optimal estimation technique used by the NEMESIS radiative transfer code \citep{irwin_2008} iteratively adjusts the vertical profile of a parameter and will converge on a solution that minimizes the cost function (Equation \ref{eq:cost}).  Thus, retrievals will favor solutions where a variable parameter is increased/decreased at altitudes of the greatest sensitivity.  This, together with the degeneracy in temperature and abundance in reproducing the observed emission features, could be driving unphysical hydrocarbon solutions.  Second, we expect the uncertainty on the spatial registration of the spectral cubes to be similar to the diffraction-limited spatial resolution, which corresponds to a $\sim$5$^\circ$ latitude-longitude footprint at 60$^\circ$N.  Spatial offsets between the different spectral settings could mean that the emission features used to constrain temperature capture a slightly different horizontal location compared to those used to constrain hydrocarbon abundances.  Third, retrievals of temperature have assumed the vertical profile of CH$_4$ is horizontally homogenous.  \citet{sinclair_2020b} found no statistically-significant variation in the vertical profile of CH$_4$ inside the northern auroral oval, however, it is possible that spatial variations do exist over a range smaller than uncertainty.  Our analysis would instead interpret these variations as temperature and not CH$_4$ abundance, which would in turn effect the hydrocarbon retrievals.  Fourth, the downwelling suggested by the presence a vortex (see Section \ref{sec:lower_strato_heating}) would also modify the vertical profiles of CH$_4$ and its photochemical by-products \citep{moses_2015}.  This may also be compounded in Jupiter's auroral regions by higher rates of ion-neutral and electron-recombination reactions.  The vertical shape of the hydrocarbon profiles in Jupiter's auroral regions in reality may therefore be very different from the photochemically-predicted profiles adopted as \textit{a priori} for our retrievals.

In addition, we note to readers that our radiative transfer software adopts the assumption of local thermodynamic equilibrium (LTE).  This describes the case where the population of upper energy states of the rotational and vibrational modes are in equilibrium with the translational/kinetic population and therefore set by the Boltzmann distribution (and dependent on the thermodynamic temperature).  Non local thermodynamic equilibrium or non-LTE is expected to become important at pressures lower than 0.1 mbar \citep{appleby_1990} due to the lower rate of intermolecular collisions. Spontaneous emission, solar pumping of lines, excitation by particle collisions and further processes (see discussion in \citet{lopez_2001}) modify the population of upper energy vibrational/rotational states away from a Boltzmann distribution and therefore, they are no longer in equilibrium with the kinetic/translational population.  The paucity of intermolecular collisions at lower pressures ($<$0.1 mbar) mean there are insufficient energy exchanges between molecules to redistribute energy gains or losses associated with the above processes.   Non-LTE effects are expected to be most noticeable for stronger lines since these correspond to a larger energy transition, which require a greater number of intermolecular collisions to redistribute the energy lost or gained by spontaneous emission, solar pumping, and so on. We believe this is one possible explanation of the inability to adequately fit the weak and strong emission lines of CH$_4$, C$_2$H$_2$, C$_2$H$_6$ (for example, Figure \ref{fig:cxhy_17mar_19mar}).   The assumption of LTE may also be contributing to the physically counterintuitive hydrocarbon results described above. Parameterising non-LTE effects in the NEMESIS forward model will be the subject of future work.


\section{Conclusions}

We presented Gemini-North/TEXES (Texas Echelon Cross Echelle Spectrograph, \citealt{lacy_2002}) spectroscopy of Jupiter's mid-infrared H$_2$ S(1) quadrupole, CH$_4$, C$_2$H$_2$, C$_2$H$_4$ and C$_2$H$_6$ emission features at mid-to-high latitudes on March 17-19th, 2017.  These observations provided a rare combination of high spectral resolving power (65000 $<$ R $<$ 85000) and the high diffraction-limited spatial resolution provided by Gemini-North's 8 meter aperture.  The data capture Jupiter's mid-infrared auroral emissions before, during and after the arrival of a solar wind compression on March 18th, which allows the modulation of stratospheric temperature and composition by the external space environment to be determined. In comparing observations on March 17th and 19th, we observe a brightening of the mid-infrared CH$_4$, C$_2$H$_2$ and C$_2$H$_4$ emissions in a region that is spatially coincident with the duskside of northern main auroral emission (MAE).  In inverting the spectra on both nights to derive atmospheric information and their variability, we find the duskside brightening of the aforementioned emission features results (in part) from an upper stratospheric (p $<$ 0.1 mbar/z $>$ 200 km) heating (e.g. $\Delta T$ = 9.1 $\pm$ 2.1 K at 9 $\upmu$bar at 67.5$^\circ$N, 162.5$^\circ$W) with negligible transient heating at pressures deeper than 0.1 mbar.  Our interpretation is that the arrival of the solar wind enhancement on March 18th drove magnetospheric dynamics by compression of the magnetosphere and/or viscous interactions on the magnetospheric flanks.  This accelerated currents and/or generated higher Poynting fluxes by Alfv\'enic waves, which ultimately heated the upper stratosphere through processes including Joule heating, chemical heating and ion-neutral collisions/drag, thereby enhancing the mid-infrared emission features of the aforementioned hydrocarbons.  We therefore suggest that mid-infrared observations of Jupiter's auroral regions also serve as a metric of magnetospheric dynamics at wavelengths accessible from ground-based telescopes.  Using observations on March 18th, retrievals of the vertical temperature profile in a region poleward of the southern MAE demonstrate auroral-related heating as deep as $\sim$10 mbar.  This is almost a decade of pressure higher than similar auroral-related heating poleward of the northern main auroral emission.  We believe the deeper heating in the south results from one or a combination of: 1) higher Pedersen conductivities, which generates stronger currents and acceleration of the neutrals, 2) the energy being concentrated to a smaller region, 3) being at a relatively higher latitude and overlapping with the rotational axis, which favours the dissipation of heat through vertical mixing instead of horizontal advection/diffusion.


\acknowledgments

he research was carried out at the Jet Propulsion Laboratory, California Institute of Technology, under a contract with the National Aeronautics and Space Administration (80NM0018D0004).  The material is based upon work supported by the NASA under Grant NNH17ZDA001N issued through the Solar System Observations Planetary Astronomy program. The High Performance Computing resources used in this investigation were provided by funding from the JPL Information and Technology Solutions Directorate.  The IRTF is operated by the University of Hawaii under contract NNH14CK55B with NASA. Moses acknowledges support from NASA Solar System Workings program 80NSSC20K0462. Fletcher was supported by a Royal Society Research Fellowship and European Research Council Consolidator Grant (under the European Union's Horizon 2020 research and innovation programme, grant agreement No 723890) at the University of Leicester. Co-investigator Tao acknowledges the support by MEXT/JSPS KAKENHI Grant 19H01948.  Hue acknowledges support from the french government under the France 2030 investment plan, as part of the Initiative d’Excellence d’Aix-Marseille Université – A*MIDEX AMX-22-CPJ-04.  The analysis is based on observations obtained at the international Gemini Observatory, a program of NSF’s NOIRLab, which is managed by the Association of Universities for Research in Astronomy (AURA) under a cooperative agreement with the National Science Foundation on behalf of the Gemini Observatory partnership: the National Science Foundation (United States), National Research Council (Canada), Agencia Nacional de Investigaci\'{o}n y Desarrollo (Chile), Ministerio de Ciencia, Tecnolog\'{i}a e Innovaci\'{o}n (Argentina), Minist\'{e}rio da Ci\^{e}ncia, Tecnologia, Inova\c{c}\~{o}es e Comunica\c{c}\~{o}es (Brazil), and Korea Astronomy and Space Science Institute (Republic of Korea).  T. Cavali\'e acknowledges funding from CNES and the Programme National de Plan\'etologie (PNP) of CNRS/INSU. 

Simulation results have been provided by the Community Coordinated Modeling Center at Goddard Space Flight Center through their publicly available simulation services (https://ccmc.gsfc.nasa.gov). The mSWiM model was developed by Kenneth Hansen at the University of Michigan."

\section{Data Availability Statement}

The Gemini-TEXES observations presented in this work are publicly available at the Gemini Observatory Archive\footnote{https://archive.gemini.edu/searchform}.  However, spatially-mapped and absolutely calibrated versions of the observations can be requested from the authors. 

\vspace{5mm}
\facilities{Gemini-North, TEXES (Texas Echelon Cross Echelle Spectrograph, \citealt{lacy_2002})}


\software{NEMESIS \citep{irwin_2008}, mSWiM \citep{zieger_2008}, \citet{tao_2005} solar wind model}


\newpage
\clearpage
\appendix

\renewcommand{\figurename}{Figure}
\renewcommand{\tablename}{Table}
\setcounter{figure}{0}
\setcounter{table}{0}
\makeatletter 
\renewcommand{\thetable}{A.\@arabic\c@table}
\makeatother

\section{Observations details}
\begin{table}[!h]
\scriptsize
\centering
\begin{tabular}{|>{\centering\arraybackslash} m{1.7cm} |>{\centering\arraybackslash} m{1.1cm}  >{\centering\arraybackslash} m{1.2cm} >{\centering\arraybackslash} m{1.0cm} >{\centering\arraybackslash} m{1.2cm}  >{\centering\arraybackslash} m{1.2cm} >{\centering\arraybackslash} m{1.2cm}  >{\centering\arraybackslash} m{1.2cm} >{\centering\arraybackslash} m{1.4cm}|} 
Date & Time & Filename & Setting & Number & Airmass & v$_{rad}$ & Hemisphere & CML \\
\multirow{32}{*}{2017-Mar-17}         &          &   (jup.X.Y)             & (cm$^{-1}$) &    of spectra        & &   (km/s)          &                     &  \\
\firsthline
&\textbf{09:13:20}& \textbf{7017.01}&\textbf{1248}&         \textbf{890}&      \textbf{1.52}&     \textbf{-12.0}&\textbf{N}&         \textbf{180}\\
&\textbf{09:18:52}& \textbf{7018.01}&\textbf{1248}&         \textbf{890}&      \textbf{1.49}&     \textbf{-11.9}&\textbf{N}&         \textbf{183}\\
&\textbf{09:18:52}& \textbf{7018.02}&\textbf{1248}&         \textbf{625}&      \textbf{1.49}&     \textbf{-11.9}&\textbf{N}&         \textbf{186}\\
&09:27:34& 7019.01&1248&        1371&      1.44&     -11.9&S&         189\\
&09:27:34& 7019.02&1248&        1091&      1.44&     -11.9&S&         192\\
&\textbf{09:40:32}& \textbf{7020.01}&\textbf{587}&        \textbf{2531}&      \textbf{1.37}&     \textbf{-11.9}&\textbf{N}&         \textbf{196}\\
&\textbf{09:40:32}& \textbf{7020.02}&\textbf{587}&        \textbf{2484}&      \textbf{1.37}&     \textbf{-11.9}&\textbf{N}&         \textbf{200}\\
&\textbf{09:51:56}& \textbf{7021.01}&\textbf{730}&        \textbf{2544}&      \textbf{1.33}&     \textbf{-11.9}&\textbf{N}&         \textbf{203}\\
&\textbf{09:51:56}& \textbf{7021.02}&\textbf{730}&        \textbf{2501}&      \textbf{1.33}&     \textbf{-11.9}&\textbf{N}&         \textbf{207}\\
&\textbf{10:03:44}& \textbf{7022.01}&\textbf{819}&        \textbf{1413}&      \textbf{1.28}&     \textbf{-11.9}&\textbf{N}&         \textbf{210}\\
&10:03:44& 7022.02&819&        1531&      1.28&     -11.9&N&         214\\
&\textbf{10:15:15}& \textbf{7023.01}&\textbf{950}&         \textbf{487}&      \textbf{1.25}&     \textbf{-11.8}&\textbf{N}&         \textbf{217}\\
&\textbf{10:15:15}& \textbf{7023.02}&\textbf{950}&         \textbf{522}&      \textbf{1.25}&     \textbf{-11.8}&\textbf{N}&         \textbf{219}\\
&10:24:01& 7024.01&950&        1005&      1.23&     -11.8&S&         223\\
&10:24:01& 7024.02&950&         729&      1.23&     -11.8&S&         225\\
&10:33:25& 7025.01&1248&         850&      1.20&     -11.8&N&         228\\
&10:38:41& 7026.01&1248&         593&      1.19&     -11.8&N&         231\\
&10:38:41& 7026.02&1248&         626&      1.19&     -11.8&N&         234\\
&10:47:21& 7027.01&1248&         737&      1.17&     -11.8&S&         237\\
&10:47:21& 7027.02&1248&         773&      1.17&     -11.8&S&         240\\
&10:57:41& 7029.01&587&        2024&      1.16&     -11.7&N&         243\\
&10:57:41& 7029.02&587&        1054&      1.16&     -11.7&N&         246\\
&11:06:38& 7030.01&587&        1683&      1.15&     -11.7&S&         249\\
&11:06:38& 7030.02&587&        1683&      1.15&     -11.7&S&         252\\
&11:17:32& 7031.01&730&        1805&      1.13&     -11.7&N&         255\\
&11:17:32& 7031.02&730&        1903&      1.13&     -11.7&N&         258\\
&11:26:59& 7032.01&730&        1759&      1.13&     -11.7&S&         261\\
&11:26:59& 7032.02&730&        1759&      1.13&     -11.7&S&         264\\
&11:37:00& 7033.01&819&         537&      1.12&     -11.7&N&         266\\
&11:37:00& 7033.02&819&         532&      1.12&     -11.7&N&         269\\
\lasthline
\end{tabular}
\caption{Details of the observations measured on March 17-19th, 2017, in chronological order.  All dates/times are UTC. CML stands for Central Meridian Longitude in System III.  Observations in bold are those chosen for coaddition and analysis, as detailed in the text.}
\label{tab:mar17_obs} 
\end{table}

\clearpage
\newpage
\setcounter{table}{0}
\begin{table}[!h]
\scriptsize
\begin{tabular}{|>{\centering\arraybackslash} m{1.7cm} |>{\centering\arraybackslash} m{1.1cm}  >{\centering\arraybackslash} m{1.2cm} >{\centering\arraybackslash} m{1.0cm} >{\centering\arraybackslash} m{1.2cm}  >{\centering\arraybackslash} m{1.2cm} >{\centering\arraybackslash} m{1.2cm}  >{\centering\arraybackslash} m{1.2cm} >{\centering\arraybackslash} m{1.4cm}|} 
Date & Time & Filename & Setting & Number & Airmass & v$_{rad}$ & Hemisphere & CML \\
\multirow{29}{*}{2017-Mar-17}         &          &   (jup.X.Y)             & (cm$^{-1}$) &    of spectra        & &   (km/s)          &                     &  \\
\firsthline
&11:46:49& 7034.01&819&         674&      1.12&     -11.6&S&         271\\
&11:46:49& 7034.02&819&         678&      1.12&     -11.6&S&         273\\
&11:56:21& 7035.01&950&         809&      1.11&     -11.6&N&         278\\
&12:01:19& 7036.01&950&         809&      1.11&     -11.6&N&         281\\
&12:06:18& 7037.01&950&         809&      1.11&     -11.6&N&         284\\
&12:11:16& 7038.01&950&         846&      1.11&     -11.6&S&         280\\
&12:11:16& 7038.02&950&         808&      1.11&     -11.6&S&         281\\
&12:20:38& 7039.01&1248&         665&      1.12&     -11.6&N&         291\\
&12:20:38& 7039.02&1248&         665&      1.12&     -11.6&N&         293\\
&12:29:19& 7040.01&1248&         774&      1.12&     -11.5&S&         284\\
&12:29:19& 7040.02&1248&         815&      1.12&     -11.5&S&         284\\
&12:39:37& 7041.01&587&        2023&      1.13&     -11.5&N&         298\\
&12:39:37& 7041.02&587&        1944&      1.13&     -11.5&N&         299\\
&12:49:34& 7042.01&587&        1245&      1.14&     -11.5&S&         288\\
&12:49:34& 7042.02&587&        1139&      1.14&     -11.5&S&         288\\
&13:02:07& 7043.01&730&        1664&      1.16&     -11.5&N&         304\\
&13:02:07& 7043.02&730&        1608&      1.16&     -11.5&N&         305\\
&13:12:07& 7044.01&730&        1826&      1.17&     -11.4&S&         291\\
&13:12:07& 7044.02&730&        1806&      1.17&     -11.4&S&         290\\
&13:22:42& 7045.01&819&         495&      1.19&     -11.4&N&         309\\
&13:22:42& 7045.02&819&         470&      1.19&     -11.4&N&         310\\
&13:31:24& 7046.01&819&         715&      1.21&     -11.4&S&         285\\
&13:31:24& 7046.02&819&         703&      1.21&     -11.4&S&         285\\
&13:40:53& 7047.01&950&         697&      1.23&     -11.4&N&         307\\
&13:40:53& 7047.02&950&         697&      1.23&     -11.4&N&         307\\
&13:49:49& 7048.01&950&         849&      1.26&     -11.4&S&         282\\
&13:49:49& 7048.02&950&         843&      1.26&     -11.4&S&         281\\
\hline
\multirow{6}{*}{2017-Mar-18}       &09:37:02& 8005.01&1248&         676&      1.37&     -11.4&N&         307\\
&09:37:02& 8005.02&1248&         784&      1.37&     -11.4&N&         305\\
&09:45:46& 8006.01&1248&         825&      1.33&     -11.4&S&         279\\
&09:45:46& 8006.02&1248&         783&      1.33&     -11.4&S&         278\\
&09:55:49& 8007.01&587&        1471&      1.30&     -11.4&N&         295\\
&09:55:49& 8007.02&587&        1513&      1.30&     -11.4&N&         281\\
\lasthline
\end{tabular}
\caption{continued from previous page}
\label{tab:mar17_obs} 
\end{table}

\clearpage
\newpage
\setcounter{table}{0}
\begin{table}[!h]
\scriptsize
\begin{tabular}{|>{\centering\arraybackslash} m{1.7cm} |>{\centering\arraybackslash} m{1.1cm}  >{\centering\arraybackslash} m{1.2cm} >{\centering\arraybackslash} m{1.0cm} >{\centering\arraybackslash} m{1.2cm}  >{\centering\arraybackslash} m{1.2cm} >{\centering\arraybackslash} m{1.2cm}  >{\centering\arraybackslash} m{1.2cm} >{\centering\arraybackslash} m{1.4cm}|} 
\multirow{19}{*}{2017-Mar-18}         &          &   (jup.X.Y)             & (cm$^{-1}$) &    of spectra        & &   (km/s)          &                     &  \\
\firsthline
&\textbf{10:05:46}& \textbf{8008.01}&\textbf{587}&        \textbf{1329}&      \textbf{1.26}&     \textbf{-11.4}&\textbf{S}&          \textbf{84}\\
&\textbf{10:05:46}& \textbf{8008.02}&\textbf{587}&        \textbf{1384}&      \textbf{1.26}&     \textbf{-11.4}&\textbf{S}&          \textbf{81}\\
&10:20:43& 8009.01&730&        1625&      1.22&     -11.3&N&          58\\
&10:20:43& 8009.02&730&        1720&      1.22&     -11.3&N&          56\\
&\textbf{10:30:44}& \textbf{8010.01}&\textbf{730}&        \textbf{1578}&      \textbf{1.20}&     \textbf{-11.3}&\textbf{S}&          \textbf{72}\\
&\textbf{10:30:44}& \textbf{8010.02}&\textbf{730}&        \textbf{1605}&      \textbf{1.20}&     \textbf{-11.3}&\textbf{S}&          \textbf{71}\\
&10:41:42& 8011.01&819&         686&      1.18&     -11.3&N&          52\\
&10:41:42& 8011.02&819&         735&      1.18&     -11.3&N&          52\\
&\textbf{10:50:24}& \textbf{8012.01}&\textbf{819}&         \textbf{625}&      \textbf{1.16}&     \textbf{-11.3}&\textbf{S}&          \textbf{75}\\
&\textbf{10:50:24}& \textbf{8012.02}&\textbf{819}&         \textbf{580}&      \textbf{1.16}  &   \textbf{-11.3}&\textbf{S}&          \textbf{75}\\
&10:59:46& 8013.01&950&         782&      1.15&     -11.2&N&          53\\
&10:59:46& 8013.02&950&         900&      1.15&     -11.2&N&          53\\
&\textbf{11:08:32}& \textbf{8014.01}&\textbf{950}&         \textbf{674}&      \textbf{1.14}&     \textbf{-11.2}&\textbf{S}&          \textbf{74}\\
&\textbf{11:08:32}& \textbf{8014.02}&\textbf{950}&         \textbf{707}&      \textbf{1.14}&     \textbf{-11.2}&\textbf{S}&          \textbf{73}\\
&\textbf{11:18:09}& \textbf{8015.01}&\textbf{1248}&         \textbf{784}&      \textbf{1.13}&     \textbf{-11.2}&\textbf{S}&          \textbf{74}\\
&\textbf{11:18:09}& \textbf{8015.02}&\textbf{124}8&         \textbf{827}&      \textbf{1.13}&     \textbf{-11.2}&\textbf{S}&          \textbf{73}\\
&11:26:55& 8016.01&1248&         751&      1.12&     -11.2&N&          58\\
\hline
\multirow{17}{*}{2017-Mar-19}&08:58:14& 9000.01&1248&         730&      1.56&     -11.0&N&         112\\
&08:58:14& 9000.02&1248&         754&      1.56&     -11.0&N&         114\\
&09:06:57& 9001.01&1248&         882&      1.50&     -11.0&S&         118\\
&09:06:57& 9001.02&1248&         843&      1.50&     -11.0&S&         121\\
&09:16:45& 9002.01&950&         877&      1.45&     -11.0&N&         123\\
&09:21:33& 9003.01&950&         837&      1.42&     -11.0&N&         126\\
&09:21:33& 9003.02&950&         839&      1.42&     -11.0&N&         129\\
&09:30:19& 9004.01&950&         872&      1.38&     -10.9&S&         133\\
&09:30:19& 9004.02&950&         875&      1.38&     -10.9&S&         135\\
&\textbf{09:40:09}& \textbf{9005.01}&\textbf{587}&        \textbf{1488}&      \textbf{1.34}&     \textbf{-10.9}&\textbf{N}&         \textbf{137}\\
&\textbf{09:40:09}& \textbf{9005.02}&\textbf{587}&        \textbf{1580}&      \textbf{1.34}&     \textbf{-10.9}&\textbf{N}&         \textbf{141}\\
&09:50:00& 9006.01&587&        1508&      1.30&     -10.9&S&         144\\
&09:50:00& 9006.02&587&        1561&      1.30&     -10.9&S&         147\\
&\textbf{10:01:31}& \textbf{9007.01}&\textbf{730}&        \textbf{1561}&      \textbf{1.26}&     \textbf{-10.9}&\textbf{N}&         \textbf{150}\\
&\textbf{10:01:31}& \textbf{9007.02}&\textbf{730}&        \textbf{1561}&      \textbf{1.26}&     \textbf{-10.9}&\textbf{N}&         \textbf{154}\\
&10:11:28& 9008.01&730&        1796&      1.24&     -10.9&S&         157\\
&10:11:28& 9008.02&730&        1849&      1.24&     -10.9&S&         160\\
\lasthline
\end{tabular}
\caption{(continued from previous page)}
\label{tab:mar17_obs} 
\end{table}

\clearpage
\newpage
\setcounter{table}{0}
\begin{table}[!h]
\scriptsize
\begin{tabular}{|>{\centering\arraybackslash} m{1.7cm} |>{\centering\arraybackslash} m{1.1cm}  >{\centering\arraybackslash} m{1.2cm} >{\centering\arraybackslash} m{1.0cm} >{\centering\arraybackslash} m{1.2cm}  >{\centering\arraybackslash} m{1.2cm} >{\centering\arraybackslash} m{1.2cm}  >{\centering\arraybackslash} m{1.2cm} >{\centering\arraybackslash} m{1.4cm}|} 
Date & Time & Filename & Setting & Number & Airmass & v$_{rad}$ & Hemisphere & CML \\
\multirow{21}{*}{2017-Mar-19}         &          &   (jup.X.Y)             & (cm$^{-1}$) &    of spectra        & &   (km/s)          &                     &  \\
\firsthline
&\textbf{10:22:15}& \textbf{9009.01}&\textbf{819}&         \textbf{667}&      \textbf{1.21}&     \textbf{-10.8}&\textbf{N}&         \textbf{163}\\
&\textbf{10:22:15}& \textbf{9009.02}&\textbf{819}&         \textbf{702}&      \textbf{1.21}&     \textbf{-10.8}&\textbf{N}&         \textbf{166}\\
&10:30:57& 9010.01&819&         748&      1.19&     -10.8&S&         169\\
&10:30:57& 9010.02&819&         779&      1.19&     -10.8&S&         172\\
&\textbf{10:40:56}& \textbf{9011.01}&\textbf{1248}&         \textbf{730}&      \textbf{1.17}&     \textbf{-10.8}&\textbf{N}&         \textbf{174}\\
&\textbf{10:40:56}& \textbf{9011.02}&\textbf{1248}&         \textbf{730}&      \textbf{1.17}&     \textbf{-10.8}&\textbf{N}&         \textbf{177}\\
&10:49:36& 9012.01&1248&         923&      1.16&     -10.8&S&         180\\
&10:49:36& 9012.02&1248&         882&      1.16&     -10.8&S&         183\\
&\textbf{10:59:13}& \textbf{9013.01}&\textbf{1248}&         \textbf{720}&      \textbf{1.14}&     \textbf{-10.8}&\textbf{N}&         \textbf{186}\\
&\textbf{10:59:13}& \textbf{9013.02}&\textbf{1248}&         \textbf{801}&      \textbf{1.14}&     \textbf{-10.8}&\textbf{N}&         \textbf{187}\\
&\textbf{11:08:09}& \textbf{9014.01}&\textbf{950}&        \textbf{1326}&      \textbf{1.13}&     \textbf{-10.7}&\textbf{N}&         \textbf{190}\\
&11:08:09& 9014.02&950&        1326&      1.13&     -10.7&N&         193\\
&11:16:53& 9015.01&950&         956&      1.13&     -10.7&S&         197\\
&11:16:53& 9015.02&950&         956&      1.13&     -10.7&S&         199\\
&\textbf{11:26:50}& \textbf{9016.01}&\textbf{587}&        \textbf{1464}&      \textbf{1.12}&     \textbf{-10.7}&\textbf{N}&         \textbf{202}\\
&\textbf{11:26:50}& \textbf{9016.02}&\textbf{587}&        \textbf{1417}&      \textbf{1.12}&     \textbf{-10.7}&\textbf{N}&         \textbf{205}\\
&11:37:48& 9017.01&730&        1513&      1.11&     -10.7&N&         208\\
&11:37:48& 9017.02&730&        1562&      1.11&     -10.7&N&         212\\
&11:48:15& 9018.01&819&         597&      1.11&     -10.6&N&         214\\
\lasthline
\end{tabular}
\caption{(continued from previous page)}
\label{tab:mar17_obs} 
\end{table}

\makeatletter 
\renewcommand{\thefigure}{B.\@arabic\c@figure}
\setcounter{figure}{0}
\setcounter{table}{0}
\makeatother

\clearpage
\newpage

\section{Individual observations}\label{sec:appendix_b}

\begin{figure*}[!th]
\begin{center}
\includegraphics[width=0.7\textwidth]{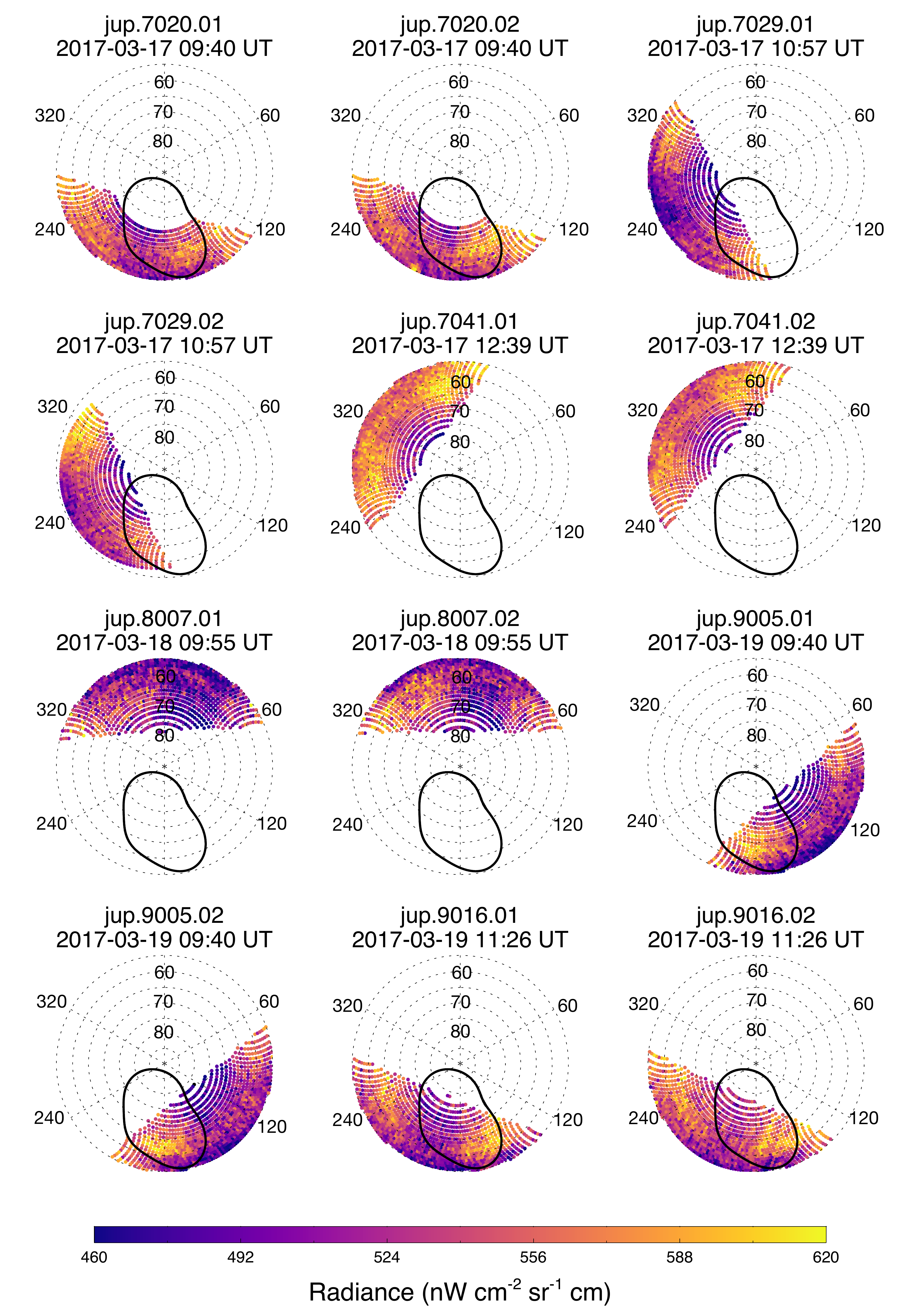}
\caption{Individual TEXES scans of high-northern latitudes.  Each point represents a spectrum and is coloured according to the radiance from 587.0275 - 587.0375 cm$^{-1}$, which captures the H$_2$ S(1) quadrupole emission feature.  Scans are shown in chronological order from left-to-right, top-to-bottom.  Solid, pink lines represent the statistical-mean position of the ultraviolet auroral ovals \citep{bonfond_2017}.    }
\label{fig:spx_587_north}
\end{center}
\end{figure*}

\begin{figure*}[!t]
\begin{center}
\includegraphics[width=0.7\textwidth]{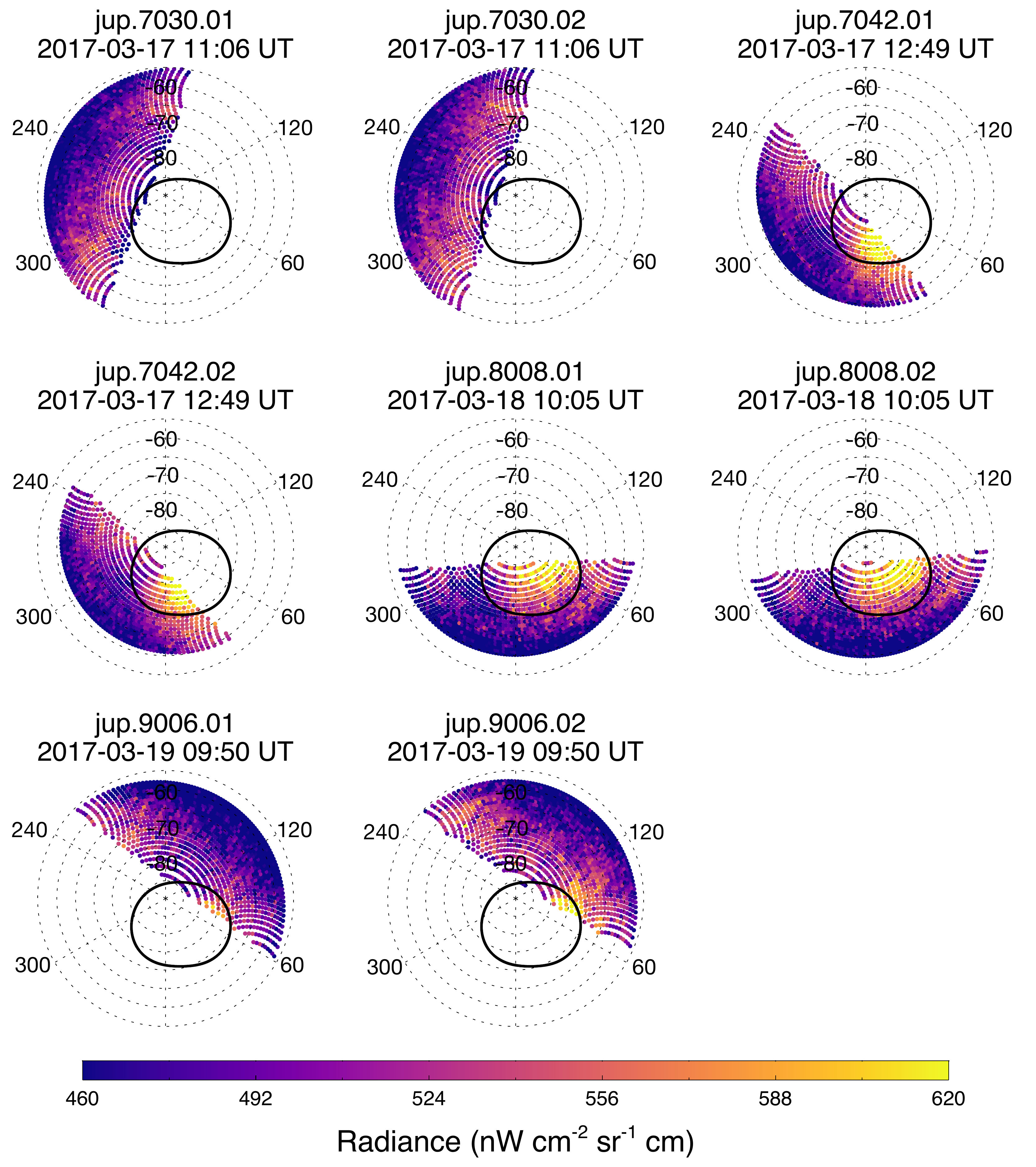}
\caption{As in Figure \ref{fig:spx_587_north} but for observations of high-southern latitudes. }
\label{fig:spx_587_south}
\end{center}
\end{figure*}

\begin{figure*}[!t]
\begin{center}
\includegraphics[width=0.7\textwidth]{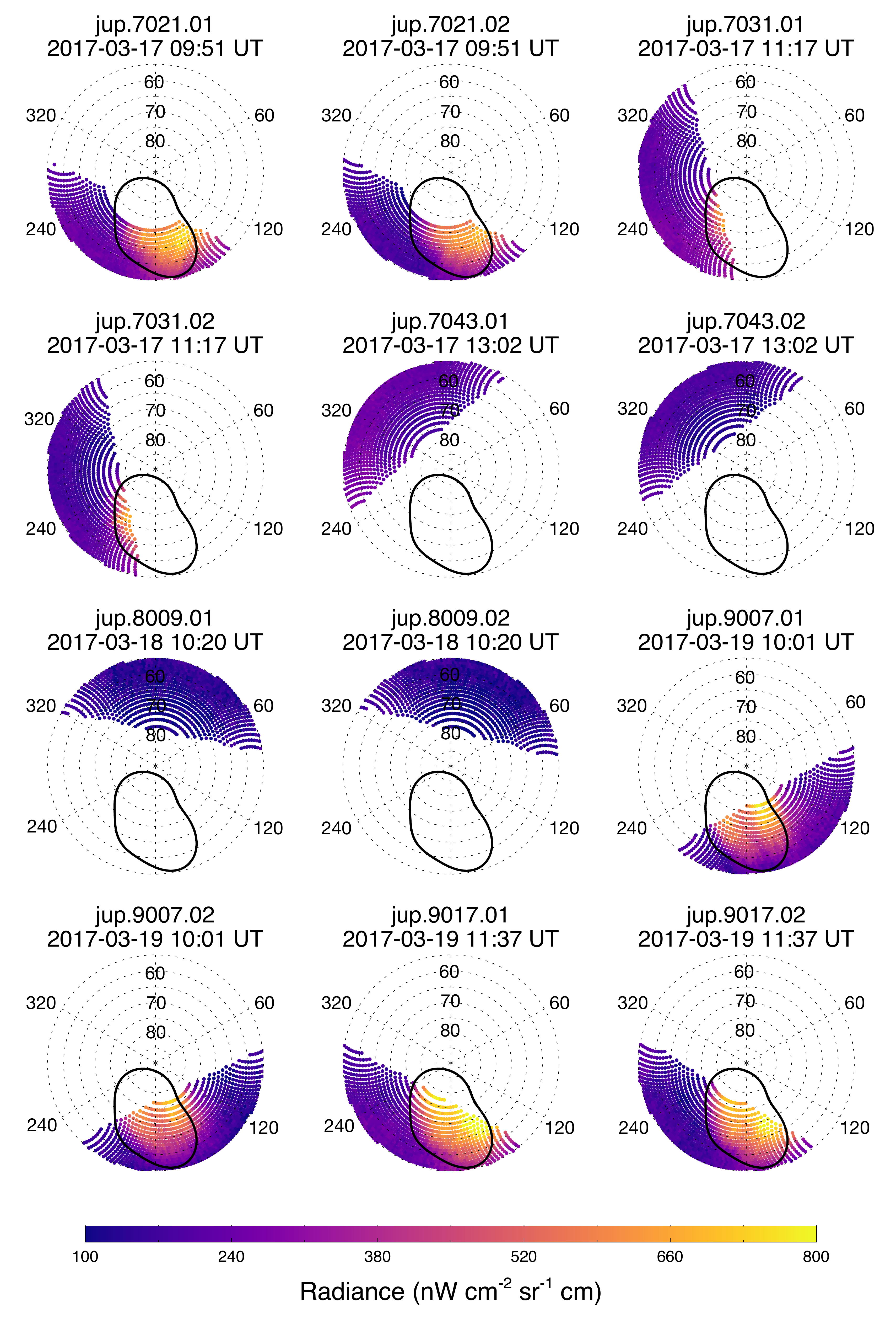}
\caption{Individual TEXES scans of high-northern latitudes.  Each point represents a spectrum and is coloured according to the mean radiance in all sampled C$_2$H$_2$ emission lines from 729.0 - 730.0 cm$^{-1}$.  Scans are shown in chronological order from left-to-right, top-to-bottom.  Solid, pink lines represent the statistical-mean position of the ultraviolet auroral ovals \citep{bonfond_2017}.    }
\label{fig:spx_730_north}
\end{center}
\end{figure*}

\begin{figure*}[!t]
\begin{center}
\includegraphics[width=0.7\textwidth]{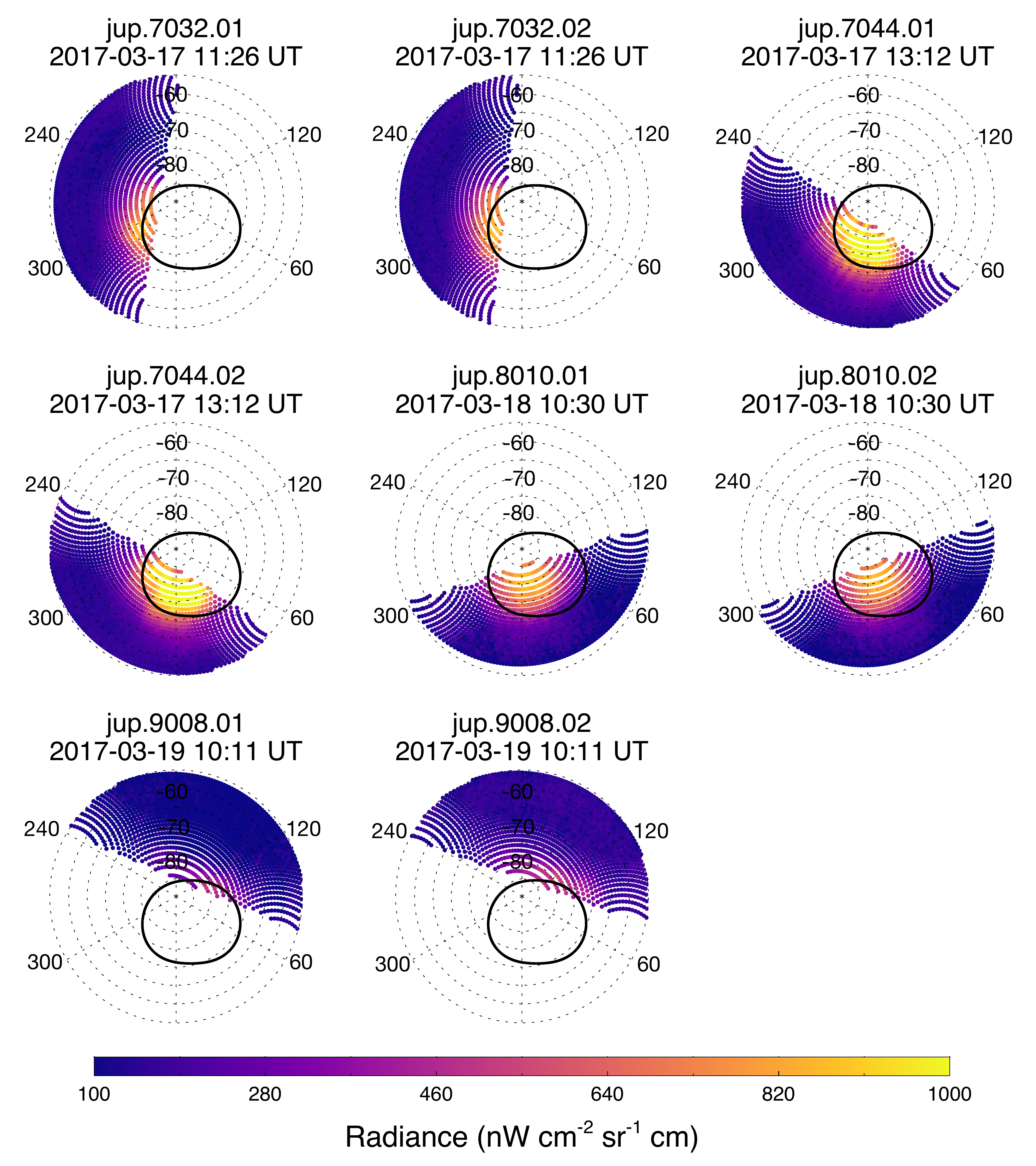}
\caption{As in Figure \ref{fig:spx_730_north} but for observations of high-southern latitudes. }
\label{fig:spx_730_south}
\end{center}
\end{figure*}

\begin{figure*}[!t]
\begin{center}
\includegraphics[width=0.7\textwidth]{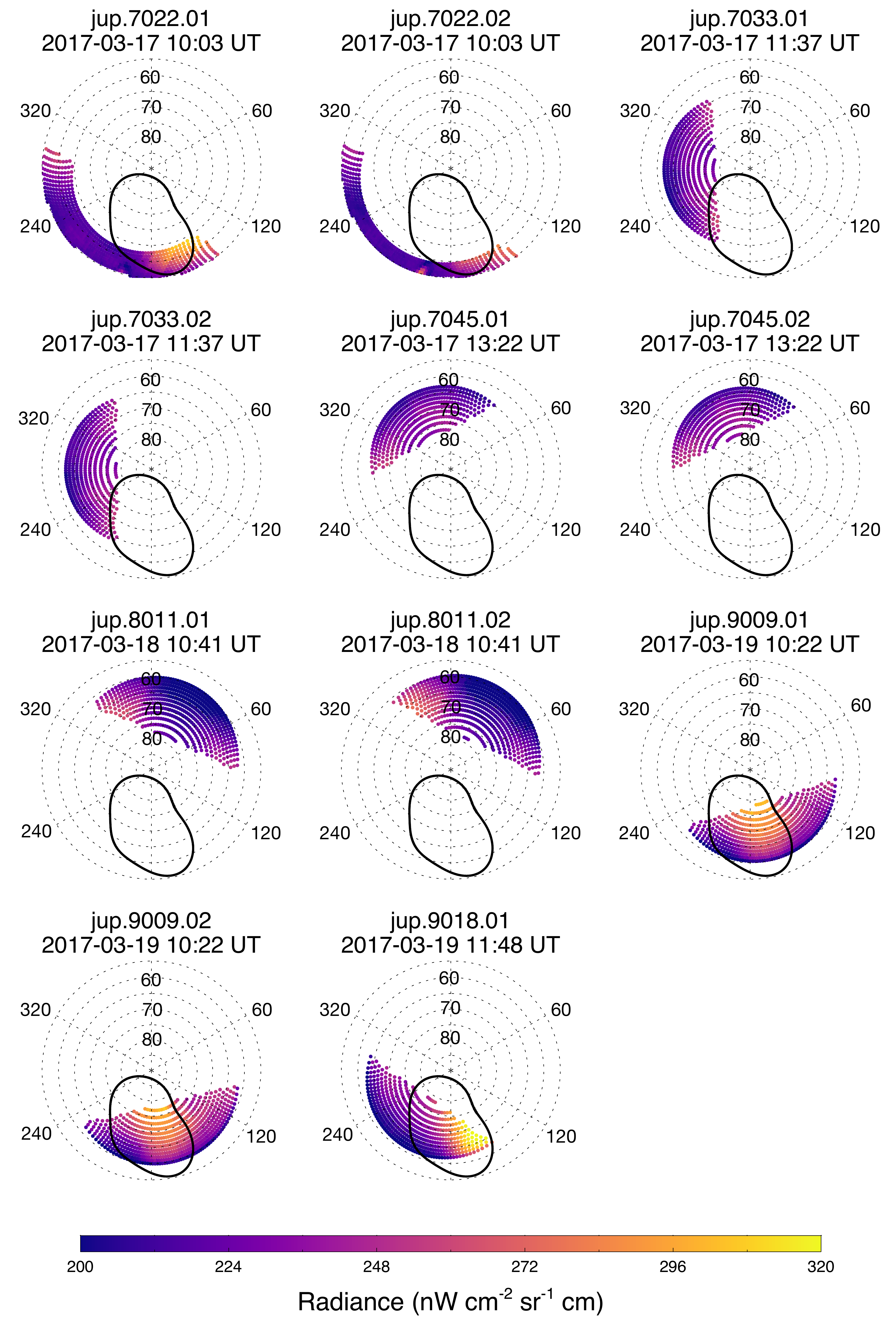}
\caption{Individual TEXES scans of high-northern latitudes.  Each point represents a spectrum and is coloured according to the mean radiance in all sampled C$_2$H$_6$ emission lines from 819.0 - 820.0 cm$^{-1}$.  Scans are shown in chronological order from left-to-right, top-to-bottom.  Solid, pink lines represent the statistical-mean position of the ultraviolet auroral ovals \citep{bonfond_2017}.    }
\label{fig:spx_819_north}
\end{center}
\end{figure*}

\begin{figure*}[!t]
\begin{center}
\includegraphics[width=0.7\textwidth]{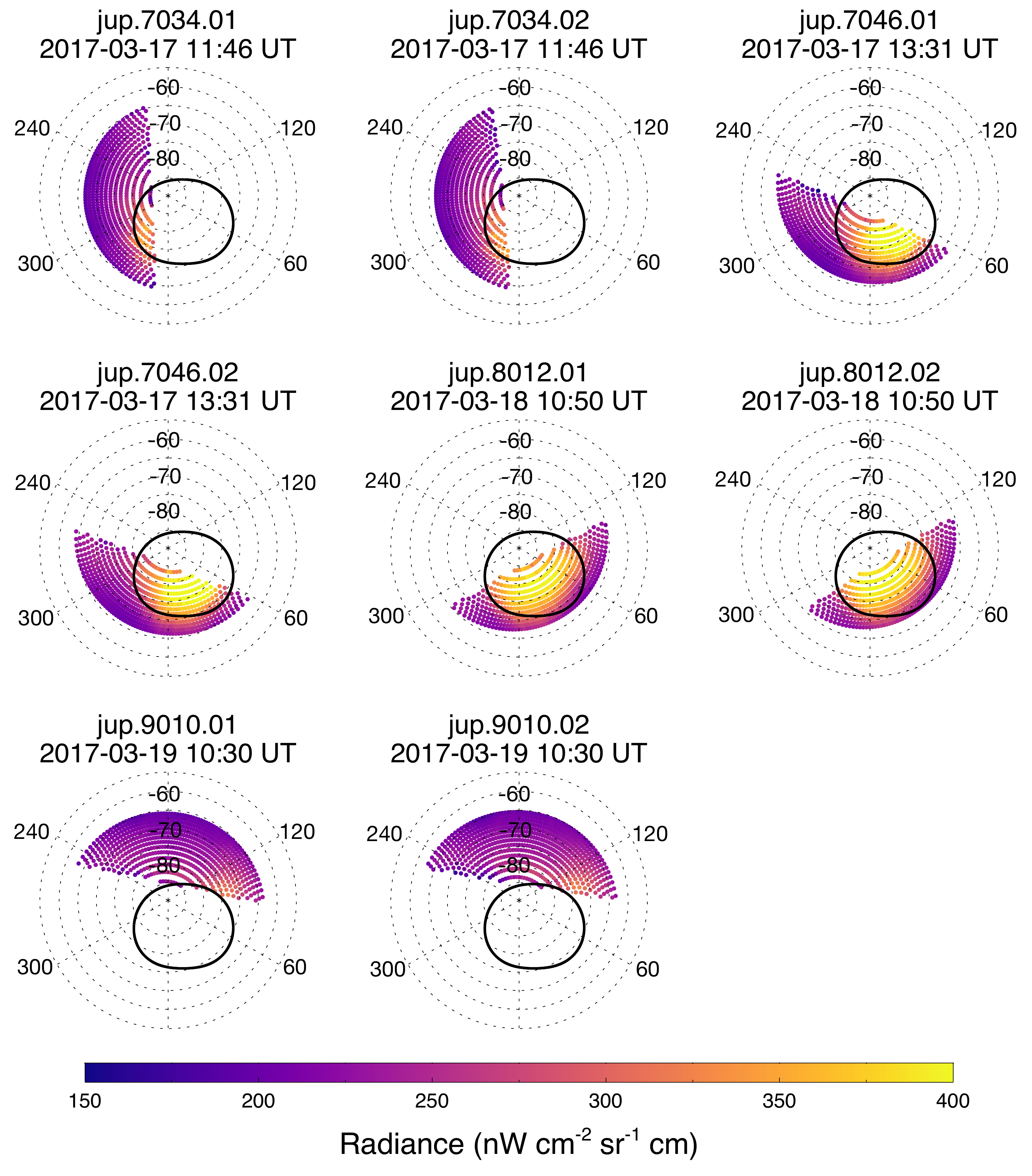}
\caption{As in Figure \ref{fig:spx_819_north} but for observations of high-southern latitudes. }
\label{fig:spx_819_south}
\end{center}
\end{figure*}

\begin{figure*}[!t]
\begin{center}
\includegraphics[width=0.7\textwidth]{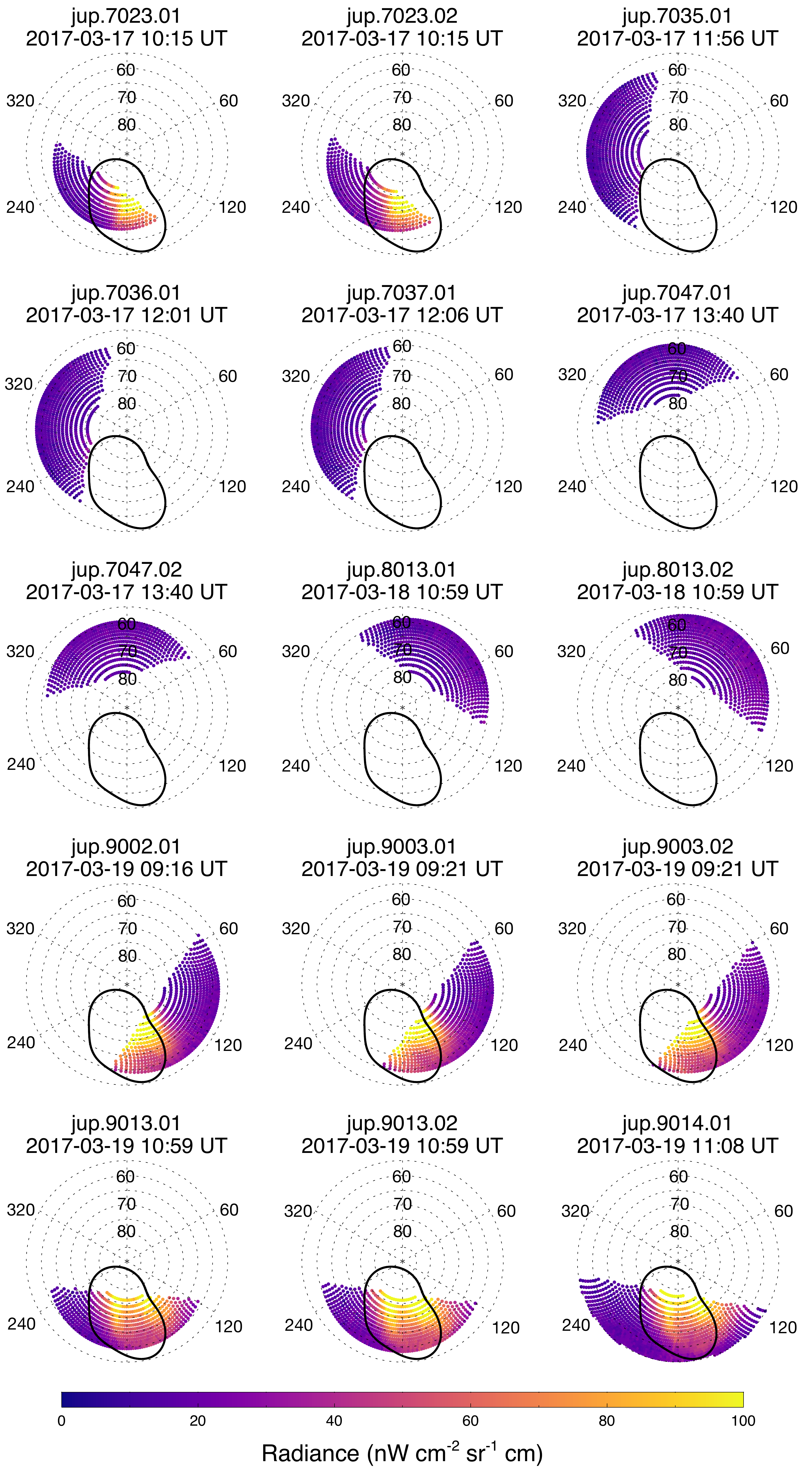}
\caption{Individual TEXES scans of high-northern latitudes.  Each point represents a spectrum and is coloured according to the mean radiance in all sampled C$_2$H$_4$ emission lines from 949.0 - 950.0 cm$^{-1}$ Scans are shown in chronological order from left-to-right, top-to-bottom.  Solid, pink lines represent the statistical-mean position of the ultraviolet auroral ovals \citep{bonfond_2017}.    }
\label{fig:spx_950_north}
\end{center}
\end{figure*}

\begin{figure*}[!t]
\begin{center}
\includegraphics[width=0.7\textwidth]{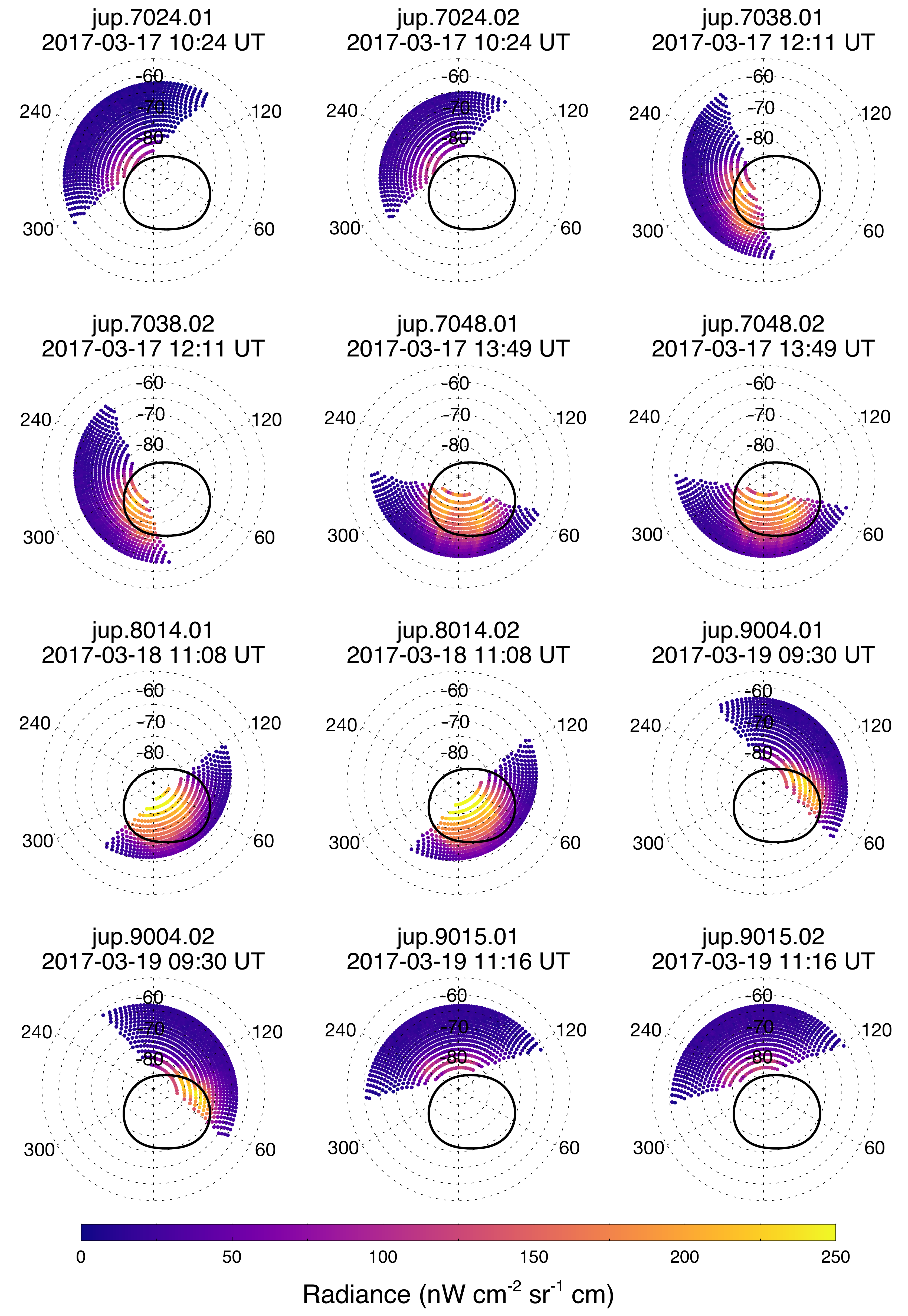}
\caption{As in Figure \ref{fig:spx_950_north} but for observations of high-southern latitudes. }
\label{fig:spx_950_south}
\end{center}
\end{figure*}

\begin{figure*}[!t]
\begin{center}
\includegraphics[width=0.7\textwidth]{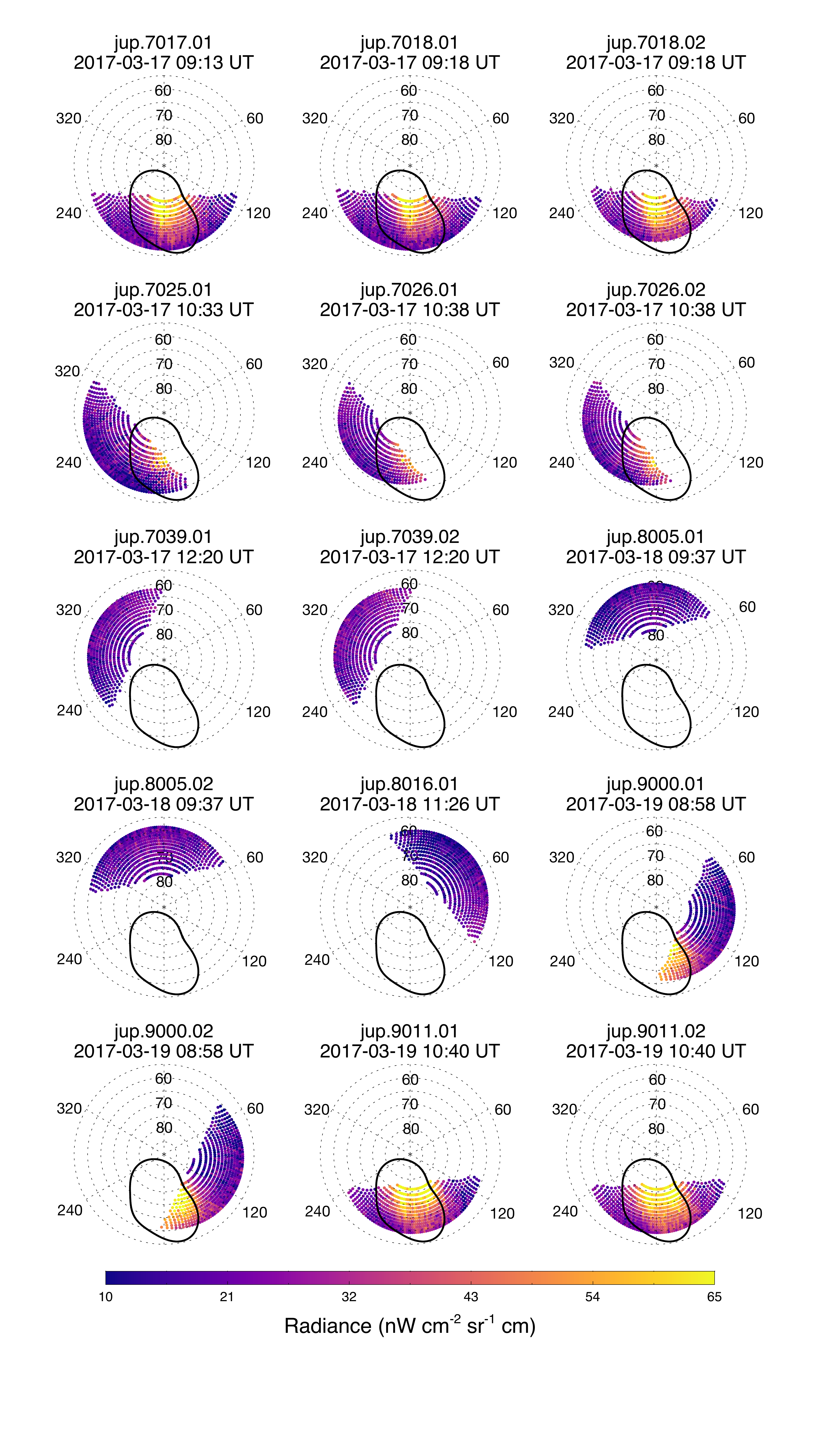}
\caption{Individual TEXES scans of high-northern latitudes.  Each point represents a spectrum and is coloured according to the mean radiance in all sampled CH$_4$ emission lines from 1245.20 - 1250.03 cm$^{-1}$. Scans are shown in chronological order from left-to-right, top-to-bottom.  Solid, pink lines represent the statistical-mean position of the ultraviolet auroral ovals \citep{bonfond_2017}.    }
\label{fig:spx_1248_north}
\end{center}
\end{figure*}

\begin{figure*}[!t]
\begin{center}
\includegraphics[width=0.7\textwidth]{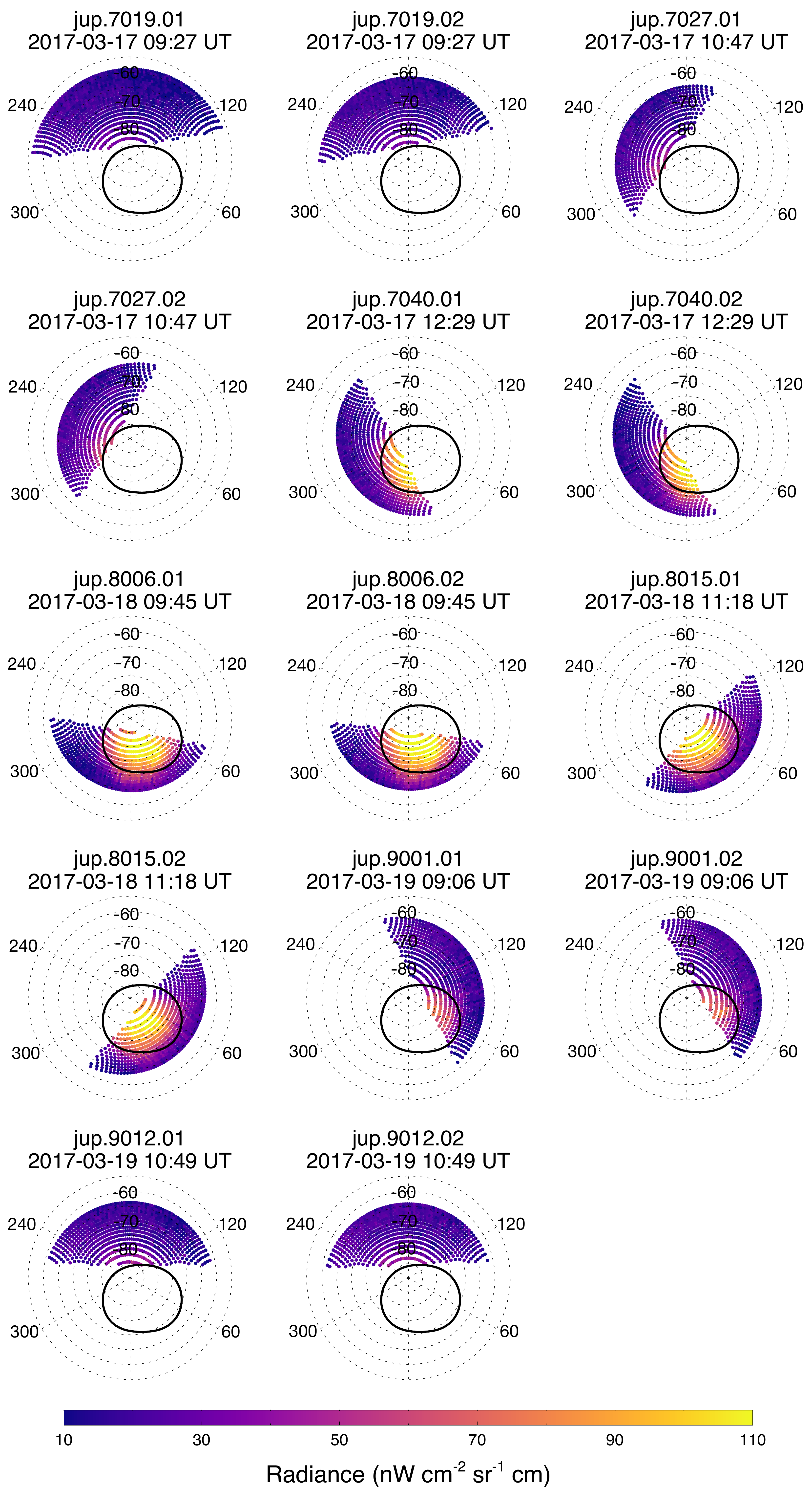}
\caption{As in Figure \ref{fig:spx_1248_north} but for observations of high-southern latitudes. }
\label{fig:spx_1248_south}
\end{center}
\end{figure*}

\clearpage
\newpage
\makeatletter 
\renewcommand{\thefigure}{C.\@arabic\c@figure}
\setcounter{figure}{0}
\setcounter{table}{0}
\makeatother
\section{C$_2$H$_4$ retrieval artefacts}

\begin{figure*}[!th]
\begin{center}
\includegraphics[width=0.65\textwidth]{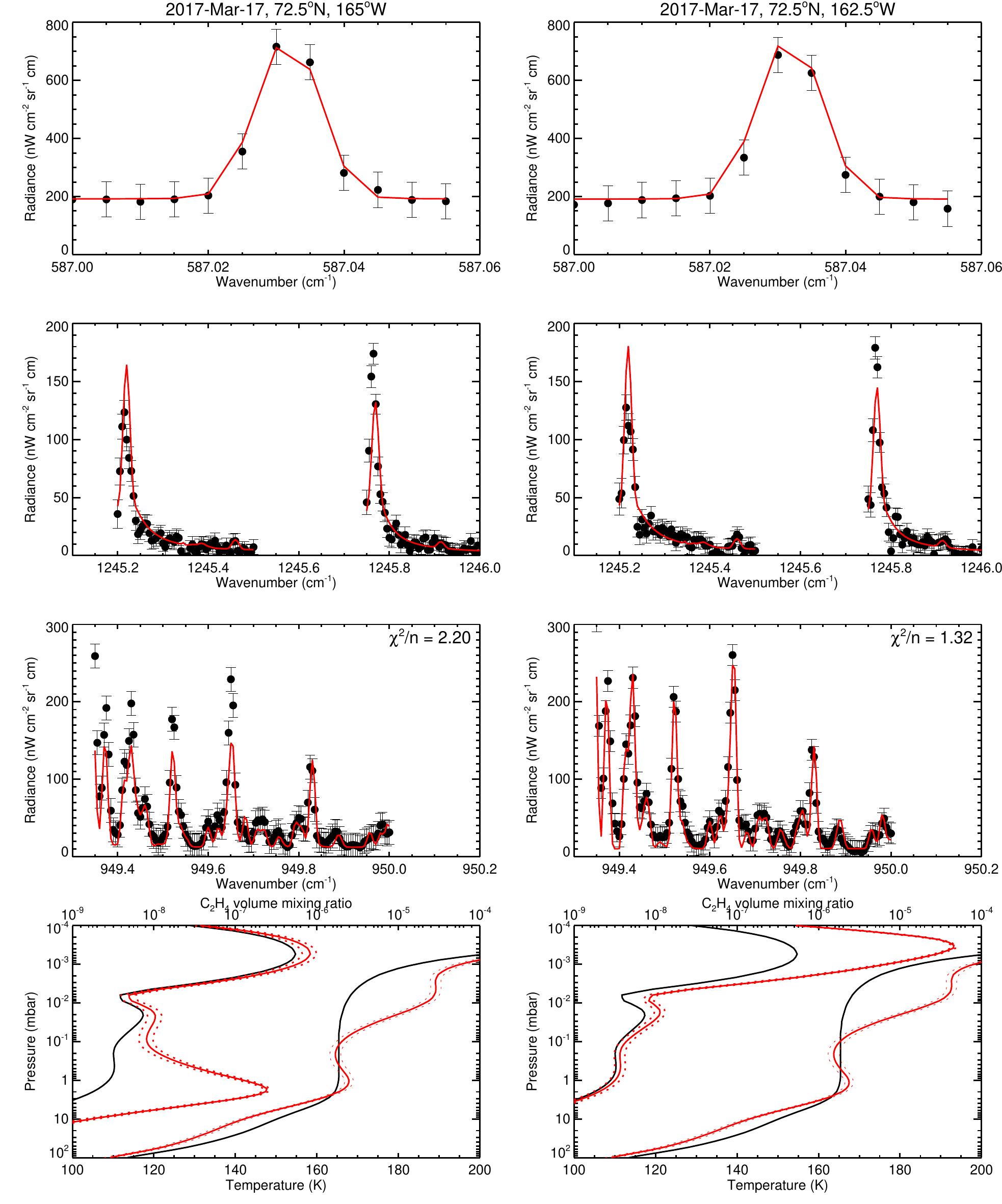}
\caption{Example retrievals of temperature and C$_2$H$_4$ at 72.5$^\circ$N, 165$^\circ$W (left column) and 162.5 $^\circ$W (right column), where the retrieval places the enhancement of C$_2$H$_4$ with respect to \textit{a priori} at different altitudes.  The 1st and 2nd row compare observed and modelled spectra at 587 cm$^{-1}$ and 1248 cm$^{-1}$, respectively, from which the temperature profile is retrieved.  The 3rd row compares observed and modelled spectra at 950 cm$^{-1}$ from which the C$_2$H$_4$ profile is retrieved.  The bottom panel shows retrieved profiles (solid red, with dotted red lines showing the 1-$\sigma$ uncertainty) of temperature, according to the lower axis, and C$_2$H$_4$, according to the upper axis.  Solid, black lines mark the \textit{a priori} profiles. }
\label{fig:bad_c2h4}
\end{center}
\end{figure*}

\begin{figure}[!th]
\begin{center}
\includegraphics[width=0.45\textwidth]{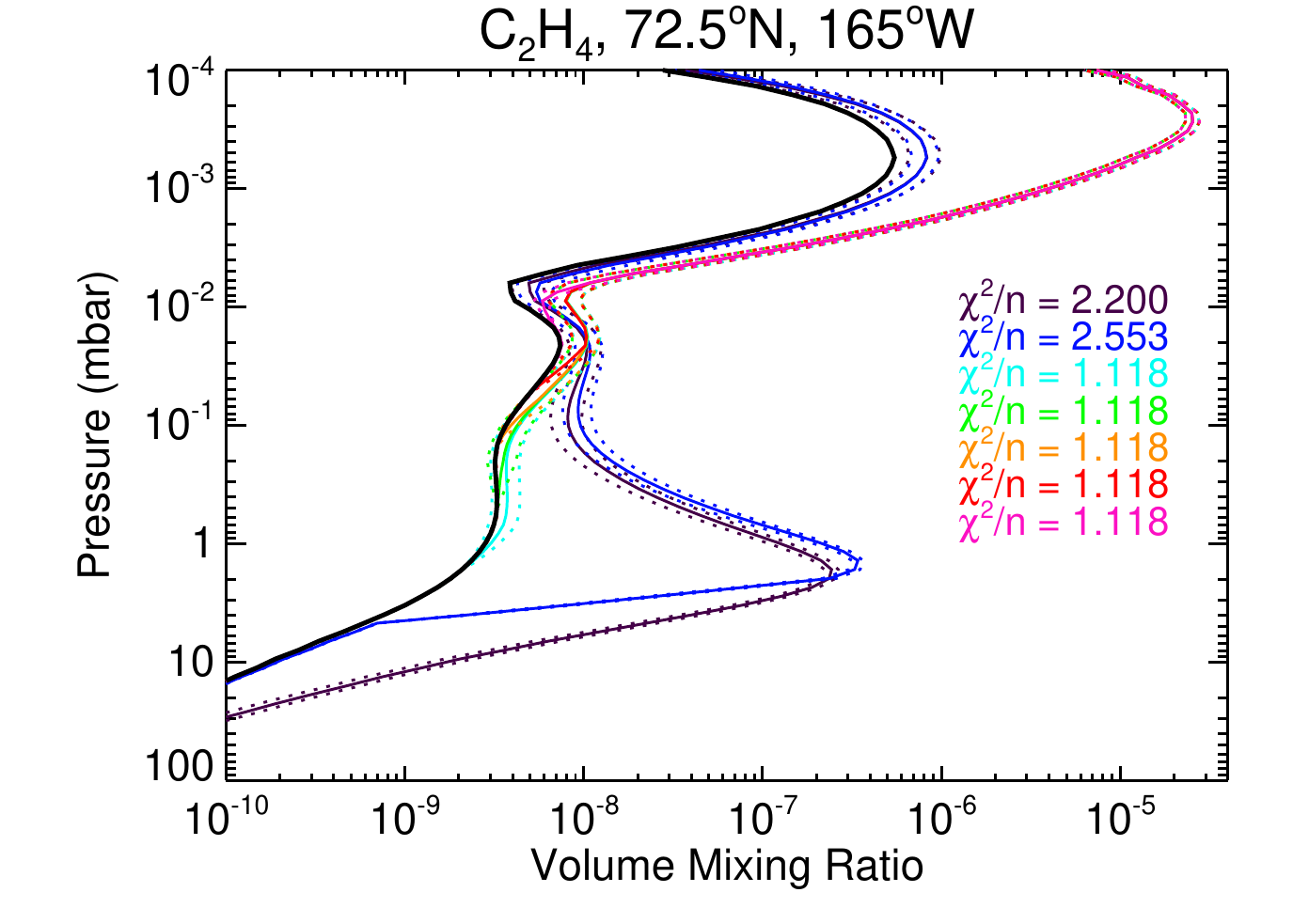}
\caption{Retrievals of the vertical profile of C$_2$H$_4$ at 72.5$^\circ$N, 165$^\circ$W using observations on March 19th, 2017.  Solid, black profiles represent the initial guess or \textit{a priori} profile.  Solid and dotted profiles represent retrieved profiles and uncertainty, respectively. Retrieved profiles are shown where the fractional uncertainty is constant at all altitudes (dark purple), and where the fractional uncertainty is reduced to 1\% abundances at pressures higher than 3 mbar (dark blue), 1 mbar (cyan), 0.3 mbar (green), 0.1 mbar (orange), 0.03 mbar (red), 0.01 mbar (pink).  The corresponding, reduced $\chi^2/n$ values are shown in the legend of each panel.   }
\label{fig:c2h4_vs_apriori}
\end{center}
\end{figure}

\newpage
\clearpage
\bibliography{ref}{}
\bibliographystyle{aasjournal}



\end{document}